\documentclass[12pt,preprint]{aastex}

\usepackage{emulateapj5}
\usepackage{aastexug}
\usepackage{amsmath}
\usepackage{graphicx}
\usepackage{subfigure}
\usepackage{apjfonts}

\begin{document}
\def\gapprox{\mathrel{\vcenter{\offinterlineskip \hbox{$>$}
    \kern 0.3ex \hbox{$\sim$}}}}
\def\lapprox{\mathrel{\vcenter{\offinterlineskip \hbox{$<$}
    \kern 0.3ex \hbox{$\sim$}}}}

\newcommand{\Dt}[0]{\bigtriangleup t}
\newcommand{\Dx}[0]{\bigtriangleup x}
\newcommand{\E}{\mathcal{E}}
\newcommand{\D}{\bigtriangleup}
\newcommand{\beq}{\begin{equation}}
\newcommand{\eeq}{\end{equation}}
\newcommand{\mm}[2]{\textrm{minmod}\left({#1},{#2}\right)}
\newcommand{\sign}{\textrm{sign}}
\newcommand{\nf}{\mathcal{F}}

\title{Athena: A New Code for Astrophysical MHD}

\author{James M. Stone, Thomas A. Gardiner\altaffilmark{1}}
\affil{Department of Astrophysical Sciences, Princeton University, Princeton,
NJ 08544}
\altaffiltext{1}{Present address: 3915 Rayado Pl NW, Albuquerque, NM 87114}
\author{Peter Teuben}
\affil{Department of Astronomy, University of Maryland, College Park, MD 20742}
\author{John F. Hawley and Jacob B. Simon}
\affil{Department of Astronomy, University of Virginia, Charlottesville VA}

\begin{abstract}
A new code for astrophysical magnetohydrodynamics (MHD) is described.
The code has been designed to be easily
extensible for use with static and adaptive mesh refinement.
It combines higher-order Godunov methods with the
constrained transport (CT) technique to enforce the divergence-free
constraint on the magnetic field.  Discretization is based on
cell-centered volume-averages for mass, momentum, and energy, and
face-centered area-averages for the magnetic field.  Novel features
of the algorithm include (1) a consistent framework for computing
the time- and edge-averaged electric fields used by CT to evolve the
magnetic field from the time- and area-averaged Godunov fluxes, (2)
the extension to MHD of spatial reconstruction schemes that involve a
dimensionally-split time advance, and (3) the extension to MHD of two
different dimensionally-unsplit integration methods.  Implementation of
the algorithm in both C and Fortran95 is detailed, including strategies
for parallelization using domain decomposition.  Results from a test suite
which includes problems in one-, two-, and three-dimensions for both
hydrodynamics and MHD are given, not only to demonstrate the fidelity
of the algorithms, but also to enable comparisons to other methods.
The source code is freely available for download on the web.

\end{abstract}


\section{Introduction}

Numerical methods are essential for the study of a very wide range
of problems in astrophysical fluid dynamics.  As such, the development
of more accurate and more capable algorithms, along with a description
of their implementation on modern parallel computer systems, is
important for progress in the field.  This paper describes a new
code for astrophysical magnetohydrodynamics (MHD) called Athena,
developed through a collaborative effort between the authors.

There are many numerical algorithms available for solving the
equations of compressible MHD.  One of the most successful is based
on operator splitting of the equations, with higher-order upwind
methods used for the advection terms, centered-differencing for the
remaining terms, and artificial viscosity for shock capturing.  This
algorithm, as implemented in for example the ZEUS code (Stone \&
Norman 1992a; b; Clarke 1996; Hayes et al. 2006), has been used for
many hundreds of applications in astrophysics.
The key advantage of the method is its simplicity, making
it easy to extend with more complex physics (for example, Stone \&
Norman 1992c; Turner \& Stone 2001; De Villiers \& Hawley 2003,
Hayes \& Norman 2003).

However, in the fifteen years since the development of ZEUS, static
and adaptive mesh refinement (SMR and AMR respectively) have emerged
as powerful techniques to resolve a large range in length scales
with grid-based methods.  Berger \& Colella (1990) have shown that
in order to prevent spurious reflections, it is important to enforce
conservation at internal boundaries between fine and coarse meshes.
Thus, operator-split methods that do not solve the dynamical equations
in conservation form such as ZEUS are unsuitable for use with SMR
or AMR.  This has been our primary motivation for the development
of Athena.

The numerical algorithms in Athena are based on directionally-unsplit,
higher-order Godunov methods, which not only are ideal for use with
both SMR and AMR, but also are superior for shock capturing and
evolving the contact and rotational discontinuities that are
typical of astrophysical flows.  Athena is neither the first nor the
only MHD code based on these methods which is designed for use with AMR;
others include RIEMANN (Balsara 2000), BATS-R-US (Powell et al. 1999;
Gombosi et al. 2004), AMRVAC (T\'{o}th 1996; Nool \& Keppens 2002), Nirvana (Ziegler 2005),
RAMSES (Fromang et al. 2006), PLUTO (Mignone
et al. 2007), and AstroBEAR (Cunningham et al. 2007).
While the wealth of papers describing AMR MHD codes demonstrates the
interest in and importance of these numerical methods, it also calls
into question the need for another paper describing yet another code.
However, it has been our experience that the precise details of the
algorithm can be important.  The numerical methods in Athena differ,
sometimes in small ways, and sometimes in substantial ways, from those
in other codes.  Our goals in developing Athena have been to write an
accurate, easy-to-use, adaptable, and maintainable code.  Our hope is
that the comprehensive description provided in this paper will be useful
to anyone who adopts, modifies, or builds upon the code, as well as for
others developing their own codes.

The development of Godunov methods for MHD has required substantial
progress over the past decade.  Most of the effort has focused on two
main areas: the multidimensional integration algorithm, and the method by
which the divergence-free constraint on the magnetic field is enforced.
Different options have been explored in different combinations, including
unconstrained directionally split integrators (Dai \& Woodward 1994),
or directionally split and unsplit integrators that use either a Hodge
projection to enforce the constraint (Zachary et al.
1994; Ryu et al. 1995; Balsara 1998; Crockett et al. 2005),
a non-conservative formulation that allows propagation and damping of
errors in the constraint (Powell 1994; Falle et al.
1998; Powell et al. 1999; Dedner et al. 2002), or some form of the
constrained transport (CT) algorithm of Evans \& Hawley (1988) to enforce
the constraint (Dai \& Woodward 1998; Ryu et al. 1998; Balsara \& Spicer
1999; T\'{o}th 2000, hereafter T2000; Pen et al. 2003; Londrillo
\& Del Zanna 2004; Ziegler 2004; Fromang et al. 2006;
Mignone et al. 2007; Cunningham et al. 2007).
T2000 provides a systematic comparison of many
of these techniques using an extensive test suite.

While the algorithms in Athena build upon this progress, they also
incorporate several innovations, including (1) the extension of two
different directionally unsplit integration algorithms to MHD, including
the corner transport upwind (CTU) method of Colella (1990 -- hereafter
the CTU+CT algorithm), and a simpler predictor-corrector method (see
the appendix in Falle 1981) similar to the MUSCL-Hancock scheme
described by van Leer (2006; Toro 1999 -- hereafter referred to as the
VL+CT algorithm), (2) the method by which the Godunov fluxes are used
to calculate the electric fields needed by CT, and (3) the extension of
the dimensionally-split spatial reconstruction scheme in the piecewise
parabolic method (PPM) of Colella \& Woodward (1984, hereafter CW) to
multidimensional MHD.  The mathematical foundations of these ingredients
for integration in two dimensions (2D) is presented in detail in
Gardiner \& Stone (2005, hereafter GS05), and for three dimensions (3D)
in Gardiner \& Stone (2008, hereafter GS08).  The focus of this paper
is on the implementation rather than the mathematics of the methods.

The use of two distinct unsplit integration algorithms in Athena,
namely the CTU+CT and the VL+CT algorithms, allows us to compare the
advantages and disadvantages of both.  We find the CTU+CT algorithm
is generally less diffusive and more accurate than VL+CT.  Thus, for
simplicity sake, the description in this paper will be based on the
CTU+CT algorithm.  However, for some applications the VL+CT algorithm has
definite advantages.  A complete description of the 3D VL+CT algorithm
implemented in Athena, including the results of tests in comparison to
the CTU+CT algorithm, is provided in a short companion paper (Stone \&
Gardiner 2008, hereafter SG08).

The primary goal of this paper is to provide a comprehensive
description of Athena that will serve as a reference for others to
adopt, modify, and extend the code for their own research.  As with
ZEUS, the source code is freely available from the web, along with
documentation and an extensive set of test problems that are useful
for any method.  The organization of this paper is as follows: \S2
introduces the equations of motion solved by Athena, while \S3
describes their finite-volume and finite-area discretizations.
Sections 4-6 describe
in detail the numerical algorithms in one, two, and three spatial
dimensions respectively, including details such as the reconstruction
algorithm, Riemann solvers used to compute upwind fluxes, and the
unsplit CTU+CT integrator used in multidimensions.  In \S7 the
implementation of the algorithms in both C and Fortran95 on parallel
computer systems is discussed.  The results of a comprehensive test
suite composed of problems in 1D, 2D, and 3D are given in \S8.
Finally, we summarize and discuss future extensions to the code in \S9.

\section{Basic Equations}

Athena implements algorithms which solve the equations of ideal MHD, which
can be written in conservative form as
\begin{eqnarray}
\frac{\partial \rho}{\partial t} +
{\bf\nabla\cdot} [\rho{\bf v}] & = & 0,
\label{eq:cons_mass} \\
\frac{\partial \rho {\bf v}}{\partial t} +
{\bf\nabla\cdot} \left[\rho{\bf vv} - {\bf BB} + {\sf P^*}\right] & = & 0,
\label{eq:cons_momentum} \\
\frac{\partial E}{\partial t} +
\nabla\cdot \left[(E + P^*) {\bf v} - {\bf B} ({\bf B \cdot v})\right] & = & 0 ,
\label{eq:cons_energy} \\
\frac{\partial {\bf B}}{\partial t} -
{\bf\nabla} \times \left({\bf v} \times {\bf B}\right) & = & 0,
\label{eq:induction} 
\end{eqnarray}
where ${\sf P^*}$ is a diagonal tensor with components
$P^* = P + B^{2}/2$ (with $P$ the gas pressure), $E$ is the total energy
density
\begin{equation}
  E = \frac{P}{\gamma -1} + \frac{1}{2}\rho v^{2} + \frac{B^{2}}{2},
\label{eq:total_energy}
\end{equation}
and $B^{2} = {\bf B} \cdot {\bf B}$.  The other symbols have their usual
meaning.  These equations are written in
units such that the magnetic permeability $\mu=1$.

An equation of state appropriate to an ideal gas, $P=(\gamma -1)e$
(where $\gamma$ is the ratio of specific heats, and $e$ is the
internal energy density), has been assumed in writing equation
\ref{eq:total_energy}.  For a barotropic equation of state $P=P(\rho)$
(for example, $P=C^{2}\rho$, where $C$
is the isothermal sound speed), both equations \ref{eq:cons_energy}
and \ref{eq:total_energy} are dropped from the system.  Of course, in this
case total energy is not conserved.  The algorithms implemented
in Athena can solve the equations of motion in four regimes: both
hydrodynamics or MHD with either an ideal-gas or barotropic equation
of state.  In each regime the system of equations to be solved is
different in number and form, however the same general numerical
techniques apply.  Extension of the numerical methods to a more
complex, e.g. tabular, equation of state is possible. 

It is useful to define vectors of the conserved and primitive variables,
${\bf U}$ and ${\bf W}$ respectively, with components
in Cartesian coordinates (for adiabatic MHD)
\begin{equation}
{\bf U}  = \left[ \begin{array}{c}
  \rho \\
  M_{x} \\
  M_{y} \\
  M_{z} \\
   E \\
   B_{x} \\
   B_{y} \\
   B_{z} \\ \end{array} \right] , \hspace*{0.5cm}
{\bf W}  = \left[ \begin{array}{c}
  \rho \\
  v_{x} \\
  v_{y} \\
  v_{z} \\
  P \\
   B_{x} \\
   B_{y} \\
   B_{z} \\ \end{array} \right] ,
\label{eq:cons+prim}
\end{equation}
where ${\bf M} = \rho {\bf v}$ is the momentum density.  The conservation
laws can now be written in a compact form (in Cartesian coordinates)
\begin{equation}
\frac{\partial {\bf U}}{\partial t} + 
    \frac{\partial \bf F}{\partial x} + \frac{\partial \bf G}{\partial y}
 + \frac{\partial\bf H}{\partial z} = 0,
\label{eq:cons_laws}
\end{equation}
where ${\bf F}$, ${\bf G}$, and ${\bf H}$ are vectors of fluxes in the
$x-$, $y-$, and $z-$directions respectively, with components
\begin{equation}
{\bf F} = \left[ \begin{array}{c}
  \rho v_{x} \\
  \rho v_{x}^{2} + P + B^{2}/2 - B_{x}^{2} \\
  \rho v_{x}v_{y} - B_{x}B_{y} \\
  \rho v_{x}v_{z} - B_{x}B_{z} \\
  (E + P^{*})v_{x} - ({\bf B}\cdot{\bf v})B_{x} \\
  0 \\
  B_{y}v_{x} - B_{x}v_{y} \\
  B_{z}v_{x} - B_{x}v_{z} \end{array} \right],
\label{eq:x-flux}
\end{equation}
\begin{equation}
{\bf G} = \left[ \begin{array}{c}
  \rho v_{y} \\
  \rho v_{y}v_{x} - B_{y}B_{x} \\
  \rho v_{y}^{2} + P + B^{2}/2 - B_{y}^{2} \\
  \rho v_{y}v_{z} - B_{y}B_{z} \\
  (E + P^{*})v_{y} - ({\bf B}\cdot{\bf v})B_{y} \\
  B_{x}v_{y} - B_{y}v_{x} \\
  0 \\
  B_{z}v_{y} - B_{y}v_{z} \end{array} \right],
\label{eq:y-flux}
\end{equation}
\begin{equation}
{\bf H} = \left[ \begin{array}{c}
  \rho v_{z} \\
  \rho v_{z}v_{x} - B_{z}B_{x} \\
  \rho v_{z}v_{y} - B_{z}B_{y} \\
  \rho v_{z}^{2} + P + B^{2}/2 - B_{z}^{2} \\
  (E + P^{*})v_{z} - ({\bf B}\cdot{\bf v})B_{z} \\
  B_{x}v_{z} - B_{z}v_{x} \\
  B_{y}v_{z} - B_{z}v_{y} \\
  0   \end{array} \right].
\label{eq:z-flux}
\end{equation}
Extension to curvilinear coordinates requires adding metric scale factors to
the definitions of the fluxes, or using a non-conservative formulation that
treats grid curvature as source terms, or a combination of these approaches.

For hydrodynamics, or for a barotropic equation of state (or for
both), the appropriate components of the vectors $\bf{U}$, ${\bf
W}$, and their fluxes are dropped.  While the last three components
of these vectors represents the induction equation in Cartesian
coordinates, the numerical algorithm actually used to evolve the magnetic
field is very different in comparison to that used for the other
components, as described in the next section.

\section{Discretization}
\label{sec:discretization}

Athena integrates the equations of motion on a regular, three-dimensional
Cartesian grid.  The continuous spatial coordinates ($x,y,z$) are
discretized into ($N_{x}, N_{y}, N_{z}$) cells within a finite
domain of size ($L_{x}, L_{y}, L_{z}$) in each direction respectively.
The cell denoted by indices ($i,j,k$) is centered at position ($x_{i},
y_{j}, z_{k}$).  For simplicity we describe the algorithm with the
assumption that the sizes of the grid cells in each direction,
$\delta x = L_{x}/N_{x}$, $\delta y = L_{y}/N_{y}$, and $\delta z =
L_{z}/N_{z}$ respectively, are uniform throughout the domain; the
numerical methods are easily extended to non-uniform grids.

Time is discretized into $N$ non-uniform steps between the initial
value $t_{0}$ and the final stopping time $t_{f}$.  Following the
usual convention, we use a superscript to denote the time level,
so $t^{n+1} -t^{n} = \delta t^{n}$.  Hereafter we drop the superscript
on $\delta t$ with the understanding that the time step may vary.

\subsection{Mass, Momentum, and Energy: Finite-Volumes}

Discretizations based on the {\em integral}, rather than the {\em
differential}, form of equations \ref{eq:cons_mass} through
\ref{eq:induction} have numerous advantages for flows that contain
shocks and discontinuities (LeVeque 2002).  Integration of equation
\ref{eq:cons_laws} over the volume of a grid cell, and over a
discrete interval of time $\delta t$ gives, after application of
the divergence theorem,
\begin{eqnarray}
  {\bf U}_{i,j,k}^{n+1} = {\bf U}_{i,j,k}^{n}
&-& \frac{\delta t}{\delta x} \left( {\bf F}^{n+1/2}_{i+1/2,j,k}
 - {\bf F}^{n+1/2}_{i-1/2,j,k} \right) \nonumber \\
&-& \frac{\delta t}{\delta y} \left( {\bf G}^{n+1/2}_{i,j+1/2,k}
 - {\bf G}^{n+1/2}_{i,j-1/2,k} \right) \nonumber \\
&-& \frac{\delta t}{\delta z} \left( {\bf H}^{n+1/2}_{i,j,k+1/2}
 - {\bf H}^{n+1/2}_{i,j,k-1/2} \right)
\label{eq:int_form}
\end{eqnarray}
where
\begin{equation}
{\bf U}_{i,j,k}^{n} = \frac{1}{\delta x \delta y \delta z}
 \int_{z_{k-1/2}}^{z_{k+1/2}} \int_{y_{j-1/2}}^{y_{j+1/2}}
 \int_{x_{i-1/2}}^{x_{i+1/2}} {\bf U}(x,y,z,t^{n})~{\rm d}x~{\rm d}y~{\rm d}z
\end{equation}
is a vector of volume-averaged variables, while
\begin{equation}
{\bf F}^{n+1/2}_{i-1/2,j,k} = \frac{1}{\delta y \delta z \delta t}
 \int_{t^n}^{t^{n+1}}
 \int_{z_{k-1/2}}^{z_{k+1/2}} \int_{y_{j-1/2}}^{y_{j+1/2}}
{\bf F}(x_{i-1/2},y,z,t)~{\rm d}y~{\rm d}z~{\rm d}t
\label{eq:average-xflux}
\end{equation}
\begin{equation}
{\bf G}^{n+1/2}_{i,j-1/2,k} = \frac{1}{\delta x \delta z \delta t}
 \int_{t^n}^{t^{n+1}}
 \int_{z_{k-1/2}}^{z_{k+1/2}} \int_{x_{i-1/2}}^{x_{i+1/2}}
{\bf G}(x,y_{j-1/2},z,t)~{\rm d}x~{\rm d}z~{\rm d}t
\label{eq:average-yflux}
\end{equation}
\begin{equation}
{\bf H}^{n+1/2}_{i,j,k-1/2} = \frac{1}{\delta x \delta y \delta t}
 \int_{t^n}^{t^{n+1}}
 \int_{y_{j-1/2}}^{y_{j+1/2}} \int_{x_{i-1/2}}^{x_{i+1/2}}
{\bf H}(x,y,z_{k-1/2},t)~{\rm d}x~{\rm d}y~{\rm d}t
\label{eq:average-zflux}
\end{equation}
are vectors of the time- and area-averaged fluxes.  We use the
convention here, and throughout this paper, that half-integer
subscripts denote the edges of the computational cells, that is
$x_{i-1/2}$ is the location of the interface between the cells
centered at $x_{i-1}$ and $x_{i}$.  Thus, the fluxes are evaluated
at (and are normal to) the faces of each grid cell (see figure 1).  Note the
half-integer superscript on the fluxes denote a time
average, rather than representing the flux evaluated at $t^{n+1/2}$.

As has been pointed out by many previous authors, equations
\ref{eq:int_form} through \ref{eq:average-zflux} are exact: to this point
no approximation has been made.  A numerical algorithm for MHD within the
finite-volume approach requires accurate and stable approximations for the
time- and area-averaged fluxes defined by equations \ref{eq:average-xflux}
through \ref{eq:average-zflux}.  In principle, one can approximate the
fluxes to any order of accuracy, although in practice most algorithms are
restricted to second-order.  A variety of authors are exploring the
use of higher than second-order accurate time- and spatial integration
(Londrillo \& Del Zanna 2000), especially in the context of
WENO schemes (Balsara \& Shu 2000; McKinney et al. 2007).  Higher-order
schemes improve the accuracy primarily in smooth flow, not in shocks or
discontinuities, and are more difficult to combine with AMR.  Based on
a set of 1D hydrodynamic test problems, Greenough \& Rider (2003)
conclude that a second-order Godunov scheme provides more accuracy per
computational cost than a fifth-order WENO scheme.  Although it is clear
that higher-order schemes will have advantages for some applications,
in Athena we shall restrict ourselves to second-order accuracy
in both space and time.

\subsection{Magnetic Field: Finite-Areas}

The last three components of equations \ref{eq:int_form} through
\ref{eq:average-zflux} are the finite-volume form of the induction
equation, which could be used to integrate the volume-averaged
components of the magnetic field.  Instead, in Athena we use an integral
form of the induction equation that is based on area- rather than
volume-averages.  In GS05, we have argued that area-averaging is the most
natural representation of the integral form of the induction equation.
This form conserves the magnetic flux through each grid cell, and as a
consequence automatically preserves the divergence free constraint on
the field (Evans \& Hawley 1988).

Integration of equation \ref{eq:induction} over the three orthogonal faces of
the cell located at ($i-1/2,j,k$), ($i,j-1/2,k$) and ($i,j,k-1/2$)
respectively, gives
\begin{eqnarray}
B^{n+1}_{x,i-1/2,j,k} = B^{n}_{x,i-1/2,j,k} & - &
 \frac{\delta t}{\delta y} (\E^{n+1/2}_{z,i-1/2,j+1/2,k} - \E^{n+1/2}_{z,i-1/2,j-1/2,k}) \nonumber \\
&+& \frac{\delta t}{\delta z} (\E^{n+1/2}_{y,i-1/2,j,k+1/2} - \E^{n+1/2}_{y,i-1/2,j,k-1/2}) 
\label{eq:CT-x}
\end{eqnarray}
\begin{eqnarray}
B^{n+1}_{y,i,j-1/2,k} = B^{n}_{y,i,j-1/2,k} & + &
\frac{\delta t}{\delta x} (\E^{n+1/2}_{z,i+1/2,j-1/2,k} - \E^{n+1/2}_{z,i-1/2,j-1/2,k}) \nonumber \\
& - & \frac{\delta t}{\delta z} (\E^{n+1/2}_{x,i,j-1/2,k+1/2} - \E^{n+1/2}_{x,i,j-1/2,k-1/2})
\label{eq:CT-y}
\end{eqnarray}
\begin{eqnarray}
B^{n+1}_{z,i,j,k-1/2} = B^{n}_{z,i,j,k-1/2} & - &
 \frac{\delta t}{\delta x} (\E^{n+1/2}_{y,i+1/2,j,k-1/2} - \E^{n+1/2}_{y,i-1/2,j,k-1/2}) \nonumber \\
&+& \frac{\delta t}{\delta y} (\E^{n+1/2}_{x,i,j+1/2,k-1/2} - \E^{n+1/2}_{x,i,j-1/2,k-1/2}) 
\label{eq:CT-z}
\end{eqnarray}
where
\begin{equation}
B^{n}_{x,i-1/2,j,k} = \frac{1}{\delta y \delta z}
 \int_{z_{k-1/2}}^{z_{k+1/2}} \int_{y_{j-1/2}}^{y_{j+1/2}}
B_x(x_{i-1/2},y,z,t^{n})~{\rm d}y~{\rm d}z
\label{eq:face-centered-Bx}
\end{equation}
\begin{equation}
B^{n}_{y,i,j-1/2,k} = \frac{1}{\delta x \delta z}
 \int_{z_{k-1/2}}^{z_{k+1/2}} \int_{x_{i-1/2}}^{x_{i+1/2}}
B_y(x,y_{j-1/2},z,t^{n})~{\rm d}x~{\rm d}z
\label{eq:face-centered-By}
\end{equation}
\begin{equation}
B^{n}_{z,i,j,k-1/2} = \frac{1}{\delta x \delta y}
 \int_{y_{j-1/2}}^{y_{j+1/2}} \int_{x_{i-1/2}}^{x_{i+1/2}}
B_z(x,y,z_{k-1/2},t^{n})~{\rm d}x~{\rm d}y
\label{eq:face-centered-Bz}
\end{equation}
are the area-averaged components of the magnetic field centered on
each of these faces, and
\begin{equation}
\E^{n+1/2}_{x,i,j-1/2,k-1/2} = \frac{1}{\delta x \delta t}
 \int_{t^n}^{t^{n+1}} \int_{x_{i-1/2}}^{x_{i+1/2}}
 \E_{x}(x,y_{j-1/2},z_{k-1/2},t)~{\rm d}x~{\rm d}t
\label{eq:emfx}
\end{equation}
\begin{equation}
\E^{n+1/2}_{y,i-1/2,j,k-1/2} = \frac{1}{\delta y \delta t}
 \int_{t^n}^{t^{n+1}} \int_{y_{j-1/2}}^{y_{j+1/2}}
 \E_{y}(x_{i-1/2},y,z_{k-1/2},t)~{\rm d}y~{\rm d}t
\label{eq:emfy}
\end{equation}
\begin{equation}
\E^{n+1/2}_{z,i-1/2,j-1/2,k} = \frac{1}{\delta z \delta t}
 \int_{t^n}^{t^{n+1}} \int_{z_{k-1/2}}^{z_{k+1/2}}
 \E_{z}(x_{i-1/2},y_{j-1/2},z,t)~{\rm d}z~{\rm d}t
\label{eq:emfz}
\end{equation}
are the components of the electric field ${\bf \E} = -{\bf v} \times
{\bf B}$ (the electromotive force, or emf) averaged along the
appropriate line element.  Note this discretization requires a
staggered grid, that is the area-averaged components of the magnetic
field are located at the faces (not the centers) of the cells.  Figure 1
shows the relative locations of the cell-centered volume-averaged
variables (${\bf U}_{i,j,k}$), the face-centered area-averaged
components of the magnetic field ($B_{x,i-1/2,j,k}, B_{y,i,j-1/2,k},
B_{z,i,j,k-1/2}$) the face-centered area-averaged fluxes (${\bf
F}_{i-1/2,j,k}, {\bf G}_{i,j-1/2,k}, {\bf H}_{i,j,k-1/2}$), and the
edge-centered line-averaged emfs ($\E_{x,i,j-1/2,k-1/2}$, etc.)..

There are many advantages to using a discretization of the induction equation
based on area- rather than volume-averages (GS05).  The most important
is that the finite-volume representation, i.e. the cell-volume average,
of the divergence-free constraint constructed using the time-advanced field
\begin{eqnarray}
(\nabla \cdot {\bf B})^{n+1}_{i,j,k} & = &
 \frac{B^{n+1}_{x,i+1/2,j,k} - B^{n+1}_{x,i-1/2,j,k}}{\delta x} \nonumber \\
&+& \frac{B^{n+1}_{y,i,j+1/2,k} - B^{n+1}_{y,i,j-1/2,k}}{\delta y} \nonumber \\
&+& \frac{B^{n+1}_{z,i,j,k+1/2} - B^{n+1}_{z,i,j,k-1/2}}{\delta z}
\label{eq:divB}
\end{eqnarray}
is kept zero by the discrete form of the induction equation, equations
\ref{eq:CT-x} through \ref{eq:CT-z}, provided of course it was zero at $t^{n}$ (Evans \& Hawley 1988).
Equivalently, the
CT algorithm conserves the magnetic flux through each grid cell.
The most serious disadvantage of using CT with face-centered fields
is that it complicates the implementation of the algorithm, and the
interface to AMR drivers.

Of course, there are many possible discretizations of the divergence-free
constraint, and the CT algorithm based on face-centered fields
described above preserves only one of them (equation \ref{eq:divB}).
T2000 has described an extension to CT which preserves the
constraint formulated using several different discretizations of
the divergence operator based on cell-centered fields.  It is
difficult to assess, for a given integration algorithm, whether
preserving one discretization is more important than any other.  We
have argued (GS05; GS08) that the discretization based on face-centered
fields is more consistent with the finite volume approach in that
it conserves the magnetic flux within each individual grid cell,
equivalently it conserves the volume integral of the density of
magnetic monopoles at the level of grid cells.  In addition, in
GS08 (see also \S\ref{sec:tests}) we describe a simple test problem based on
the advection of a field loop that is
sensitive to whether the discretization of the divergence-free
constraint that is preserved is consistent with the numerical
algorithm used to update the induction equation.  If not, growth
of net magnetic flux will be observed.

In Athena, the primary description of the magnetic field is taken to be
the face-centered area-averages equations \ref{eq:face-centered-Bx}
through \ref{eq:face-centered-Bz}.  However, cell-centered values for
the field are needed to construct the fluxes of momentum and energy
(equations \ref{eq:x-flux} through \ref{eq:z-flux}).  Here, we adopt
the second-order accurate averages
\begin{equation}
B_{x,i,j,k} = \frac{1}{2}(B_{x,i+1/2,j,k} + B_{x,i-1/2,j,k}),
\label{eq:face-to-cell-x}
\end{equation}
\begin{equation}
B_{y,i,j,k} = \frac{1}{2}(B_{y,i,j+1/2,k} + B_{y,i,j-1/2,k}),
\label{eq:face-to-cell-y}
\end{equation}
\begin{equation}
B_{z,i,j,k} = \frac{1}{2}(B_{z,i,j,k+1/2} + B_{z,i,j,k-1/2}).
\label{eq:face-to-cell-z}
\end{equation}
Operationally, the face-centered fields are {\em evolved} using equations
\ref{eq:CT-x} through \ref{eq:CT-z}, and at the end of each integration
step the cell-centered fields are {\em computed} using equations
\ref{eq:face-to-cell-x} through \ref{eq:face-to-cell-z}.  As shown
in GS05 (and discussed further in \S5.3), the relationship between
the face- and cell-centered components of the field given above
determines how their fluxes
(the time- and line-averaged emfs in equations \ref{eq:emfx} through
\ref{eq:emfz} and last three components of the time- and area-averaged
fluxes in equations \ref{eq:average-xflux} through \ref{eq:average-zflux}
respectively) are computed from one another.

\section{One-dimensional integration algorithm}
\label{sec:oned-integration}

It is useful (and standard) pedagogy to describe the algorithm for
integration of the equations of motion in 1D first, before introducing
methods for multidimensions.   However, for MHD, this approach can
be misleading.  In 1D the divergence-free constraint reduces to the
condition that the longitudinal component of the magnetic field be
constant, the CT algorithm is not needed, and the discrete forms
of the induction equation for the area- and volume-averaged fields
are identical.  As a consequence, 1D algorithms for MHD are a simple
extension of those for hydrodynamics.  Moreover, 1D test problems
for MHD will not reveal errors associated with the development of
a non-solenoidal field.  Any rigorous test suite for MHD must be
based on multidimensional problems.

Nonetheless, we begin a description of the algorithms in Athena
with the 1D integrator as it allows us to introduce basic components,
such as Riemann solvers and methods for spatial reconstruction, required in
multidimensions.  We emphasize that the integrators for 2D and 3D
MHD, described in detail in \S5 and \S6 respectively, are substantially
different and more complex than the 1D integrator introduced here.

In 1D, the equations of adiabatic MHD can be written in Cartesian 
coordinates as
\begin{equation}
\frac{\partial {\bf q}}{\partial t} + \frac{\partial \bf f}{\partial x} = 0
\label{eq:1D-MHD}
\end{equation}
where the vectors of conserved variables and their fluxes are
\begin{equation}
{\bf q}  = \left[ \begin{array}{c}
  \rho \\
  M_{x} \\
  M_{y} \\
  M_{z} \\
   E \\
   B_{y} \\
   B_{z} \\ \end{array} \right] , \hspace*{0.5cm}
{\bf f} = \left[ \begin{array}{c}
  \rho v_{x} \\
  \rho v_{x}^{2} + P + B^{2}/2 - B_{x}^{2} \\
  \rho v_{x}v_{y} - B_{x}B_{y} \\
  \rho v_{x}v_{z} - B_{x}B_{z} \\
  (E + P^{*})v_{x} - ({\bf B}\cdot{\bf v})B_{x} \\
  B_{y}v_{x} - B_{x}v_{y} \\
  B_{z}v_{x} - B_{x}v_{z} \end{array} \right],
\label{eq:cons1D}
\end{equation}
Note these are identical to equations \ref{eq:cons+prim} and
\ref{eq:x-flux} with the sixth component dropped.  We introduce the
notation that vectors denoted by lower-case letters are in one
spatial dimension (and therefore contain 7 components for adiabatic
MHD, rather than 8 for the same vectors written in full 3D).
It is important
to remember that the components of the 1D vectors defined in equation
\ref{eq:cons1D} will change depending on direction.  For example, in the
$y-$direction for ideal MHD, the order of the three components are permuted (so
the second to fourth components become $M_{y}$, $M_{z}$, and $M_{x}$
respectively), and the sixth and seventh components
become $B_{z}$ and $B_{x}$ respectively.

The finite-volume discretization of equation
\ref{eq:1D-MHD} proceeds as described in \S 3.1, giving
\begin{equation}
  {\bf q}_{i}^{n+1} = {\bf q}_{i}^{n} 
- \frac{\delta t}{\delta x} \left( {\bf f}^{n+1/2}_{i+1/2}
 - {\bf f}^{n+1/2}_{i-1/2} \right)
\label{eq:1D_int_form}
\end{equation}
where
\begin{equation}
{\bf q}_{i}^{n} = \frac{1}{\delta x}
 \int_{x_{i-1/2}}^{x_{i+1/2}} {\bf q}(x,t^{n})~{\rm d}x
\label{eq:1D-volume-averages}
\end{equation}
is a vector of volume-averaged variables, while
\begin{equation}
{\bf f}^{n+1/2}_{i-1/2} = \frac{1}{\delta t}
 \int_{t^n}^{t^{n+1}}
{\bf f}(x_{i-1/2},t)~{\rm d}t
\label{eq:1D-average-xflux}
\end{equation}
are the time-averaged fluxes at the interface located at $x_{i-1/2}$.

In a Godunov method, the time-averaged fluxes (equation
\ref{eq:1D-average-xflux}) are computed using a Riemann solver (see
Toro 1999 for an introduction to the subject).  Figure 2 illustrates
the process (see also LeVeque 2002).  Starting from the 1D
volume-averages stored at cell-centers ${\bf q}^{n}_{i}$ a spatial
reconstruction scheme is used to construct the conserved quantities to
the left- and right-sides of the interface, ${\bf q}_{L,i-1/2}$ and ${\bf
q}_{R,i-1/2}$ respectively.  For the CTU+CT integrator, the reconstruction is performed in
the primitive variables, and includes a time-advance using characteristic
variables, with ${\bf q}_{L,i-1/2}$ and ${\bf q}_{R,i-1/2}$
computed from the resulting interpolants (this step will be described
in detail in \S \ref{sec:reconstruction}).  Due to the slope-limiters
used to keep the interpolants non-oscillatory, the left- and right-states
${\bf q}_{L,i-1/2}$ and ${\bf q}_{R,i-1/2}$ will not be equal, except in
smooth flow.  Thus, they define a Riemann problem, the solution to which
is the time evolution of the various waves, and the intermediate states
that connect them, that propagate away from the interface.  The solution
to the Riemann problem, evaluated at the location of the interface, can
be used to construct the time-averaged flux (details of the calculation
of fluxes using Riemann solvers is given in \S \ref{sec:fluxes}).

\subsection{Steps in the 1D Algorithm}
\label{sec:oned-steps}

The 1D algorithm outlined above can be summarized by the following steps:

\noindent

{\em Step 1.}  From ${\bf q}^{n}_{i}$, the volume averages at time level $n$,
compute the left- and right-states ${\bf q}_{L,i-1/2}$ and
 ${\bf q}_{R,i-1/2}$ at every interface using one of the spatial
reconstruction algorithms described below in \S \ref{sec:reconstruction}.

{\em Step 2.} Compute the time-averaged fluxes at every interface ${\bf
f}_{i-1/2}^{n+1/2} = \nf({\bf q}_{L,i-1/2},{\bf q}_{R,i-1/2},B_{x,i-1/2})$
using one of the Riemann solvers described in \S \ref{sec:fluxes}.
Note the face-centered longitudinal component of the magnetic field is
passed to the Riemann solver as a parameter.

{\em Step 3.} Update the cell-centered conserved variables and the transverse
components of the magnetic field using
the finite-volume difference equation in 1D, equation \ref{eq:1D_int_form}.

{\em Step 4.} Increment the time: $t^{n+1} = t^{n} + \delta t$.  Compute
a new timestep that satisfies an estimate of the CFL stability condition
based on wavespeeds at cell centers
\begin{equation}
\delta t = C_{\circ} \delta x / \max ( \vert v^{n+1}_{x,i} \vert + C^{n+1}_{fx,i} )
\end{equation}
where $C_{\circ} \le 1$ is the CFL number, $C^{n+1}_{fx,i}$ is the fast
magnetosonic speed in the $x-$direction, evaluated using the updated
quantities, and the maximum is taken over all grid cells.
Note this is only an estimate of the CFL
stability condition, since the wavespeeds used in the Riemann solver
can be different from those computed from the cell-centered values.

{\em Step 5.} Repeat steps 1-4 until the stopping criterion is reached,
i.e.. $t^{n+1} \ge t_f$ 

The entire 1D integration algorithm is summarized by the flow chart shown
in figure 3.

\subsection{MHD Interface States}
\label{sec:reconstruction}

The first step in the 1D algorithm is to compute the left- and
right-states ${\bf q}_{L,i-1/2}$ and ${\bf q}_{R,i-1/2}$ that define the
Riemann problem at the interface located at $x_{i-1/2}$.  (Note that in
our notation the left-state ${\bf q}_{L,i-1/2}$ is actually on the {\em
right} side of the cell center at $x_{i-1}$, while the right-state ${\bf
q}_{R,i-1/2}$ is on the {\em left} side of the cell center at $x_i$, see
figure 2).  The reconstruction is inherently 1D, and therefore is based on
the vector of conserved variables in 1D (equation \ref{eq:cons1D}).  This
vector contains only the transverse components of the field: in 1D these
are cell-centered quantities.  For reconstruction in multidimensions,
the cell-centered averages of the face-centered transverse components
of the field (for example, equations \ref{eq:face-to-cell-y} and
\ref{eq:face-to-cell-z} for reconstruction in the $x-$direction) would
be used.  When the longitudinal component of the field is needed, the
area-averaged value stored at the appropriate interface is adopted.
The fact that the longitudinal component of the field does not need to
be reconstructed from cell-centered values is a further advantage of
the CT algorithm based on staggered (face-centered) fields; it avoids
the problem of the longitudinal component being discontinuous at the
interface due to slope-limited reconstruction from cell centers.

When the CTU+CT unsplit integrator is used in Athena, the second- and
third-order reconstruction algorithms described below include both spatial
interpolation with slope-limiting in the characteristic variables, and
a characteristic evolution of the linearized system in the primitive
variables.  We have found these steps help to make the reconstruction
less oscillatory.  However, they also require an eigenvalue decomposition
of the linearized equations of motion in the primitive variables.
Appendix A catalogs the eigenvalues and left- and right-eigenvectors
for adiabatic and isothermal hydrodynamics and MHD in the primitive
variables needed for this approach.  For more complex physics (e.g.,
relativistic MHD) this eigenvalue decomposition may be difficult.
One advantage of the VL+CT integrator described in SG08 is that it
does not require a characteristic evolution in the reconstruction step.
This avoids the need for an eigenvalue decomposition in the primitive
variables, and therefore this integrator may be a better choice for
more complex physics.  The interface state algorithm used in the VL+CT
algorithm is described more fully in SG08.

\subsubsection{Piecewise constant (first-order) reconstruction}

The simplest possible reconstruction algorithm is to assume the primitive
variables are piecewise constant
within each cell (implying the conserved variables are also piecewise
constant), leading to the first-order method
\begin{eqnarray}
{\bf q}_{L,i-1/2} & = & {\bf q}_{i-1} \\
{\bf q}_{R,i-1/2} &=& {\bf q}_{i} \nonumber
\end{eqnarray}
First-order reconstruction is far too diffusive for applications,
however it is useful for testing, or in those circumstances when extra
diffusion is in fact desired.

\subsubsection{Piecewise linear (second-order) reconstruction}
\label{sec:2nd-order-recon}

A better approximation is to assume the primitive variables vary linearly
within each cell (meaning that the profile of the conserved variables
within a cell may be steeper than linear).  This approximation leads to
the second-order reconstruction algorithm used with the CTU+CT unsplit
integrator that is given by the following steps:

{\em Step 1.}  Compute the eigenvalues and eigenvectors of the
linearized equations in the primitive variables using ${\bf w}_i$,
the cell-centered primitive variables in 1D (which differs from
${\bf W}_i$ defined in equation \ref{eq:cons+prim} only in that it
lacks the longitudinal component of the magnetic field).  Explicit
expressions for these are given in Appendix A.

{\em Step 2.}  Compute the left-, right-, and centered-differences of
the primitive variables ${\bf w}_i$
\begin{eqnarray}
\delta {\bf w}_{L,i} & = & {\bf w}_{i}-{\bf w}_{i-1}, \nonumber \\
\delta {\bf w}_{R,i}& =& {\bf w}_{i+1}-{\bf w}_{i}, \\
\delta {\bf w}_{C,i} & = & ({\bf w}_{i+1}-{\bf w}_{i-1})/2 \nonumber
\end{eqnarray}
(Note that in these equations the subscripts $L$, $R$, and $C$ refer to
locations relative to the cell-center at $x_{i}$.)

{\em Step 3.}  Project the left, right, and centered differences onto the
characteristic variables
\begin{eqnarray}
\delta {\bf a}_{L,i}  & = & {\sf L}({\bf w}_i)\cdot \delta {\bf w}_{L,i}, \nonumber \\
\delta {\bf a}_{R,i} & = & {\sf L}({\bf w}_i)\cdot \delta {\bf w}_{R,i}, \\
\delta {\bf a}_{C,i} & = & {\sf L}({\bf w}_i)\cdot \delta {\bf w}_{C,i} \nonumber
\end{eqnarray}
where ${\sf L}({\bf w}_i)$ is a matrix whose rows are the appropriate
left-eigenvectors computed in Step 1.

{\em Step 4.}  Apply monotonicity constraints to the differences in the
characteristic variables, so that the characteristic reconstruction is total
variation diminishing (TVD), e.g. see LeVeque (2002).
\begin{equation}
\delta {\bf a}^{m}_i = {\rm SIGN}(\delta {\bf a}_{C,i}) \min (2 \vert \delta  {\bf a}_{L,i} \vert, 2 \vert \delta {\bf a}_{R,i} \vert , \vert\delta {\bf a}_{C,i} \vert )
\end{equation}

{\em Step 5.} Project the monotonized difference in the characteristic variables
back onto the primitive variables
\begin{equation}
\delta {\bf w}^{m}_i = \delta {\bf a}^{m}_i \cdot {\sf R}({\bf w}_i)
\end{equation}
where ${\sf R}({\bf w}_i)$ is a matrix whose columns are the
appropriate right-eigenvectors computed in Step 1.

{\em Step 6.}  Compute the left- and right-interface values using the
monotonized difference in the primitive variables
\begin{equation}
\hat{\bf w}_{L,i+1/2} = {\bf w}_{i} + \left[ \frac{1}{2} - 
\max (\lambda^{M}_{i},0)\frac{\delta t}{2 \delta x} \right] \delta {\bf w}_{i}^{m}
\end{equation}
\begin{equation}
\hat{\bf w}_{R,i-1/2} = {\bf w}_{i} - \left[ \frac{1}{2} -
\min (\lambda^{0}_i,0)\frac{\delta t}{2 \delta x} \right] \delta {\bf w}_{i}^{m}
\end{equation}
where $\lambda_{i}^{M}$ and $\lambda_{i}^0$ are the largest and
smallest eigenvalues computed in Step 1 respectively, at the appropriate cell
center.
Note these values are at different cell faces, with $\hat{\bf w}_{L,i+1/2}$
($\hat{\bf w}_{R,i-1/2}$) located to the {\em right} ({\em left})
of the cell center at $x_{i}$.

{\em Step 7.}  Perform the characteristic tracing, that is subtract from the
integral performed in step 6 that part of each wave family that does not
reach the interface in $\delta t/2$, using (CW; Colella 1990)
\begin{equation}
{\bf w}_{L,i+1/2} = \hat{\bf w}_{L,i+1/2} + \frac{\delta t}{2 \delta x}
\sum_{\lambda^{\alpha} > 0} \left( (\lambda_{i}^{M} - \lambda_{i}^{\alpha})
{\sf L}^{\alpha} \cdot \delta {\bf w}_{i}^{m} \right) {\sf R}^{\alpha}
\label{eq:tracing-l}
\end{equation}
\begin{equation}
{\bf w}_{R,i-1/2} = \hat{\bf w}_{R,i-1/2} + \frac{\delta t}{2 \delta x}
\sum_{\lambda^{\alpha} < 0} \left( (\lambda_{i}^{0} - \lambda_{i}^{\alpha})
{\sf L}^{\alpha} \cdot \delta {\bf w}_{i}^{m} \right) {\sf R}^{\alpha}
\label{eq:tracing-r}
\end{equation}
where the sums are taken only over those waves that propagate towards the
interface (i.e., whose eigenvalue has the appropriate sign), and 
${\sf L}^{\alpha}$ and ${\sf R}^{\alpha}$ are the rows and columns of the left-
and right-eigenmatrices respectively corresponding to $\lambda^{\alpha}$.

When using approximate Riemann solvers that average over intermediate
states (like the HLL family of solvers), it is also necessary to include
a correction for waves which propagate away from the interface in order
to make the algorithm higher than first-order.  This is because either the
right-interface state (if the wavespeed is positive) or the left-interface
state (if the wave speed is negative) will not include the half-timestep
predictor evolution in the reconstruction, and will thus be first-order.
Since the numerical flux in the HLL solver is given by a weighted average
of the flux in the left-interface state and the right-interface state
for such waves, the flux itself will be first-rder.  Specifically, an
additional term $\Delta {\bf w}_{L,i+1/2}$ and $\Delta {\bf w}_{R,i-1/2}$
is added to each of equations \ref{eq:tracing-l} and \ref{eq:tracing-r}
respectively, where these terms are
\begin{equation}
\Delta{\bf w}_{L,i+1/2} = - \frac{\delta t}{2 \delta x}
\sum_{\lambda^{\alpha} < 0} \left( (\lambda_{i}^{\alpha} - \lambda_{i}^{M})
{\sf L}^{\alpha} \cdot \delta {\bf w}_{i}^{m} \right) {\sf R}^{\alpha}
\label{eq:HLL-correction-wl}
\end{equation}
\begin{equation}
\Delta{\bf w}_{R,i-1/2} = - \frac{\delta t}{2 \delta x}
\sum_{\lambda^{\alpha} > 0} \left( (\lambda_{i}^{\alpha} - \lambda_{i}^{0})
{\sf L}^{\alpha} \cdot \delta {\bf w}_{i}^{m} \right) {\sf R}^{\alpha}
\label{eq:HLL-correction-wr}
\end{equation}
We emphasize these terms are {\em not} added when the Roe or exact solvers
are used.

{\em Step 8.}  Finally, convert the left- and right-states in the primitive
to the conserved variables, ${\bf q}_{L,i-1/2}$ and ${\bf q}_{R,i-1/2}$.

\subsubsection{Piecewise parabolic (third-order) reconstruction}
\label{sec:3rd-order-recon}

Although the numerical algorithms in Athena are formally only
second-order accurate, we have found that using third-order accurate
spatial reconstruction can lower the amplitude of the truncation
error and increase the accuracy of
the solution.  Thus, we have implemented the PPM interface state
algorithm of CW in Athena.  In \S \ref{sec:tests}, we provide a
quantitative comparison of both the second-order (PLM) and
third-order (PPM) reconstruction algorithms for smooth and discontinuous
solutions in 1D, 2D and 3D.

The PPM reconstruction algorithm 
consists of the following steps.

{\em Steps 1 through 5}.  These steps are identical to the first five
steps in the second-order algorithm, see \S \ref{sec:2nd-order-recon}.

{\em Step 6.} Use parabolic interpolation to compute values at the
left- and right-side of each cell center
\begin{eqnarray}
{\bf w}_{L,i} & = & ({\bf w}_{i} + {\bf w}_{i-1})/2 - (\delta {\bf w}^{m}_{i} + \delta {\bf w}^{m}_{i-1})/6  \nonumber \\
{\bf w}_{R,i} & = & ({\bf w}_{i+1} + {\bf w}_{i})/2 - (\delta {\bf w}^{m}_{i+1} + \delta {\bf w}^{m}_{i})/6 
\end{eqnarray}
where in the above, the subscript $L$ ($R$) refers to the
left (right) side of cell center at $x_{i}$.

{\em Step 7.} Apply further monotonicity constraints to ensure the
values on the left- and right-side of cell center lie between
neighboring cell-centered values (CW equation 1.10).  These can
be written as a series of conditional statements:
\begin{equation}
{\rm if}~~({\bf w}_{R,i}-{\bf w}_{i})({\bf w}_{i} - {\bf w}_{L,i}) \le 0 \nonumber
\end{equation}
\begin{equation}
{\bf w}_{L,i} = {\bf w}_{i} \nonumber
\end{equation}
\begin{equation}
{\bf w}_{R,i} = {\bf w}_{i} \nonumber
\end{equation}
\begin{equation}
{\rm if}~~6({\bf w}_{R,i}-{\bf w}_{L,i})({\bf w}_{i} - ({\bf w}_{L,i}+{\bf w}_{R,i})/2) > ({\bf w}_{R,i}-{\bf w}_{L,i})^2 \nonumber
\end{equation}
\begin{equation}
{\bf w}_{L,i} = 3{\bf w}_{i} - 2{\bf w}_{R,i} \nonumber
\end{equation}
\begin{equation}
{\rm if}~~6({\bf w}_{R,i}-{\bf w}_{L,i})({\bf w}_{i} - ({\bf w}_{L,i}+{\bf w}_{R,i})/2) < -({\bf w}_{R,i}-{\bf w}_{L,i})^2 \nonumber
\end{equation}
\begin{equation}
{\bf w}_{R,i} = 3{\bf w}_{i} - 2{\bf w}_{L,i} \nonumber
\end{equation}
These conditions are applied independently to each component of ${\bf w}$.

{\em Step 8.} Compute the coefficients for the monotonized
parabolic interpolation function,
\begin{equation}
\delta {\bf w}^{m}_{i} = {\bf w}_{R,i} - {\bf w}_{L,i}, \hspace*{0.5cm} 
{\bf w}_{6,i} = 6({\bf w}_{i} - ({\bf w}_{L,i} + {\bf w}_{R,i})/2)
\end{equation}

{\em Step 9.} Compute the left- and right-interface values using
monotonized parabolic interpolation (CW equation 1.12)
\begin{equation}
\hat{\bf w}_{L,i+1/2} = {\bf w}_{R,i} - 
 \lambda^{\rm max} \frac{\delta t}{2 \delta x} \left[ \delta {\bf w}_{i}^{m}
- \left( 1 - \lambda^{\rm max} \frac{2 \delta t}{3 \delta x} \right)
{\bf w}_{6,i} \right]
\end{equation}
\begin{equation}
\hat{\bf w}_{R,i-1/2} = {\bf w}_{L,i} + 
 \lambda^{\rm min} \frac{\delta t}{2 \delta x} \left[ \delta {\bf w}_{i}^{m}
+ \left( 1 - \lambda^{\rm min} \frac{2 \delta t}{3 \delta x} \right)
{\bf w}_{6,i} \right]
\end{equation}
where $\lambda^{\rm max} = \max (\lambda_{i}^{M},0)$ and
$\lambda^{\rm min}= \min (\lambda_{i}^{0},0)$ respectively, and
$\lambda^{M}_{i}$ and $\lambda^{0}_{i}$ are the largest and smallest eigenvalues
computed in Step 1 respectively.
Note these values are at different cell faces, with $\hat{\bf w}_{L,i+1/2}$
($\hat{\bf w}_{R,i-1/2}$) located to the {\em right} ({\em left})
of the cell center at $x_{i}$.

{\em Step 10.}  Perform the characteristic tracing, that is subtract from the
integral performed in step 9 that part of each wave family that does not
reach the interface in $\delta t/2$ (CW; Colella 1990), using
\begin{equation}
{\bf w}_{L,i+1/2}  =  \hat{\bf w}_{L,i+1/2} +
\sum_{\lambda^{\alpha} > 0} \left[ {\sf L}^{\alpha}  \left(
A (\delta {\bf w}_{i}^{m} - {\bf w}_{6,i})  + B {\bf w}_{6,i} \right) \right] 
 {\sf R}^{\alpha}
\end{equation}
\begin{equation}
{\bf w}_{R,i+1/2}  =  \hat{\bf w}_{R,i+1/2} +
\sum_{\lambda^{\alpha} < 0} \left[ {\sf L}^{\alpha} \left(
C (\delta {\bf w}_{i}^{m} + {\bf w}_{6,i}) + D {\bf w}_{6,i} \right) \right]
 {\sf R}^{\alpha}
\end{equation}
where in the above
\begin{equation}
A = \frac{\delta t}{2 \delta x}(\lambda^{M} - \lambda^{\alpha}) \hspace*{.5cm}
B = \frac{1}{3} \left[ \frac{\delta t}{\delta x} \right]^{2}
(\lambda^{M}\lambda^{M} - \lambda^{\alpha}\lambda^{\alpha})
\nonumber
\end{equation}
\begin{equation}
C = \frac{\delta t}{2 \delta x}(\lambda^{0} - \lambda^{\alpha}) \hspace*{.5cm}
D = \frac{1}{3} \left[ \frac{\delta t}{\delta x} \right]^{2}
(\lambda^{0}\lambda^{0} - \lambda^{\alpha}\lambda^{\alpha})
\nonumber
\end{equation}
where the sums are taken only over those waves that propagate towards the
interface (i.e., whose eigenvalue has the appropriate sign), and 
${\sf L}^{\alpha}$ and ${\sf R}^{\alpha}$ are the rows and columns of the left-
and right-eigenmatrices respectively corresponding to $\lambda^{\alpha}$.

Once again, when using the HLL family of solvers, it is necessary to
add a correction for waves which propagate away from the interface (as
was required in step 7 of the PLM integration).  These
terms are identical to those in equations \ref{eq:HLL-correction-wl}
and \ref{eq:HLL-correction-wr}, which are correct to second-order.
Again, we emphasize
these terms are not added when the Roe or exact solvers are used.

{\em Step 11.}  Finally, convert the left- and right-states in the primitive
to the conserved variables, ${\bf q}_{L,i-1/2}$ and ${\bf q}_{R,i-1/2}$.

An important ingredient of the reconstruction algorithm is the slope
limiters used in steps 4 and 7.  It is well-known that these limiters
clip extrema in the solutions.  We have also implemented the limiters
described in Colella \& Sekora (2007, hereafter CS), which are designed
to prevent clipping of extrema.  We find for some tests, the CS limiters
significantly improve the solution compared to the original PPM limiters
used above.  For the test results shown in \S\ref{sec:tests} we will
always indicate if the CS limiters are used.  The lesson, however, is
that improving the convergence rate of the reconstruction algorithm
is not always the best way to improve the overall accuracy of the solution.

\subsection{Godunov Fluxes}
\label{sec:fluxes}

The second step in the 1D algorithm is to compute time-averaged
fluxes using a Riemann solver.  Exact Riemann solvers for MHD (e.g.
Ryu \& Jones 1995) are generally too expensive for practical
computations with current hardware.
Moreover, since the full solution to the Riemann
problem over all space-time is not required, but only the time-integral
of the solution along the line $x=x_{i-1/2}$ (which gives the flux
through the interface), approximate solvers which provide an accurate
estimate of the {\em flux} are all that is needed.  In fact, it is
not even necessary to use the same solver to compute the flux at
every interface in the grid.  Instead, simple solvers can
be used in smooth regions, while more robust (and expensive)
solvers are adopted only when needed, for example in highly nonlinear
flow where simple solvers fail (such as strong rarefactions).
Since the latter generally
occupy only a tiny fraction of the total number of interfaces over
the whole grid, this strategy can be very cost effective.

A wide variety of approximate Riemann solvers for MHD are possible,
including nonlinear solvers such as the HLL flux (Harten et al.
1983), the HLLD flux (Miyoshi \& Kusano 2005), Toro's FORCE flux
(Toro 1999), Roe's linear solver (Roe 1991) extended to
MHD (Cargo \& Gallice 1997), as well as MHD solvers based
on other approximations (e.g, Dai \& Woodward 1994; 1995; Zachary et al.
1994).  A range of solvers is implemented in Athena,
including exact solvers in the simplest cases (isothermal hydrodynamics).
In the subsections below we describe some of the most useful.

Finally, it is important to emphasize that Godunov methods do not {\em
require} expensive solvers based on complex characteristic decompositions.
Simple solvers based on the local Lax-Friedrichs (LLF) or HLL fluxes
that are typically adopted in other methods can
also be used.  Generally, the reason for adopting more complex and
expensive Riemann solvers is that they reduce dissipation, especially
in the neighborhood of discontinuities in the intermediate waves.

\subsubsection{HLL Solvers}

The simplest Riemann solver implemented in Athena uses the HLL fluxes
as described by Einfeldt et al. (1991), hereafter termed the HLLE solver.
The HLLE flux at the interface $x_{i-1/2}$ is defined as
\begin{equation}
\nf^{\rm HLLE}_{i-1/2} =
\frac{b^{+}{\bf f}_{L,i-1/2} - b^{-}{\bf f}_{R,i-1/2}}{b^{+}-b^{-}}
+ \frac{b^{+} b^{-}}{b^{+}-b^{-}}({\bf q}_{i} - {\bf q}_{i-1})
\label{eq:HLL-flux}
\end{equation}
where ${\bf f}_{L,i-1/2} = {\bf f}({\bf q}_{L,i-1/2})$ and
${\bf f}_{R,i-1/2} = {\bf f}({\bf q}_{R,i-1/2})$ are the fluxes evaluated using
the left- and right-states of the conserved variables (using equation
\ref{eq:cons1D}), and
\begin{equation}
b^+ = {\rm max}[{\rm max}(\lambda^{M},v_{x,R}+c_{R}),0]
\label{eq:HLL+speed}
\end{equation}
\begin{equation}
b^- = {\rm min}[{\rm min}(\lambda^{0},v_{x,L}-c_{L}),0]
\label{eq:HLL-speed}
\end{equation}
Here $\lambda^{M}$ and $\lambda^{0}$ are the maximum and minimum
eigenvalues of Roe's matrix ${\sf A}$ (see \S \ref{sec:Roe} and Appendix
B), $v_{x,L}$ and $v_{x,R}$ are the velocity component normal to the
interface in the left- and right-states respectively, and $c_{L}$ and
$c_{R}$ are the maximum wavespeeds (the fast magnetosonic speed in MHD, or
the sound speed in hydrodynamics) computed from the left- and right-states.
The HLLE solver does not require a characteristic decomposition of the MHD
equations; the eigenvalues of Roe's matrix ${\sf A}$ are given by simple,
explicit formulae (see Appendix B).  Note that if both $\lambda^{M} < 0$
and $v_{x,R}+c_{R} < 0$ (or both $\lambda^{0} > 0$ and $v_{x,L}-c_{L}
> 0$), the HLLE flux will be ${\bf f}_{R,i-1/2}$ (or ${\bf f}_{L,i-1/2}$),
as expected.

The HLLE solver approximates the solution to the Riemann problem using a
single constant intermediate state computed from a conservative average,
bounded using an estimate for the maximum and minimum wavespeeds.
Thus, for hydrodynamics it neglects the contact wave, and for MHD it
neglects the Alfv\'{e}n, slow magnetosonic, and contact waves.  For this
reason, the HLLE is extremely diffusive for these waves (in fact, even
if $v_{x}=0$, contact discontinuities are diffused with the method).
Thus, in practice, the HLLE solver is of limited use for applications.
However, a distinct advantage of the HLLE solver is that the intermediate
state is positive-definite, that is the pressure and density in the
intermediate state can never
be negative.  Thus, in 1D it can be used to construct a positive-definite
integration algorithm (Einfeldt et al 1991).  This is in contrast to
linearized solvers such as Roe's method, in which the Riemann solver
itself can produce negative densities and pressures for one or more
of the intermediate states.  The HLLE flux is therefore an excellent
alternative in the rare circumstance that a more accurate solver fails.
In multidimensions, however, use of the HLLE flux at higher than first
order does not necessarily guarantee the method is positive definite:
this depends on the details of the multidimensional integrator being used.

For hydrodynamics, the HLL solver has been extended to include the contact
wave, resulting in a solution consisting of two constant intermediate
states bounded by shocks and separated by a contact discontinuity.
The resulting method is termed the HLLC solver.  A basic description
of the method is given in \S10.4 of Toro (1999) and will not be repeated here;
although it is important to note in Athena we
choose the wavespeeds following the suggestion in Batten et al. (1997).
This choice has the attractive property that the pressure in the
intermediate states computed from the Rankine-Hugoniot relations across
the left and right shocks is the same.  We find that for hydrodynamics,
this implementation of the HLLC solver produces results that are as,
if not more, accurate than
Roe's method (see below), but at much lower computational cost.
For 1D problems, it also is a positive definite method (although again,
this is not guaranteed in multidimensions).  Thus, the HLLC solver is
highly recommended for adiabatic hydrodynamic simulations with Athena.

Recently, Miyoshi \& Kusano (2005) have described an extension of the HLL
solver to MHD which includes the fast magnetosonic, Alfv\'{e}n, and contact
waves.  The resulting solver approximates the solution of the Riemann
problem with four constant intermediate states.  It reduces exactly to
the HLLC solver when the longitudinal component of the magnetic field is
zero, and is a positive definite method.  The implementation of the solver
is detailed in Miyoshi \& Kusano (2005), and will not be repeated here.
Tests using Athena indicate that this solver, termed HLLD, is typically as
accurate as the MHD extension of Roe's method, although it is much faster.
Thus, the HLLD solver is the best choice for many
MHD applications using Athena.

\subsubsection{Roe's Method}
\label{sec:Roe}

The HLL fluxes are based on an {\em approximate} solution to the nonlinear
equations of MHD.  Instead, Riemann solvers can be constructed from {\em exact}
solutions to an {\em approximate} (linearized) form of the MHD equations,
for example
\begin{equation}
\frac{\partial {\bf q}}{\partial t} = {\sf A}({\bf \bar{q}})
\frac{\partial {\bf q}}{\partial x}.
\end{equation}
The matrix ${\sf A}({\bf \bar{q}})$ is the Jacobian
$\partial {\bf f}/ \partial {\bf q}$ evaluated at some appropriate,
constant mean state ${\bf \bar{q}}$ (treating this matrix as
constant is what makes the system linear).  Finding the exact solution
to linear hyperbolic systems is less difficult because
only discontinuities (no rarefactions) are allowed.

Of course, the challenge in developing linearized solvers is finding
the appropriate representation for ${\sf A}({\bf \bar{q}})$.  Roe
(1981) proposed one particularly useful linearization, which
has subsequently been extended to adiabatic MHD by
Cargo \& Gallice (1997).  In this linearization, 
the Jacobian is evaluated using an average
state defined in the primitive variables ${\bf \bar{w}} =
(\bar{\rho},{\bf \bar{v}},\bar{P},\bar{B_y},\bar{B_z})$ as follows
\begin{eqnarray}
\bar{\rho} & = & \sqrt{\rho_{L}} \sqrt{\rho_{R}} \nonumber \\
\bar{\bf v} & = & (\sqrt{\rho_{L}}{\bf v_{L}} + \sqrt{\rho_{R}} {\bf v_{R}})/
(\sqrt{\rho_{L}} + \sqrt{\rho_{R}}) \nonumber \\
\bar{H} & = & (\sqrt{\rho_{L}}H_{L} + \sqrt{\rho_{R}} H_{R})/
(\sqrt{\rho_{L}} + \sqrt{\rho_{R}}) \\
\bar{B_y} & = & (\sqrt{\rho_{R}}B_{y,L} + \sqrt{\rho_{L}} B_{y,R})/
(\sqrt{\rho_{L}} + \sqrt{\rho_{R}}) \nonumber \\
\bar{B_z} & = & (\sqrt{\rho_{R}}B_{z,L} + \sqrt{\rho_{L}} B_{z,R})/
(\sqrt{\rho_{L}} + \sqrt{\rho_{R}}) \nonumber
\end{eqnarray}
where $H=(E+P^{*})/\rho$ is the enthalpy (used to compute the
pressure), and the subscripts $L$ and $R$ denote the left- and
right-states of each variable at the interface (computed using  one
of the reconstruction schemes described in \S \ref{sec:reconstruction}).
Explicit forms for the matrix ${\sf A}$, and its eigenvalues and
eigenvectors for isothermal and adiabatic hydrodynamics and MHD are
given in Appendix B.

Given the eigenvalues $\lambda^{\alpha}$ and left- and right-eigenmatrices
${\sf L}({\bf \bar{w}})$ and ${\sf R}({\bf \bar{w}})$ respectively, where 
$\alpha =1,M$ denotes the $M$
characteristics in the solution, the Roe fluxes are simply
\begin{equation}
\nf^{\rm Roe}_{i-1/2} = \frac{1}{2} \left( {\bf f}_{L,i-1/2} 
+ {\bf f}_{R,i-1/2} + \sum_{\alpha} a^{\alpha} \vert \lambda^{\alpha} \vert {\sf R}^{\alpha} \right)
\label{eq:roe-flux}
\end{equation}
where as before ${\bf f}_{L,i-1/2}={\bf f}({\bf q}_{L,i-1/2})$, 
${\bf f}_{R,i-1/2}={\bf f}({\bf q}_{R,i-1/2})$, and
\begin{equation}
  a^{\alpha} = {\sf L}^{\alpha} \cdot \delta {\bf q}_{i-1/2}
\end{equation}
\begin{equation}
  \delta {\bf q}_{i-1/2} = {\bf q}_{L,i-1/2} - {\bf q}_{R,i-1/2}
\end{equation}
and the ${\sf L}^{\alpha}$ and ${\sf R}^{\alpha}$ are the rows and columns of
the left- and right-eigenmatrices corresponding to $\lambda^{\alpha}$.

The primary advantage of Roe's method is that it includes all of
the characteristics in the problem, and therefore is less diffusive
and more accurate than the HLLE solver for intermediate waves such
as contact discontinuities.  Moreover, Roe (1981) showed that it
gives the flux exactly if the solution to the full nonlinear Riemann
problem contains only an isolated discontinuity.  However,
because it is based on a linearization of the MHD equations, for
some values of the left- and right-states Roe's method will fail
(Einfeldt et al. 1991);  it will return negative densities and/or
pressures in one or more of the intermediate states.  In Athena,
if this occurs we replace the calculation of the fluxes at that
interface with the HLLE solver (which is a positive-definite method)
or some other more accurate (e.g. an exact) solver.  Tests indicate this is
only required very rarely.

\section{Two-dimensional integration algorithm}
\label{sec:twod-integration}

Probably the most popular method for constructing a 2D integration
algorithm from the 1D method described in \S \ref{sec:oned-integration}
is based on dimensional splitting (Strang 1968).  Unfortunately,
dimensional splitting cannot be used for MHD if
the equations are to be solved in the conservative form.  This is
because during each one-dimensional update, only the transverse
components of the magnetic field evolve (e.g., from equation
\ref{eq:cons1D} it is clear that $B_{x}$ is non-evolutionary during
an update in the $x-$direction).  However, the divergence-free
constraint can only be maintained if all three components of the
field evolve simultaneously.  Thus, during the update in the
$x-$direction, $B_{x}$
must evolve.  The terms that describe this evolution cannot be
written in conservative form, leading to for example the $\nabla
\cdot {\bf B}$ source term formulations of Powell (1994) and Powell
et al. (1999).  However, there are significant advantages to
maintaining the conservative form (T2000), thus in Athena
we adopt dimensionally-unsplit integrators for MHD, either based
on the CTU+CT method (described below), or the VL+CT method (SG08).
The use of directionally unsplit integrators
in multidimensions is one of the most important components of the
MHD algorithms in Athena.

Even after adopting an unsplit integration algorithm, combining it
with the CT method to enforce the divergence-free constraint presents
challenges.  In particular, the method by which the corner-centered,
line-averaged emfs are constructed from the face-centered, area-averaged
fluxes returned by the Riemann solver is non-trivial.  In GS05,
we showed that simple arithmetic averaging does not work for the
unsplit integrators adopted here.  Instead, we developed several
methods for constructing the emfs from the Godunov fluxes, the
version actually used in Athena is described in \S\ref{sec:CT_Alg}.
The resulting method reduces exactly to the 1D algorithm described in
\S\ref{sec:oned-integration} for plane-parallel, grid-aligned flow,
and preserves the flux normal to the plane of the calculation.

\subsection{Steps in the 2D Algorithm}
\label{sec:twod-steps}

The 2D CTU+CT integration algorithm is based on the method of Colella (1990),
and is described in detail in GS05; below
we provide an overview of the main steps.

{\em Step 1.} Compute and store the left- and right-states at cell interfaces
in {\em both} the $x-$direction (${\bf q}_{L,i-1/2,j}, {\bf q}_{R,i+1/2,j}$)
and the $y-$direction (${\bf q}_{L,i,j-1/2}, {\bf q}_{R,i,j+1/2}$)
simultaneously, using any of the 1D spatial reconstruction algorithms
described in \S \ref{sec:reconstruction}, for all the interfaces
over the entire grid.  Since the 1D reconstruction algorithms in
Athena include a characteristic tracing step, when applied in
multidimensions the 1D reconstruction must include $\nabla \cdot
{\bf B}$ source terms as described in \S3.1 in GS05, and briefly in
\S \ref{sec:2D-reconstruction}.  Note that the components of ${\bf
q}_L$ (and ${\bf q}_R$) are different on the $x-$ and $y-$interfaces.

{\em Step 2.} Compute 1D fluxes of the conserved variables using any
one of the Riemann solvers
described in \S \ref{sec:fluxes} at interfaces in {\em both} the
$x-$ and $y-$directions simultaneously
\begin{eqnarray}
{\bf f}_{i-1/2,j}^* &=& \nf(q_{L,i-1/2,j},q_{R,i-1/2,j},B_{x,i-1/2,j}) \\
{\bf g}_{i,j-1/2}^* &=& \nf(q_{L,i,j-1/2},q_{R,i,j-1/2},B_{y,i,j-1/2})
\end{eqnarray}
where the appropriate longitudinal component of the magnetic field has
been passed to the Riemann solver as a parameter.

{\em Step 3.}
Using the algorithm of GS05, described in \S\ref{sec:CT_Alg},
calculate the emf at
cell corners $\E_{z,i-1/2,j-1/2}^*$ from the appropriate components
of the face-centered fluxes returned by the Riemann solver in step
2, and the $z-$component of a cell center reference electric field
${\bf \E}^{r,n}_{i,j,}$
calculated using the initial data at time level $n$, i.e. 
${\bf \E}^{r,n}_{z,i,j} = -(v^{n}_{x,i,j}B^{n}_{y,i,j}-v^{n}_{y,i,j}B^{n}_{x,i,j})$.

{\em Step 4} Evolve the left- and right-states at each interface
by $\delta t/2$ using transverse flux gradients.  For example,
the mass density, momentum density, energy density, and
$B_{z}$ at the $x-$interface located at $x_{i-1/2}$ are advanced
using
\begin{eqnarray}
{\bf q}_{L,i-1/2,j}^{n+1/2} &=& {\bf q}_{L,i-1/2,j}+ \frac{\delta t}{2 \delta y}
\left( {\bf g}_{i-1,j+1/2}^* - {\bf g}_{i-1,j-1/2}^* \right)
+ \frac{\delta t}{2}{\bf s}_{x,i-1,j}
\label{eq:trans-flux-x-l} \\
{\bf q}_{R,i-1/2,j}^{n+1/2} &=& {\bf q}_{R,i-1/2,j}+ \frac{\delta t}{2 \delta y}
\left( {\bf g}_{i,j+1/2}^* - {\bf g}_{i,j-1/2}^* \right) 
+ \frac{\delta t}{2}{\bf s}_{x,i,j}
\label{eq:trans-flux-x-r}
\end{eqnarray}
Since the components of 1D vectors on the $x-$ and $y-$interfaces differ,
care must be taken to associate the components of the left- and right-states
with the appropriate components of the transverse fluxes (for example, the
components of ${\bf q}_{L,i-1/2,j}$ with the components of
${\bf g}_{i-1,j+1/2}^*$).
The updates in equations \ref{eq:trans-flux-x-l} and
\ref{eq:trans-flux-x-r} are directionally split (only the
transverse flux gradient is used) and are based on the conservative
form, therefore $\nabla \cdot {\bf B}$ source terms
must be added to the momentum density, energy, and $B_z$.
These are represented by the source term vector ${\bf s}_{x}$,
the last term in both equations.  For the left- and right-states
on the $x-$interface, the source term vector has components
${\bf s}_{x} = (0, {\bf s^{M}},s^{E},0,s^{B_{z}})$ where
\begin{eqnarray}
{\bf s^{M}}_{x,i,j} & = & {\bf B}_{i,j}(B_{x,i+1/2,j}-B_{x,i-1/2,j} )/\delta x 
\nonumber \\
s^{E}_{x,i,j} & = & (B_z v_z)_{i,j} (B_{x,i+1/2,j} - B_{x,i-1/2,j})/\delta x 
\label{eq:2D-divB-src-terms} \\
s^{B_{z}}_{x,i,j} & = & v_{z,i,j} (B_{x,i+1/2,j} - B_{x,i-1/2,j})/\delta x
\nonumber
\end{eqnarray}
Expressions similar to equations \ref{eq:trans-flux-x-l} and 
\ref{eq:trans-flux-x-r} are used to update the $y-$interface states
located at $y_{j-1/2}$, that is ${\bf q}_{L,i,j-1/2}$ and ${\bf q}_{R,i,j-1/2}$,
for $\delta t/2$ using
the flux gradient in the $x-$direction.  Source terms analogous to
those in equation \ref{eq:2D-divB-src-terms}, but proportional to
$\delta B_{y}/\delta y$, also are necessary (see \S4.1.2
in GS05).  The in-plane components of the magnetic field are evolved using CT,
\begin{eqnarray}
B_{x,i-1/2,j}^{n+1/2} & = & B_{x,i-1/2,j} - \frac{\delta t}{2 \delta y}
\left(\E_{z,i-1/2,j+1/2}^* - \E_{z,i-1/2,j-1/2}^* \right) 
\label{eq:CT-half-x} \\
B_{y,i,j-1/2}^{n+1/2} & = & B_{y,i,j-1/2} + \frac{\delta t}{2 \delta x}
\left(\E_{z,i+1/2,j-1/2}^* - \E_{z,i-1/2,j-1/2}^* \right)
\label{eq:CT-half-y}
\end{eqnarray}
using the emfs computed in step 3.

{\em Step 5.} Calculate a cell-centered reference electric field at
$t^{n+1/2}$, ${\bf \E}^{r,n+1/2}_{i,j,}$, which is needed as a reference
state for the CT algorithm in step 7.  The cell-centered velocities
at the half-timestep needed to compute ${\bf \E}^{r,n+1/2}_{i,j,}$
come from a conservative finite-volume update of the initial
mass and momentum density, using the fluxes ${\bf f}^{*}_{i-1/2,j}$ and
${\bf g}^{*}_{i,j-1/2}$.  The cell-centered components of the magnetic
field at the half-timestep come from averaging the face centered
fields at the half-timestep computed by equations \ref{eq:CT-half-x}
and \ref{eq:CT-half-y} in step 4 to cell-centers.

{\em Step 6.}
Compute new fluxes at cell interfaces using the corrected left- and 
right-states from step 4 using one of the Riemann solvers described in
\S \ref{sec:fluxes}, giving
\begin{eqnarray}
{\bf f}_{i-1/2,j}^{n+1/2} &=& \nf({\bf q}_{L,i-1/2,j}^{n+1/2},{\bf q}_{R,i-1/2,j}^{n+1/2},B^{n+1/2}_{x,i-1/2,j})  \\
{\bf g}_{i,j-1/2}^{n+1/2} &=& \nf({\bf q}_{L,i,j-1/2}^{n+1/2},{\bf q}_{R,i,j-1/2}^{n+1/2},B^{n+1/2}_{y,i,j-1/2})
\end{eqnarray}
Note the appropriate face-centered fields updated to the half-timestep computed
in step 4 are passed as parameters to the Riemann solver.
If needed, the H-correction is used
in this step to eliminate the carbuncle instability (see Appendix C).

{\em Step 7.}  Apply the algorithm of \S\ref{sec:CT_Alg} to calculate the
CT electric fields $\E_{z,i-1/2,j-1/2}^{n+1/2}$
using the numerical fluxes from step 6 and the cell center reference
electric field calculated in step 5.
 
{\em Step 8.} Update the solution from time level $n$ to $n+1$,
using the 2D version of the finite-volume difference discretization
(equation \ref{eq:int_form}) for the mass density,
momentum density, energy density, and $B_z$, and the CT formulae (equations
\ref{eq:CT-x} and \ref{eq:CT-y}) for the in-plane components of the
field $B_x$ and $B_y$.

{\em Step 9.} Compute the cell-centered components of the magnetic field from
the updated face-centered values using equations \ref{eq:face-to-cell-x} and
\ref{eq:face-to-cell-y}.

{\em Step 10.} Increment the time: $t^{n+1} = t^{n} + \delta t$.  Compute
a new timestep that satisfies an estimate of the CFL stability condition
based on wavespeeds at cell centers
\begin{equation}
\delta t = C_{\circ} \min \left( 
\frac{\delta x}{\vert v^{n+1}_{x,i,j} \vert + C^{n+1}_{fx,i,j}},
\frac{\delta y}{\vert v^{n+1}_{y,i,j} \vert + C^{n+1}_{fy,i,j}} \right)
\end{equation}
where $C_{\circ} \le 1$ is the CFL number, $C^{n+1}_{fx,i,j}$ and
$C^{n+1}_{fy,i,j}$ are the fast magnetosonic speeds in the $x-$ and
$y-$directions respectively, evaluated using the updated quantities, and
the minimum is taken over all grid cells.  Note this is only an estimate
of the CFL stability condition, since the wavespeeds used in the Riemann
solver can be different from those computed from the cell-centered values.

{\em Step 11.} Repeat steps 1-10 until the stopping criterion is reached,
i.e.. $t^{n+1} \ge t_f$

The entire 2D integration algorithm is summarized by the flow chart shown in
figure 4.

\subsection{MHD Interface States in 2D}
\label{sec:2D-reconstruction}

In step 1 of the 2D algorithm discussed above, source terms must
be added to the left- and right-states in the primitive variables
that arise due to the characteristic tracing step in the reconstruction
algorithms (see \S \ref{sec:reconstruction}).  These terms
are necessary for a proper accounting of all the evolutionary terms that
form the characteristic tracing step in multidimensioal MHD
(see GS05 and GS08 for a complete discussion of the origin of these 
terms).  Since the reconstruction
is performed in the primitive variables, the only terms required are
for the transverse components of the magnetic field (in contrast to
step 4 in the 2D algorithm above, where the directional splitting is performed
on the equations in conservative form, and therefore source terms were needed
for ${\bf M}$, $E$, and $B_z$).  Thus, for the left-state at the $x-$interface
located at $x_{i-1/2}$, the change to the transverse fields due to the source
term is
\begin{eqnarray}
\delta B_{y,L,i-1/2,j} & = & \frac{\delta t}{2\delta x} v_{y,i-1,j}
\left( B_{x,i-1/2,j} - B_{x,i-3/2,j} \right) \nonumber 
\end{eqnarray}
while for the left- and right-interface values at the $y-$interface
located at $y_{j-1/2}$, the change to the transverse field due to the
source term is
\begin{eqnarray}
\delta B_{x,L,i,j-1/2} & = & \frac{\delta t}{2\delta y} v_{x,i,j-1}
\left( B_{y,i,j-1/2} - B_{y,i,j-3/2} \right) \nonumber
\end{eqnarray}
Similar expressions are needed for the right-state values at each interface
(GS05).
In both cases the terms are added to the primitive variables after the
reconstruction, and before converting back to the conserved variables.

\subsection{Calculating the emfs}
\label{sec:CT_Alg}

As discussed in \S \ref{sec:discretization}, the CT update of the
magnetic field requires the line-averaged emfs at cell corners,
whereas the Riemann solver returns area-averaged electric fields
at cell faces.  For example, figure 5 shows the relative positions
of the fluxes returned by the Riemann solver, and the emfs needed
by CT, for the 2D grid cell with indices $(i,j)$.  In GS05, it was
shown that the relationship between the two is determined by the
averaging formulae used to convert between the face-centered
area-averages of the magnetic field, and the cell-centered
volume-averages.  A variety of different algorithms were explored,
and the best compromise between accuracy and simplicity
was found to be
\begin{eqnarray}
\E_{z,i-1/2,j-1/2} & = &
\frac{1}{4} \left(
\E_{z,i-1/2,j} + \E_{z,i-1/2,j-1} + \E_{z,i,j-1/2} + \E_{z,i-1,j-1/2}
\right) \nonumber \\
& + & \frac{\delta y}{8} \left(
\left(\frac{\partial \E_z}{\partial y}\right)_{i-1/2,j-1/4} -
\left(\frac{\partial \E_z}{\partial y}\right)_{i-1/2,j-3/4}
\right) \nonumber \\
& + & \frac{\delta x}{8} \left(
\left(\frac{\partial \E_z}{\partial x}\right)_{i-1/4,j-1/2} -
\left(\frac{\partial \E_z}{\partial x}\right)_{i-3/4,j-1/2}
\right) ~.
\label{eq:2d_CT_Ez}
\end{eqnarray}
where the 
derivative of $\E_z$ on each grid cell face is computed by
selecting the ``upwind'' direction according to the contact mode, e.g. 
\begin{equation}
\left(\frac{\partial \E_z}{\partial y}\right)_{i-1/2} = \left \{
\begin{array}{ll}
(\partial \E_z/\partial y)_{i-1} & \textrm{for}~ v_{x,i-1/2} > 0 \\
(\partial \E_z/\partial y)_{i} & \textrm{for}~ v_{x,i-1/2} < 0 \\
\frac{1}{2}\left(
\left( \frac{\partial \E_z}{\partial y} \right)_{i-1} + 
\left( \frac{\partial \E_z}{\partial y} \right)_{i}
\right) & \textrm{otherwise}
\end{array}
\right .
\label{eq:Ndup_CT_Ez}
\end{equation}
(where the subscript $j$ has been suppressed)
with an analogous expression for the $(\partial \E_z/\partial x)$.
The
derivatives of the electric field in equation (\ref{eq:Ndup_CT_Ez}) are computed
using the face centered electric fields (Godunov fluxes) and a cell
center ``reference'' value $\E^r_{z,i,j}$, e.g.
\begin{equation}
\left(\frac{\partial \E_z}{\partial y}\right)_{i,j-1/4} =
2\left(\frac{\E^r_{z,i,j} - \E_{z,i,j-1/2}}{\delta y}\right) ~.
\label{eq:CT_Ez_der}
\end{equation}
where the cell center reference electric field $\E^r_{z,i,j}$ is
computed at the appropriate time level (either $t^{n}$ for step 3
of the 2D algorithm, or $t^{n+1/2}$ for step 7).  To help clarify
the above, figure 5 diagrams the relative locations of the Godunov
fluxes, corner-centered emf, cell-centered reference states, and
the derivatives of the electric field.  Further details are provided
in GS05 (and GS08 for the 3D case).  

Note for the 3D CTU+CT algorithm,
analogous expressions to the above are required to convert the $x-$
and $y-$components of the electric field to the appropriate cell
corners (see figure 1).  These expressions follow directly from
equations \ref{eq:Ndup_CT_Ez} and \ref{eq:CT_Ez_der} using a cyclic
permutation of the $(x,y,z)$ and $(i,j,k)$.

\section{Three-Dimensional Integration Algorithm}
\label{sec:threed-integration}

The extension of the dimensionally unsplit CTU integrator due to
Colella (1990) used in Athena from 2D to 3D is in fact quite complex.
In particular, for stability with a CFL number $C_{\circ} \leq 1$ requires
12 Riemann solves per cell per timestep, and multiple fractional timesteps
are required to correct the left- and right-states with transverse flux
gradients in a genuinely multidimensional fashion.  This extension of
CTU to 3D has been described by Saltzman (1994) for hydrodynamics.

In GS08, we explored the use of the 12-solve CTU+CT algorithm for
MHD, as well as a simpler variant that uses only 6-solves per
timestep, but formally is only stable for CFL numbers $C_{\circ}
\leq 0.5$.  The tests presented in GS08 show that the 6-solve
algorithm is as accurate as the 12-solve method, and requires about
the same computational cost.  However, the 6-solve algorithm is
dramatically simpler to implement, and therefore is the primary 3D
integrator used in Athena.

The 6-solve CTU+CT 3D algorithm is designed in such a way that for
grid aligned flows it reduces exactly to the 2D CTU+CT algorithm
described in \S\ref{sec:twod-integration}, or the 1D algorithm
described in \S\ref{sec:oned-integration}, depending on the symmetry
of the problem.  Perhaps even more importantly, in GS08 we introduced
a test problem to demonstrate the 3D CTU+CT algorithm preserves 
a discrete representation of the divergence-free constraint
that prevents anomalous growth of magnetic flux for problems with certain
symmetries.  The test involves
advection of a cylindrical column of 2D field loops in the $x-y$
plane, with $B_z=0$, and a constant but fully 3D velocity field.  In this case
the $z-$component of the induction equation reduces to
\begin{equation}
\frac{\partial B_z}{\partial t}
- v_{z} \left( \frac{\partial B_x}{\partial x}+\frac{\partial B_y}{\partial y}
\right) = 0 \nonumber
\end{equation} 
Clearly, the second term is proportional to $\nabla \cdot {\bf B}$.  Thus,
if the {\em discrete form} of the induction equation used to update
the field components in 3D is able to preserve $B_z=0$ exactly,
then the algorithm must preserve the appropriate discrete representation
of $\nabla \cdot {\bf B} =0$.  We present results of this field
loop test in \S\ref{sec:2D-mhd-tests} in 2D, and \S\ref{sec:3D-mhd-tests}
in 3D.

\subsection{Steps in the 3D Algorithm}

The 6-solve version of the dimensionally unsplit 3D CTU+CT algorithm
can be described by the following steps (see GS08 for details).
It may also be useful to compare and contrast the steps in the 3D algorithm
with those in the 2D method (\S\ref{sec:twod-steps}).

{\em Step 1.}  Compute and store the left- and right-states at
cell interfaces in the $x-$direction (${\bf q}_{L,i-1/2,j,k},
{\bf q}_{R,i-1/2,j,k}$), the $y-$direction (${\bf q}_{L,i,j-1/2,k},
{\bf q}_{R,i,j-1/2,k}$), and the $z-$direction (${\bf q}_{L,i,j,k-1/2},
{\bf q}_{R,i,j,k-1/2}$) simultaneously, using any of the 1D spatial
reconstruction schemes described in \S \ref{sec:reconstruction}, for all
the interfaces over the entire grid.  This
requires adding $\nabla \cdot {\bf B}$ source terms to the primitive
variables, as discussed in GS08 and \S \ref{sec:3D-reconstruction}.

{\em Step 2.}  Compute 1D fluxes of the conserved variables
using any one of the Riemann solvers described
in \S \ref{sec:fluxes} at interfaces in all three dimensions
\begin{eqnarray}
{\bf f}_{i-1/2,j,k}^* &=& \nf({\bf q}_{L,i-1/2,j,k},{\bf q}_{R,i-1/2,j,k},B_{x,i-1/2,j,k}) \\
{\bf g}_{i,j-1/2,k}^* &=& \nf({\bf q}_{L,i,j-1/2,k},{\bf q}_{R,i,j-1/2,k},B_{y,i,j-1/2,k}) \\
{\bf h}_{i,j,k-1/2}^* &=& \nf({\bf q}_{L,i,j,k-1/2},{\bf q}_{R,i,j,k-1/2},B_{z,i,j,k-1/2}) .
\end{eqnarray}
using the appropriate longitudinal component of the magnetic field passed
as a parameter to the Riemann solver.

{\em Step 3.}  Apply the algorithm of \S\ref{sec:CT_Alg} to calculate the CT
electric fields at cell-corners,
$\E_{x,i,j-1/2,k-1/2}^*$, $\E_{y,i-1/2,j,k-1/2}^*$
and $\E_{z,i-1/2,j-1/2,k}^*$, from the appropriate components of the
face-centered fluxes returned by the Riemann solver in step 2,
and a cell center reference electric field calculated using the
initial data at time level $n$, i.e. ${\bf \E}^{r,n}_{i,j,k} =
- ({\bf v}^{n}_{i,j,k} \times {\bf B}^{n}_{i,j,k})$.  (Note
the algorithms for computing the $x-$ and $y-$components of the
emf are a straightforward extension of the algorithm to compute the
$z-$component described in \S\ref{sec:CT_Alg}, see GS08.)

{\em Step 4.}  Update the face-centered magnetic field by $\delta t/2$
using the CT difference equations \ref{eq:CT-x} through \ref{eq:CT-z}, and
the emfs computed in step 3.

{\em Step 5.} Evolve the left- and right-states at each interface by $\delta
t/2$ using transverse flux gradients.  For example, the hydrodynamic
variables (mass, momentum and energy density) are advanced using
\begin{eqnarray}
{\bf q}_{L,i-1/2,j,k}^{n+1/2} &=& {\bf q}_{L,i-1/2,j,k} - \frac{\delta t}{2 \delta y}
\left({\bf g}_{i,j+1/2,k}^* - {\bf g}_{i,j-1/2,k}^* \right) \nonumber \\
& - & \frac{\delta t}{2 \delta z} \left({\bf h}_{i,j,k+1/2}^* - {\bf h}_{i,j,k-1/2}^* \right) + \frac{\delta t}{2}{\bf s}_{x,i-1,j,k} \\
{\bf q}_{R,i-1/2,j,k}^{n+1/2} &=& {\bf q}_{R,i-1/2,j,k} - \frac{\delta t}{2 \delta y}
\left({\bf g}_{i+1,j+1/2,k}^* - {\bf g}_{i+1,j-1/2,k}^* \right) \nonumber \\
 & - & \frac{\delta t}{2\delta z}
\left({\bf h}_{i+1,j,k+1/2}^* - {\bf h}_{i+1,j,k-1/2}^*  \right)
+ \frac{\delta t}{2}{\bf s}_{x,i,j,k}
\end{eqnarray}
Once again, care must be taken to associate the components of the
vectors of interface states (e.g. ${\bf q}_{L,i-1/2,j,k}$) with the
appropriate components of the transverse fluxes (e.g., ${\bf
g}_{i,j-1/2,k}^*$ and ${\bf h}_{i,j,k-1/2}^*$).  Moreover, since
these updates are directionally split, $\nabla \cdot {\bf B}$ source
terms must be added.  These are represented by the source 
term vector ${\bf s}_{x}$,
the last term in both equations.  For the left- and right-states
on the $x-$interface, the source term vector has components
${\bf s}_{x} = (0, {\bf s^{M}},s^{E},0,0)$ where
\begin{eqnarray}
{\bf s^{M}}_{x,i,j,k} & = & {\bf B}_{i,j,k}
\left( \frac{\partial B_x}{\partial x} \right)_{i,j,k} \nonumber \\
{\bf s}^{E}_{i,j,k} & = & -(B_y v_y)_{i,j,k}
\mm{\frac{\partial B_z}{\partial z}}{-\frac{\partial B_x}{\partial x}}_{i,j,k}
\nonumber \\
 & & -(B_z v_z)_{i,j,k}
\mm{\frac{\partial B_y}{\partial y}}{-\frac{\partial B_x}{\partial x}}_{i,j,k}
~.
\end{eqnarray}
where the minmod function is defined as 
\beq
\mm{x}{y} = \left \{
\begin{array}{ll}
\sign(x) \min(\vert x \vert,\vert y \vert) & \textrm{if }xy > 0 \\
0 & \textrm{otherwise.}
\end{array}
\right .
\eeq
The use of the minmod operator to limit the source terms according to the
magnitude of the terms in the divergence of ${\bf B}$ is discussed in GS08, it
is needed because there are now {\em two} terms that arise from
transverse gradients, instead of only one as in 2D.
The transverse components of the magnetic field stored at each of the
interfaces is evolved using a combination of the emfs computed in step 3, and
$\nabla \cdot {\bf B}$ source terms.  For example, the right-state value
of the $y-$ and $z-$components of the magnetic field at the $x-$interface
at $x_{i-1/2}$ are evolved using
\begin{eqnarray}
(B_y)_{R,i-1/2,j,k}^{n+1/2} &=& (B_y)_{R,i-1/2,j,k} -
 \frac{\delta t}{4 \delta z}
\left(\E_{x,i,j+1/2,k+1/2}^* - \E_{x,i,j+1/2,k-1/2}^* \right) \nonumber \\
 & & -\frac{\delta t}{4 \delta z}
\left(\E_{x,i,j-1/2,k+1/2}^* - \E_{x,i,j-1/2,k-1/2}^* \right) \nonumber \\
 & & - \frac{\delta t}{2} (v_y)_{i,j,k}
\mm{\frac{\partial B_z}{\partial z}}{-\frac{\partial B_x}{\partial x}}_{i,j,k}
\end{eqnarray}
\begin{eqnarray}
(B_z)_{R,i-1/2,j,k}^{n+1/2} &=& (B_z)_{R,i-1/2,j,k} +
 \frac{\delta t}{4 \delta y}
\left(\E_{x,i,j+1/2,k+1/2}^* - \E_{x,i,j-1/2,k+1/2}^* \right) \nonumber \\
 & & + \frac{\delta t}{4 \delta y}
\left(\E_{x,i,j+1/2,k-1/2}^* - \E_{x,i,j-1/2,k-1/2}^* \right) \nonumber \\
 & & - \frac{\delta t}{2} (v_z)_{i,j,k}
\mm{\frac{\partial B_y}{\partial y}}{-\frac{\partial B_x}{\partial x}}_{i,j,k}
\end{eqnarray}
with similar expressions for the left-state values (but using quantities at
$i-1$ on the right hand side of the above equations as appropriate).
The origin of these MHD source terms for the transverse components of the
magnetic field is discussed further in GS08.  The $y$- and $z$-interface
states are advanced in an equivalent manner by cyclic
permutation of $(x,y,z)$ and $(i,j,k)$ in the above expressions.

{\em Step 6.} Calculate a cell-centered electric field at $t^{n+1/2}$
by using the fluxes ${\bf f}^{*}_{i-1/2,j,k}$, ${\bf g}^{*}_{i,j-1/2,k}$,
and ${\bf h}^{*}_{i,j,k-1/2}$ to compute the cell-centered velocities
at the half-timestep using a conservative finite volume update for
the momentum and density, and by averaging the face centered fields
at the half-timestep computed in step 4.
This is needed as a reference state for the CT algorithm in step 8.

{\em Step 7.}  Compute new fluxes at cell interfaces using the
corrected left- and right-states from step 5, and the interface magnetic
fields at $t^{n+1/2}$ computed in step 4, using one of the Riemann solvers
described in \S\ref{sec:fluxes}
\begin{eqnarray}
{\bf f}_{i-1/2,j,k}^{n+1/2} &=& \nf({\bf q}_{L,i-1/2,j,k}^{n+1/2},{\bf q}_{R,i-1/2,j,k}^{n+1/2},B^{n+1/2}_{x,i-1/2,j,k}) \\
{\bf g}_{i,j-1/2,k}^{n+1/2} &=& \nf({\bf q}_{L,i,j-1/2,k}^{n+1/2},{\bf q}_{R,i,j-1/2,k}^{n+1/2},B^{n+1/2}_{y,i,j-1/2,k})\\
{\bf h}_{i,j,k-1/2}^{n+1/2} &=& \nf({\bf q}_{L,i,j,k-1/2}^{n+1/2},{\bf q}_{R,i,j,k-1/2}^{n+1/2},B^{n+1/2}_{z,i,j,k-1/2})
\end{eqnarray}
using the appropriate longitudinal component of the magnetic field passed
as a parameter to the Riemann solver.  If needed, the H-correction is used
in this step to eliminate the carbuncle instability (see Appendix C).

{\em Step 8.}  Apply the algorithm of \S\ref{sec:CT_Alg} to calculate the
CT electric fields $\E_{x,i,j-1/2,k-1/2}^{n+1/2}$,
$\E_{y,i-1/2,j,k-1/2}^{n+1/2}$ and $\E_{z,i-1/2,j-1/2,k}^{n+1/2}$
using the appropriate components of the numerical fluxes from step 7 and the
cell center reference electric field calculated in step 6.
 
{\em Step 9.} Update the solution from time level $n$ to $n+1$ using the
conservative finite volume update (equation \ref{eq:int_form}) for the
hydrodynamic variables (mass, momentum and energy density)
and the CT formulae (equations \ref{eq:CT-x} through \ref{eq:CT-z})
to update the area-averaged face-centered components of the magnetic field.

{\em Step 10.} Compute the cell-centered components of the magnetic field from
the updated face-centered values using equations \ref{eq:face-to-cell-x} 
through \ref{eq:face-to-cell-z}.

{\em Step 11.} Increment the time: $t^{n+1} = t^{n} + \delta t$.  Compute
a new timestep that satisfies an estimate of the CFL stability condition
based on wavespeeds at cell centers
\begin{equation}
\delta t = C_{\circ} \min \left( 
\frac{\delta x}{\vert v^{n+1}_{x,i,j,k} \vert + C^{n+1}_{fx,i,j,k}},
\frac{\delta y}{\vert v^{n+1}_{y,i,j,k} \vert + C^{n+1}_{fy,i,j,k}},
\frac{\delta z}{\vert v^{n+1}_{z,i,j,k} \vert + C^{n+1}_{fz,i,j,k}} \right)
\end{equation}
where $C_{\circ} \le 1/2$ is the CFL number, $C^{n+1}_{fx,i,j,k}$,
$C^{n+1}_{fy,i,j,k}$, and $C^{n+1}_{fz,i,j,k}$ are the fast
magnetosonic speeds in the $x-$, $y-$, and $z-$directions respectively,
evaluated using the updated
quantities, and the minimum is taken over all grid cells.
Note this is only an estimate of the CFL
stability condition, since the wavespeeds used in the Riemann solver
can be different from those computed from the cell-centered values.

{\em Step 12.} Repeat steps 1-11 until the stopping criterion is reached,
i.e., $t^{n+1} \ge t_f$.

The steps in the 3D integration algorithm are very similar to those
summarized by the flow chart in
figure 4 for the 2D algorithm.

\subsection{MHD Interface States in 3D}
\label{sec:3D-reconstruction}

As with the 2D integrator, source terms must be added to the left- and
right-states in the primitive variables calculated using the 1D spatial
reconstruction schemes described in \S\ref{sec:reconstruction}.
Since the reconstruction is in the primitive
variables, only the transverse components of the magnetic field
require these terms.  For the right-state at the $x-$interface located
at $x_{i-1/2}$, the change to the transverse fields due to the source
terms are
\begin{eqnarray}
(\delta B_y)_{R,i-1/2,j,k} &=& - \frac{\delta t}{2} (v_y)_{i,j,k}
\mm{\frac{\partial B_z}{\partial z}}{-\frac{\partial B_x}{\partial x}}_{i,j,k} \\
(\delta B_z)_{R,i-1/2,j,k} &=& - \frac{\delta t}{2} (v_z)_{i,j,k}
\mm{\frac{\partial B_y}{\partial y}}{-\frac{\partial B_x}{\partial x}}_{i,j,k}
\end{eqnarray}
Similar expressions are needed for the left-state values, while the
equations for the left- and right-state values at the $y-$ and $z-$interfaces
follow from cyclic permutation of the $(x,y,z)$.  These terms are added
to the primitive variables after reconstruction, and before converting back
to the conserved variables.

\section{Implementation}

The implementation of the numerical algorithms described in the
previous sections into a functioning computer code can be complex,
and warrants at least some discussion.

Athena was developed in C, but many applications scientists prefer to
work with Fortran.  Hence, we
have written two different versions of Athena: the original C code, and
another in Fortran95.  These two versions provide the community with
implementations of the Athena algorithm in the two most popular languages
used for scientific computing in astrophysics.  The most important design
criteria we have adopted for both versions are
\begin{enumerate}
\item modularity,
\item documentation,
\item strict adherence to ANSI standards,
\item simple control of physics and runtime options
\end{enumerate}
We briefly discuss each of these below.

By far the most important design priority is modularity.  Thus,
the Riemann solvers, 1D reconstruction algorithms, conversion from
conserved to primitive variables, boundary conditions, data output, and
the integrators themselves are all broken into individual functions, with
a common interface specific to each class.  This makes adding everything
from a new Riemann solver to a new data output format simply a matter
of writing a new function which conforms to the appropriate interface.
Moreover, all problem-specific code is contained in a single file, with
functionality provided that makes it easy to add new boundary conditions 
or new source terms in the equations.

Although writing documentation is never enjoyable, it is critical if
anyone other than the developer is to use the code.  We have found this to
be true even amongst members of our own research groups.  The C version
of Athena comes with an extensive {\em User's Guide} which describes
installing, compiling, and running the code, and a {\em Programmer's
Guide} which explains the grid, data structures, and program control
and flow.  Both are included with the source code in the download from
the web.  The Fortran95 version has its own {\em User's Guide}.
Ample comments are also embedded within the source files.

By adhering to ANSI standards, we ensure Athena can be compiled and
run on any machine with a C or Fortran95 compiler, as appropriate.
To avoid reliance on external libraries, we do not use special
purpose output formats.  The philosophy is that data can always be
converted into other format by post-processing software if needed,
or by writing a new user-defined output routine.  Athena is written
to run either as a serial code on one CPU or in parallel using domain
decomposition through MPI calls.  The only external libraries needed
by Athena are for parallelization with MPI (using any version of the
MPICH or OpenMPI libraries).  As algorithms become more complex, the
use of external libraries for I/O may become unavoidable.  For example,
the HDF5 library has proved to be useful in organizing the complex data
structures associated with AMR grids.

The compile and runtime options in the C version of Athena are documented
in the {\em User's Guide}.  Physics and algorithm options are set at
compile time using a configure script generated by the {\tt autoconf}
toolkit.  In the Fortran95 version, these options are determined by selecting
which modules to {\tt USE}.  A perl build script {\tt buildathena} is
included to simplify the choice of problem module, physics, and parallel
or serial version.  A separate user guide is provided for the Fortran code.
Both codes use a simple block-structured input file
with runtime parameter values.  The Fortran95 version uses {\tt NAMELIST}
and the the C version uses a flexible format that emulates {\tt NAMELIST}
functionality.  Although there is nothing special about the specific
way compiler and run options are set in Athena, the key point is that
simple and extensible mechanisms to control both are provided.

Two final important aspects of code implementation are the single
processor performance, and parallelization on distributed memory clusters.
Aggressive optimization requires mature and static algorithms, and often
comes at the cost of clarity and adaptability in the code.  Athena is
intended to be a community code, and we plan that Athena will continue to
be developed and extended.  Thus, optimization has been limited to the
basic concepts guided by the rules of data locality and vectorization.
In the C version, for example, to optimize cache use we define all
variables within a cell as a data structure, and then create 3D arrays of
this structure.  This ensures values for each variable associated with a
given cell are contiguous in memory.  To promote vectorization, as much
computational work as is possible is done on 1D pencils drawn from the
grid (for example, the spatial reconstruction step).  The Fortran95
version is designed to take advantage of Fortran array syntax where
possible.  One drawback of dimensionally unsplit algorithms is that the
left- and right-states and fluxes must be computed and stored for every
interface over the entire 3D grid.  This requires many 3D arrays, which
increases the memory footprint of the code and reduces cache-performance.
However, unsplit algorithms are essential for MHD.

Although Athena requires many more floating point operations per cell
than algorithms such as ZEUS (as much as ten times more), the primary
bottleneck on modern processors is generally accessing cache and
interprocess communication for parallel problems.  Thus, the
performance of Athena in comparison to ZEUS is not decreased in proportion
to the amount of work per cell in the two codes.  One of the most useful
measures of performance is the number of cells updated per cpu second.
This depends on many factors, including the algorithm, the size of the
grid, and the processor speed.  Table 1 lists the performance of the C
version of Athena on a 2.2 GHz Opteron processor, compiled with {\tt
gcc} using an optimization level of {\tt -O3} for various physics and
algorithm options and using a 3D $128^{3}$ grid.  For comparison, a 3D
version of ZEUS written in F77 by one of us (Stone) and run on the same
processor gives $404\times 10^{3}$ cell-updates/sec for adiabatic MHD
on a $128^{3}$ grid.  Thus, while the algorithms in Athena typically
require $10\times$ the work of those in ZEUS, the code is only four
times slower when using the HLLD fluxes.

Parallelization is achieved in Athena using domain decomposition with
MPI calls to swap data in ghost cells at grid boundaries.  The number of
ghost cells required depends on the type of physics used and the order
of the reconstruction.  For example, MHD with third-order reconstruction
requires four ghost cells at every boundary (more are required if the
H-correction is used, see Appendix C).  By sequential exchange of boundary
conditions in the $x-$, $y-$, and $z-$directions, we avoid the need for
extra MPI calls to swap values across diagonal domains at the corners
of the grid.  Two factors contribute to making Athena very efficient on
distributed memory clusters.  First, the unsplit direct Eulerian update
in Athena requires communication of ghost zones only once per timestep,
greatly reducing the number of MPI calls compared to split methods.
Second, the ratio of computational work to data communicated is large
in Athena due to the complexity of the algorithms.  Figure 6 plots the
efficiency of the C version of Athena, defined as the speed per processor
in a parallel calculation normalized by the speed of a single processor
calculation, on Red Storm, a Cray XT-3 at Sandia National Laboratory.
Even up to 20,000 processors the efficiency of Athena remains above 85\%,
and is nearly flat indicating essentially perfect weak scaling.

\section{Tests}
\label{sec:tests}

Tests are an integral part of the code development process, used
not only to find bugs in the implementation, but also to measure the
fidelity of the method in comparison to other techniques.  In this
section we present a selection of tests that we have found useful in the
development of Athena for both hydrodynamics and MHD in 1D, 2D and 3D.
A more comprehensive set of tests is published on the web.  Many of
the problems are drawn from test suites of our own codes (Stone et
al. 1992) or from those published by other authors (Woodward \& Colella
1984, hereafter WC; Ryu \& Jones 1995, hereafter RJ; T2000; Liska \&
Wendroff 2003, hereafter LW).  Although we begin by showing 1D tests
for hydrodynamics and MHD, our focus will be on the multidimensional
results that follow, since multidimensinal tests are so critical for MHD.

In only a few of the tests do we show the results from more than one
Riemann solver.  In general, we find the most accurate (and often nearly
identical) results are obtained with either the Roe and HLLC solvers
in hydrodynamics, or the Roe and HLLD solver in MHD.  Thus, we use
these solvers interchangeably.  If one solver fails on a particular
test, it will be mentioned in the discussion.

\subsection{One-Dimensional Hydrodynamics}
\label{sec:1D-hydro-tests}

{\em Linear wave convergence.}  One of the simplest, yet most
discriminating tests is to follow the propagation of linear modes of
each wave family in a periodic domain to measure the amplitude of both
diffusion and dispersion errors.  Exact eigenfunctions of sound, contact,
and shear waves are initialized in a uniform medium with $\rho_{0}=1$,
$P_{0}=3/5$, and $\gamma=5/3$. The wave
amplitude $A=10^{-6}$, and the wavelength is equal to the size of
the domain $L=1$.  For sound waves, the background
medium is initially at rest.  (It is also useful to try a test in which
$v_{x,0} = -c_{s}$, where $c_{s}^{2} = \gamma P/\rho$ is the sound speed,
so that the right-propagating sound waves are standing waves.)  For the
contact and shear waves, the background medium has a constant velocity
$v_{x,0} = 1$.  The solution is then evolved for 1 crossing time, or
until $t_{f}=1$.  Figure 7 shows the norm of the L$_{1}$ error vector
for each wave, defined as
\begin{equation}
\delta {\bf q} = \frac{1}{N} \sum_{i} \vert {\bf q}_{i} - {\bf q}_{i}^{0} \vert
\end{equation}
where ${\bf q}_{i}^{0}$ is the initial solution, as a
function of the numerical resolution up to 1024 zones, using third-order
reconstruction and the HLLE, HLLC, or Roe fluxes.  The errors for the HLLC
and Roe fluxes are nearly identical, and converge at second-order for
each wave family.  The errors for the HLLE solver are slightly larger,
and converge at a slightly lower rate.  By plotting profiles of the
waves, we find the errors are dominated primarily by diffusion error;
with 16 or more grid points per wavelength the plots show almost no
dispersion in any of the waves.  A number of
very sensitive tests of the coding can be designed.  Firstly, the L$_{1}$
errors should be identical (to {\em every} digit of accuracy) for left- and
right-propagating waves.  Secondly, convergence should continue until
either the limits of round-off error are reached, or nonlinear steeping
becomes important (when L$_{1} \sim A^{2}$).  We
have found that both double precision, and very small initial amplitudes,
are necessary to see convergence out to 1024 cells.  This suggests that
round-off error can dominate truncation error in very high resolution 
simulations with higher-order methods such as Athena.

{\em Sod shocktube.}  Long a standard test for hydrodynamic codes, the Sod
shocktube consists of two constant states separated by a discontinuity (a
Riemann problem).  Table 2 lists the values in the left- and right-states
for this test.  Figure 8 shows the results for the density, pressure,
velocity, and $P/\rho$ (which is
proportional to the specific internal energy density) at
$t_{f}=0.25$ when run on a grid of 100 cells in the domain $-0.5 <
x < 0.5$ using third-order reconstruction, the HLLC Riemann solver, and
an adiabatic index $\gamma=1.4$.   When configured for 1D
hydrodynamics, Athena reduces to a direct Eulerian PPM code (e.g. \S4
of CW), thus we expect the results should be similar to those published
by e.g., Greenough \& Rider (2003).  As is typical of a PPM code, Athena resolves the shock front and
contact discontinuity with only 2-3 zones.  Although we show this test
for posterity, in our opinion the 1D Sod shocktube should no longer be
considered a discriminating test of algorithms.

{\em Two-interacting blast waves.}  Introduced as a test by WC, this
problem consists of an initially constant density $\rho_{0}=1$
in a stationary medium in a domain of size $L_{x}=1$ with
reflecting boundary conditions, and $\gamma=1.4$.  For $x<0.1$, the
initial pressure is $P=1000$, for $x>0.9$ $P=100$, while $P=0.01$
everywhere else.  The solution is evolved to an arbitrary time of
$t_{f}=0.038$, at which point the shocks and rarefactions generated at
the two discontinuities in the initial state have interacted multiple
times in the domain.  The test is quite sensitive of the ability of the
method to capture the interaction of shocks with contact discontinuities
and rarefactions.  Figure 9 shows the solution computed with Athena
using 400 grid points, third-order reconstruction, the CS limiters,
and the HLLC Riemann
solver, with a reference solution computed using 9600 grid points shown
as a solid line.  In addition, the solution can be compared to figure 2h
of WC.  Note that the contact discontinuity near $x=0.6$ is quite smeared
out in the Athena solution, this seems to be a common property of direct
Eulerian methods (see figures 18 and 19 in Greenough \& Rider 2003), 
the Lagrange-plus-remap version of PPM
seems to capture this feature more sharply (WC, LW).

{\em Shu \& Osher shocktube.}  Introduced by Shu \& Osher (1989),
this test measures the ability of a scheme to capture the interaction
of shocks with smooth flow.  The initial conditions are a strong shock,
initially located at $x=-0.8$, propagating into a background medium
with a sinusoidally varying density in a domain $-1 \leq x \leq 1$ with
adiabatic index $\gamma=1.4$.  Table 2 lists the initial conditions for
this test.  Figure 10 shows the result at $t=0.47$ computed with both
200 and 800 cells using third-order reconstruction, the CS limiters,
and the HLLC solver.  Comparison of this plot with, e.g. figure 5 in
Balsara \& Shu (2000), shows the Athena solution is similar to a 3rd-order
WENO scheme.   The use of the CS limiters significantly improves the
solution in comparison to the original PPM limiters, since with only 200
cells many of the extrema in the postshock gas are unresolved, and are
clipped with the PPM limiters.  We conclude that low-order (less than
5th-order) WENO schemes are not more accurate than 2nd order Godunov
methods like Athena for this test.  A more comprehensive comparison of
Godunov and higher-order WENO schemes is provided by Greenough \& Rider
(2003).  In particular they conclude for problems involving shocks and
discontinuities, second-order Godunov schemes are more accurate per
fixed computational cost.

{\em Einfeldt strong rarefaction tests.} Einfeldt et al. (1991) described
several test problems designed to reveal shortcomings of various Riemann
solvers for hydrodynamics.  In particular, the Roe solver will always fail
on these tests, in the sense that it will produce negative densities and
pressures in the intermediate states for the initial discontinuity in the
first timestep.  For this reason, when using the Roe solver in Athena we
test the intermediate states, and if the density or pressure is negative,
we replace the Roe flux with the HLLE flux for {\em that interface
only}.  As an example, figure 11 shows the results for the density,
pressure, velocity, and $P/\rho$ (which is proportional to the specific
internal energy)
for test 1-2-0-3 in Einfeldt et al. (1991) at $t=0.1$, computed using
200 grid points, $\gamma=1.4$, and second-order spatial reconstruction (the 
initial left- and right-states for this test are given in Table 2).
The profiles of density and pressure are captured accurately.  We find
that the HLLE solver is only needed for one interface in the first
timestep, thereafter the Roe solver returns positive states.  We have also
run the 1-1-2-5 test in Einfeldt et al. (1991); we find this test is less
challenging.

\subsection{One-Dimensional MHD}
\label{sec:1D-mhd-tests}

{\em Linear wave convergence.}
As in hydrodynamics, the convergence of errors in the propagation of linear
amplitude MHD waves is a sensitive test.
For MHD waves, we use a uniform medium with $\rho_{0}=1$, $P_{0}=3/5$,
${\bf B} = (1,\sqrt{2},1/2)$ and $\gamma=5/3$ in a domain of size $L=1$.
These choices give well separated wave speeds: $C_{f}=2$, $C_{A,x}=1$,
and $C_{s}=1/2$ for the fast, Alfv\'{e}n, and slow magnetosonic speeds
respectively.  Exact eigenfunctions for fast and slow magnetosonic,
Alfv\'{e}n, and contact waves for this background state are given in GS05.
These are used to initialize each wave family with amplitude $A=10^{-6}$
and exactly one wavelength in the domain.  Figure 12 shows the norm of the
L$_{1}$ error vector for each wave family as a function of the numerical
resolution up to 1024 zones, using third-order reconstruction and the
HLLE, HLLD, or Roe fluxes.  The errors using the HLLD or Roe fluxes
are nearly identical, converge at second-order, and are slightly lower
than the HLLE fluxes.  As before, this problem can be used as the basis
for a number of very sensitive tests.  For example, standing waves in
each family can be initialized by setting $v_{x,0}$ to the appropriate
wave speed, the L$_{1}$ error should be {\em identical} for left- and
right-propagating waves, and convergence should continue until the limits
of round-off error or wave-steepening effects are reached.

{\em Brio \& Wu shocktube.}
An MHD analog to the Sod shocktube was introduced by Brio \& Wu (1988),
and has now become a standard test for MHD codes.  Table 2 gives
the values of the primitive variables in the left- and right-states.
The longitudinal component of the magnetic field is $B_{x}=0.75$, and is
of course constant everywhere.  The solution is computed with $\gamma=2$.
Figure 13 shows results computed with second-order spatial
reconstruction and the Roe fluxes, on a grid of 800 zones at time
$t_{f}=0.08$.  A reference solution, computed using $10^{4}$ grid points,
is shown as a solid line.  Once again, shocks and contacts are captured
in only 2-3 zones.  Small oscillations are present in the velocity if
third-order reconstruction is used, indicating our TVD limiters could
be improved.  Recently, Torrilhon (2003) has performed a careful study
of the convergence of finite-volume schemes for MHD Riemann problems
similar (but not identical) to the Brio \& Wu shocktube.  We have run
the regular, nearly coplanar problem defined in \S4.2 of that paper.  The 
left- and right-states for this test are given in Table 2, in addition
$B_{x}=1$.
The results, computed using
third-order reconstruction and the Roe solver, are nearly
identical to those shown in figure 7 of that paper, although the Athena
solution with $10^{4}$ grid points is comparable to the solution with
twice as many points in that paper.  At lower resolution (800 grid points) the Athena
solution shows the compound wave structure which appears in dissipative MHD
(similar to figure 6 of Torrilhon 2003).  As the numerical resolution is
increased, the solution converges to the the exact solution for ideal
MHD, which does not contain this structure.  The fact that Athena
shows more rapid convergence to the exact solution for ideal MHD than
the central scheme tested in Torrilhon (2003) is indicative of lower
numerical dissipation.

{\em RJ shocktube 2a.}
RJ introduced a large number of MHD shocktube problems as tests of a 1D
algorithm they developed.  Figure 14 shows the results for the problem
shown in their figure 2a, which we refer to as the RJ2a test.  Table 2
lists the left- and right-states for this test, in addition $B_{x}=2$.  The results in figure 14
are computed using third-order reconstruction and the Roe fluxes on a grid
of 512 cells.  This test is of particular interest because discontinuities
in each MHD wave family are produced from the initial conditions, that
is both left- and right-propagating fast and slow magnetosonic shocks,
left- and right-propagating rotational discontinuities, and a contact
discontinuity.  The results in figure 14 show that Athena captures each
of these discontinuities with 2-4 cells.

{\em RJ shocktube 4d.}
A second test introduced by RJ is shown in their figure 4d, hereafter
we refer to this problem as test RJ4d.  The left- and right-states are
given in table 2, with $B_{x}=0.7$.  The solution at $t_{f}=0.16$ is shown in figure 15
computed with third-order reconstruction and the HLLD fluxes.  The problem
is interesting because it involves a switch-on slow rarefaction and a
slow shock.  Although the HLLD solver does not include the slow wave
explicitly, figure 15 shows these features are all captured well in the
Athena solution using this solver.

\subsection{Two-Dimensional Hydrodynamics}
\label{sec:2D-hydro-tests}

{\em Double Mach reflection.}
Another classic test of hydrodynamic algorithms introduced by WC,
this problem follows the oblique reflection of a Mach 10 shock in air
($\gamma=1.4$).  The interaction of the reflected and incident shocks
produces a triple-point, and between the resulting contact discontinuity
and the reflected shock a short jet is formed along the wall.  The
structure of this jet is very sensitive to the numerical diffusion of
contact waves.  This test requires a time-dependent boundary condition
be applied along the top edge to follow the propagation of the incident
shock; this is easily achieved in Athena using function pointers.  The
problem is initialized following the description in WC.  Figure 16 shows
contour plots of the solution at $t=0.2$ computed with both second- and
third-order reconstruction, and at two different numerical resolutions.
The H-correction described in Appendix C is used for all the calculations
to reduce small amplitude noise in the postshock flow.
The low-resolution ($260\times 80$) results (first and third panels)
show small but distinct changes in the jet between the reconstruction
algorithms.  The third-order reconstruction is slightly less diffusive.
Comparison of the results with those in WC (their figure 4) demonstrate
the differences between the Lagrange-plus-remap version of PPM, and the
direct Eulerian version implemented in Athena.  The results can also
be compared with those from ZEUS shown in figures 15 and 16 of Stone \& Norman
(1992a).

{\em LW implosion test.}
LW have provided an extensive comparison of a wide variety of
hydrodynamic codes using 1D and 2D problems (including some of the 1D
problems presented in \S\ref{sec:1D-hydro-tests}).  We have found
the problem discussed in \S4.7 in LW, hereafter the implosion test, to
be one of the most informative.  It consists of initial states identical
to the Sod shocktube problem separated by a discontinuity inclined at
$45^{\circ}$ in a 2D domain of size $(L_{x},L_{y}) = (0.3,0.3)$ with
reflecting boundary conditions everywhere (a more precise description
of the initial conditions and grid is given in LW).  It produces a
shock wave which initially propagates into the lower left corner, and
a rarefaction which propagates in the opposite direction.  Along the
bottom and left-side walls, the initial evolution is nearly identical to
the double Mach reflection test described above.  The jets along each wall
produced in this interaction collide in the lower left corner, and produce
vortices which propagate outwards along the diagonal.  In the meantime,
a succession of reflected shocks interact with the vortices and contact
discontinuity, driving the Richtmyer-Meshkov instability, and complex
shock reflections and rarefactions (animations of the evolution, available
on the Athena web page, are useful for interpreting the evolution).
Figure 17 shows contours of the density at two times (the same two times
shown in LW) for a solution computed using third-order reconstruction and
the HLLC fluxes.  The key result of the test is the production of the
jet along the diagonal.  Whether this is the correct dynamics was left
uncertain in the discussion in LW: some codes produce it and others do not.
However, we have found the jet is reliant on maintaining symmetry in the
problem.  In directionally-split algorithms, perfect symmetry is lost,
and the collision of the jets in the lower left corner does not eject
vortices along the diagonal.  In dimensionally unsplit algorithms such as
the CTU method in Athena, the jet is clearly formed.  We conclude the
jet is the correct result, and that it is a sensitive test of symmetry.
We consider the preservation of symmetry a further advantage of the
unsplit integrators used in Athena, however the primary motivation for
their use is the preservation of the divergence-free constraint in MHD.

{\em LW Rayleigh-Taylor instability test.}  Another test introduced by
LW in their \S4.6 is the nonlinear evolution of a single mode of the
Rayleigh-Taylor instability.  Two fluids, with densities two and one
respectively, are initialized at rest in a domain of size $(L_{x},L_{y})
= (1/3,1)$ with constant vertical gravitational acceleration $g=0.1$,
and the heavier fluid on top of the light.
The pressure is computed so that the fluids are in hydrostatic
equilibrium, with the sound speed equal to one in the light fluid
at the interface, with $\gamma=1.4$.  The interface between the two
is perturbed with a vertical velocity $v_y = 0.01 \sin (6 \pi x)$.
Running this test requires adding gravitational
source terms to the equations of motion.  In Athena, the source terms for
a fixed gravitational potential are added in such a way as to conserve
total energy exactly,  This extension to the algorithms, along with
the addition of self-gravity in a way that conserves total momentum
exactly, is described in Gardiner \& Stone (in preparation).  Without explicit
viscosity, or surface tension at the interface, there is no one correct
solution to this problem to which all codes should converge.  Instead,
the resulting structure of the interface between the light and heavy
fluids is sensitive to the numerical diffusion of the method, and to
the numerical perturbations introduced by the grid that seed secondary
Kelvin-Helmholtz instability.  Figure 18 shows the results at time $t_{f}=8.5$
computed with
Athena using third-order spatial reconstruction, the HLLC fluxes, and
a grid of $200 \times 400$ cells.  It can be compared directly to the
results of other codes shown in figure 4.8 in LW.  The Athena solution
shows more fine-scale structure than many other methods, but less than
the Lagrange-plus-remap PPM codes.  This may indicate greater diffusion
of contacts in a direct Eulerian PPM code like Athena, or it may also
indicate the effect of a contact steepener (which tends to seed more KH
instability in multidimensions) in the other codes.

\subsection{Two-Dimensional MHD}
\label{sec:2D-mhd-tests}

{\em Circularly polarized Alfv\'{e}n waves.}
Circularly polarized Alfv\'{e}n waves are an exact nonlinear solution to
the equations of MHD.  T2000 introduced the propagation of these
waves as a sensitive test of dispersion properties of MHD algorithms.
Although such waves are subject to a parametric instability (Del Zanna
et al. 2001), for the parameters adopted by T2000 no
instability should be present.  A complete description of this test,
including the procedure for initializing the solution at an oblique
angle to the mesh, is presented in GS05.  This test has proved
extremely useful for developing Athena.
Figure 19 shows profiles of the waves after propagating 5 crossing times
as a function of resolution, computed using third-order reconstruction,
the CS limiters,
and the Roe fluxes, for both traveling and standing waves.
Dispersion error is seen to be important only at the lowest resolution,
diffusion error generally dominates (this is also true for the linear
wave convergence tests described in \S\ref{sec:1D-hydro-tests}
and \S\ref{sec:1D-mhd-tests}).  Even with only 8 grid points per
wavelength, the wave profile is captured well with an amplitude at
least 0.8 of the original.  With 16 or more grid points per wavelength,
the amplitude is better than 0.95 the original in both cases.  The CS
limiter greatly improves the solution at low resolution, as it prevents the
clipping of extrema in the wave profile.  Figure 20 
shows the norm of the L$_{1}$ error vector as a function of resolution
for traveling waves, after propagating one wavelength, for both second-
and third-order reconstruction.  For comparison, the errors on both a
1D and 2D grid are shown.  In all cases, second-order convergence is
evident, with the 2D errors larger by a factor of about two.

{\em Advection of a field loop.}
This test was introduced and discussed extensively in GS05; it consists of
the advection of a circular field loop by a constant velocity inclined
to the grid in a periodic 2D domain.  For the CT algorithm, solving
field advection problems is non-trivial.  This test demonstrates the
importance of constructing the line-averaged corner-centered emfs used
by CT from the area-averaged face-centered electric fields returned by
the Riemann solver using the technique outlined in \S\ref{sec:CT_Alg}
with the CTU integrator.  Along with the circularly polarized
Alfv\'{e}n wave test described above, this test has been critical to
the development of the algorithms.   Figure 21 shows the magnetic field
lines and contours of the out-of-plane component of the
current density ${\bf J} = \nabla \times {\bf
B}$ after advection of the loop twice around the domain.  The current
density is particularly sensitive to diffusion or oscillations in the
field.  The figure shows the CTU+CT algorithm in Athena preserves the
shape of the field loop extremely well.  We have also checked that if this
test is performed with a uniform $v_{z} \ne 0$, the code keeps $B_{z}=0$
to round-off error (provided it was zero to begin with).  As discussed at
the beginning of section \ref{sec:threed-integration},
this confirms our formulation of CT
preserves the appropriate discretization of the divergence-free constraint.

{\em Orszag-Tang vortex.}
A 2D MHD test which has now become a standard is the evolution of
the vortex of Orszag \& Tang (1979).  There is some confusion in the
literature as to the time at which comparisons between solutions are made.
The results shown here are computed with constant initial densities and
pressure, $\rho_{0} = 25/(36\pi)$ and $P_{0} = 5/(12\pi)$, in a
periodic domain of size $(L_{x},L_{y}) = (1,1)$, with an initial velocity
$(v_{x}, v_{y}) = (-\sin(2\pi y),\sin (2\pi x))$, and a magnetic field
computed from the vector potential $A_{z} = (B_{0}/4\pi) \cos (4\pi x) +
(B_{0}/2\pi) \cos (2 \pi y)$, with $B_{0}=1/\sqrt{4\pi}$.  Figure 22 shows
contour plots of the density, pressure, magnetic pressure, and specific
kinetic energy density at time $t_{f}=1/2$ computed on a grid of $192 \times
192$ cells, which can be compared directly to the results in, e.g.,
T2000 at a time of $t_{f}=\pi$.
Of particular note is the symmetry in the solutions.
Figure 23 shows horizontal slices of the pressure at $y=0.3125$ and
$y=0.427$ (shown by the horizontal lines in the upper right panel of
figure 22), with the solution on a $512^2$ grid shown as a solid line
for reference.  This test does not seem to be extremely discriminating
for MHD algorithms.  (We consider linear wave convergence
(see \S\ref{sec:3D-mhd-tests}), circularly polarized Alfv\'{e}n waves,
and field loop advection to be more quantitative MHD tests.)  The
most stringent comparison between methods is provided by the slices
shown in figure 23.
Finally, figure 24 plots contours of the density, magnetic pressure,
specific kinetic energy density, and total pressure $P^{*}$ for an
isothermal version of the Orszag-Tang vortex test.  Comparison to results
shown previously by Balsara (1998, see his figure 8) appear to show significant differences.

{\em MHD Rotor.}
The test suite of Stone et al. (1992) contained tests based on the
propagation of nonlinear amplitude shear Alfv\'{e}n waves in 1D generated
by rotating disks in axisymmetry.  Since analytic solutions are available
for this problem, it was possible to provide quantitative measure of
the errors in ZEUS.  (We have confirmed Athena reproduces these tests
accurately, with second-order convergence on the version of the test that uses
continuous initial conditions.)  Following the suggestion of Brackbill
(1986), Balsara \& Spicer (1999) introduced a 2D version of this test
consisting of a rotating disk located in the plane of the computation,
with an initial magnetic field perpendicular to the rotation axis.
Strong rotational discontinuities are generated in the field due to the
shear at the surface of the disk, and shocks and rarefactions are produced
by the radial expansion of the disk due to unbalanced centrifugal forces.
We use the initial conditions as described by T2000.  We present results
only for the problem labeled ``Rotor Test \# 1", as it involves higher
initial velocities and is therefore more difficult.  No smoothing is used
at the surface of the disk.  Figure 25 shows contours of the density,
pressure, Mach number, and magnetic pressure at $t_{f}=0.15$ on a grid
of $400 \times 400$ cells, computed using third-order reconstruction
and the Roe fluxes.  Figure 26 plots slices of the $y-$component of the
magnetic field taken along $x=0$, and the $x-$component of the magnetic
field taken along $y=0$.  Of note is the near-perfect symmetry maintained
in the solutions, with no oscillations.  In particular, contours of the
Mach number remain concentric circles in the rarefaction at the center
all the way to the origin.  Similarly, the slices show constant field
strength within the central rarefaction, and sharp discontinuities.

{\em Magnetic Rayleigh-Taylor instability.}
To show the effect of magnetic fields on the nonlinear evolution of
the 2D RT instability, and to demonstrate the use of AMR with Athena,
figure 27 shows the results of a single mode RT instability computed
with 5 levels of refinement.  A base grid of $8 \times 16$ cells is used,
giving an effective resolution on the finest grid of $256 \times 512$.
The parameters for this calculation are not identical to those used for
the LW hydrodynamic RT test shown in figure 16.  In particular, for the MHD
test we use a domain of size $(L_{x},L_{y}) = (0.1,0.2)$ with $g=0.1$,
an adiabatic index $\gamma=5/3$, and densities in the light and heavy
fluids of $\rho_{l}=1$ and $\rho_{h}=3$ respectively.
The magnetic field is initially uniform and horizontal, with initial amplitude
$B_{0}$ compared to the critical value $B_{c}=[Lg(\rho_{h} - \rho_{l})]^{1/2}=0.14$
which suppresses instability of $B_{0}/B_{c} = 0.05$.
The figure shows the distribution of a passive contaminant
advected with the flow at a final time $t_{f}=3$ in order to show mixing,
as well as the grid levels used in the AMR calculation.  For reference,
the identical calculation but without the magnetic field is shown as well.
Note the suppression of secondary KH instabilities at the interface 
in the MHD case.  An extensive discussion of the nonlinear evolution of the
magnetic RT instability is presented in Stone \& Gardiner (2007a; 2007b).
The use of an AMR grid is very efficient for this problem, since the
refinement is predominantly required near the interface.

{\em Blast wave in a strongly magnetized medium.}
In order to demonstrate the propagation of strong MHD shocks in
multidimensions, we show the results of an MHD blast wave problem.
Many authors have performed similar versions of this test, we adopt the
initial conditions
used in Londrillo \& Del Zanna (2000).  The results
are shown at time $t_{f}=0.2$ in figure 28 using a domain of size
$(L_{x},L_{y}) = (1,3/2)$
with a grid of $200 \times 300$ cells, third-order reconstruction, and
the HLLC (hydro) and HLLD (MHD) fluxes.  The top row shows contour plots from
a hydrodynamic version of this test, while the lower row shows the MHD results
with an initial magnetic field inclined at $45^{\circ}$ to the grid
(${\bf B} = (B_{0}/\sqrt{2}, B_{0}/\sqrt{2})$ where $B_{0}=1$.
By using periodic boundary conditions, the flow becomes more complex as
the outgoing blast wave re-enters the grid on the opposite side, and
interacts with the contact discontinuity that bounds the evacuated bubble at
the center.  Figure 29 shows the result at $t_{f}=1$ for both the
hydrodynamic and MHD problem.  Note the CTU integrator preserves perfect
symmetry (most noticable in the fingers at the contact discontinuity 
generated by the Richtmyer-Meshkov instability in the unmagnetized problem).
Also note the magnetic field suppresses the R-M instability (Wheatley et al.
2005).  Finally, figure 30 plots contours of the MHD blast problem using an
isothermal equation of state and both $B_{0}=1$ (top row)
and $B_{0}=10$ (bottom row).  The plasma $\beta = 2P/B^{2} = 2$ for
$B_{0}=1$, and $\beta = 0.02$ for $B_{0}=10$ in the external medium initially.
GS05 shows results for adiabatic MHD with $B_{0}=10$.
This problem demonstrates the CTU+CT algorithm is robust for
low$-\beta$ flows.

\subsection{Three-Dimensional Hydrodynamics}
\label{sec:3D-hydro-tests}

{\em Noh's strong shock.}
As a fully 3D hydrodynamical test, we present results from the strong
shock test described by Noh (1987).  This is a very difficult test.
A uniform ($\rho_{0}=1$), cold ($P=0$) medium converges in a spherically
symmetric radial inflow $v_{r}=-1$ onto the origin.  This generates
a very strong (formally, $M = \infty$) spherical shock wave which
propagates away from the origin at constant velocity $V_{s}=1/3$.  Due to
the spherical convergence, the preshock density increases everywhere in time
according to $\rho(r,t) = \rho_{0} (1 + t/r)^{2}$.  However, the density
immediately upstream of the shock location is always 16,
thus the postshock gas is uniform with $\rho = 64$ for $\gamma=5/3$.
A similar test is often run in planar (1D) and cylindrical (2D)
symmetry, however when run with a Cartesian grid the 3D test presented
here is probably the most difficult.  In practice Athena cannot be run
with pressure identically zero, thus initially we set $P_{0}=10^{-6}$
everywhere.  The problem is run until $t_{f}=2$ in a
domain of size $(L_{x},L_{y},L_{z}) = (1,1,1)$ computed only
in the positive octant with $200^{3}$ cells.
The inner boundary condition in each dimension is reflecting.  At the
outer boundary the density is evolved according to the analytic solution
for the preshock flow, the radial velocity is held fixed at $v_{r}=-1$,
and the entropy is evolved identically to the density, i.e. $P(R_{B},t)
= P_{0}(1+t/R_{B})^{2(1+\gamma)}$, where $R_{B}$ is the spherical radius
of the boundary.  Figure 31 shows contours of the
density at $t=t_{f}$ computed using second order reconstruction, the Roe
flux, and the H-correction (see Appendix C).  Note the contours are smooth
and spherical, with virtually no noise in the postshock gas.  Also shown
is a scatter plot of $\rho(r)$ versus $r$ for every eighth grid cell
in the computation.   The solution has the correct density jump and
shock speed.  The small scatter behind the shock demonstrates that
with the H-correction, the shock remains sharp, smooth, and spherically
symmetric.  Near the origin, the small dip in the density is a signature
of wall-heating (Noh 1987).  These plots can be compared to figure 4.7
in LW, who ran the same test but in 2D cylindrical symmetry.  Only a few of
the algorithms tested by LW were able to run the test at all.  The 3D results
shown in figure 31 are similar to the best result in LW (for PPM).  Without the
H-correction, Athena still runs this test but the shock develops strong
perturbations along the grid directions, similar to but somewhat stronger
than those evident
in the results for the VH-1 code shown in LW.  Finally, at low resolutions
(less than $50^{3}$), this test can cause Athena to crash when the
Roe solver is used, even with the H-correction, unless the CFL number
is reduced.

\subsection{Three-Dimensional MHD}
\label{sec:3D-mhd-tests}

{\em Linear wave convergence.}
We have argued that tests of MHD codes must be multidimensional, yet
the most quantitative tests generally involve plane-wave (1D) solutions.
Sensitive multidimensional tests can be constructed by simply inclining
the plane wave to the grid at an arbitrary angle.  Here, we measure
the convergence rate of Athena for each MHD wave family in 3D by
initializing a plane waves solution at an oblique angle to a grid of size
$(L_{x},L_{y},L_{z}) = (3,3/2,3/2)$, using the same initial conditions
as in the 1D test described in \S\ref{sec:1D-mhd-tests} and a grid with
resolution of $2N \times N \times N$ cells, with $N=4, 8, 16, 32, 64$ and
128.  The angle of the wavevector is chosen so that it does not lie
along the diagonal of a grid cell, that is there are no symmetries in the
fluxes in any direction.  Details of the initialization of this problem
in 3D, which requires care to prevent grid noise along the
wave front, are given in GS08.  Figure 32 shows the norm of the L$_{1}$
error vector for each wave family using both second- and third-order
reconstruction computed with the HLLD solver, as a function of resolution
along $L_{x}$.  For comparison, the errors for this same problem in 1D are
shown as a dashed line.  Again, we see second order convergence in all
wave families.  The amplitude of the errors in the fast wave are higher
than the 1D case by about a factor of two, but for all other waves
the errors are comparable.  The fact that the errors in 3D are not
significantly larger than those in 1D reflects the fidelity of the
multidimensional CTU+CT algorithm.

{\em Circularly polarized Alfv\'{e}n waves.}
We initialize a 1D plane wave solution corresponding to the parameter
values given by T2000 on a grid of size $(L_{x},L_{y},L_{z}) =
(3,3/2,3/2)$, with the wave front oblique to the grid, following the
procedure given in GS08.  The technique for initializing the wave
solution at an oblique angle is similar to that used above for linear waves.
Figure 33 plots profiles of the traveling wave at different resolutions
using third-order reconstruction, the CS limiters, and the HLLD fluxes.
Also shown are the norm of the L$_{1}$ error vector computed using
both second- and third-order reconstruction.  These results can be
compared directly to the 2D results shown in Figure 19.  Once again,
the solution in 3D compares extremely favorably with the 2D solution,
for example the L$_{1}$ errors are nearly identical to the 2D errors
for an adiabatic equation of state.

{\em Advection of a field loop.}
On a 3D grid, we have found there are two challenging versions of
this test that can be attempted.  The first is the 3D analogue of the
test described in \S\ref{sec:2D-mhd-tests}, that is a cylindrically
symmetric field loop with $B_{z}$ = 0, but with a constant advection
velocity along the grid diagonal so that $v_{z} \ne 0$.  As discussed
in \S\ref{sec:threed-integration}, the numerical algorithm should
maintain $B_{z}=0$, which can only be achieved if the code maintains
the balance between the two nonzero terms in the $z-$component of the
induction equation, that is $v_{z}(\partial B_{x}/\partial x + \partial
B_{y}/\partial y)=0$.  In turn, for constant $v_{z}$, this requires the
code to maintain the divergence-free constraint properly.  Since the 3D
CTU+CT algorithm in Athena has been designed to reduce exactly to the
2D version for problems with symmetry in $z$, we obtain the identical
results for the profile of the field loop in an $x-y$ slice in this
test as shown in figure 21.  Moreover, we confirm that Athena maintains
$B_{z}=0$ to round-off.  A second sensitive test is to incline the field
loop at an oblique angle to the grid, and advect it with a velocity
along a perpendicular diagonal (see GS08 for details).  The resulting
current density after advecting the loop twice around the grid for both
second- and third-order reconstruction is shown in Figure 34 for a grid
of size $(L_{x},L_{y},L_{z}) = (1,1,1)$ with $128^{3}$ grid points, and
the HLLD fluxes.  In this case, the component of the field along the
axis of the cylinder should remain zero.  Although it is not possible
to enforce this constraint to round-off error (as was the case when the
axis of the field loop is parallel to a grid direction), nonetheless we
find that this component is zero to truncation error (see GS08).

{\em MHD shocktube inclined to the grid.}
To demonstrate the ability of the 3D algorithm to capture shocks
and discontinuities that propagate at an oblique angle to the mesh, we have
repeated the RJ2a test described in \S\ref{sec:1D-mhd-tests} on a 3D
grid of size  $(L_{x},L_{y},L_{z}) =
(3/2,1/64,1/64)$, with the initial discontinuity oblique to the grid, using
a mesh of $768 \times 8 \times 8$ grid points.  This gives an effective
resolution along the direction of shock propagation which is equivalent
to the 1D test.  Initializing the discontinuity so as to avoid introducing waves
transverse to the interface requires care: for more detail see GS08.
The results, at a time of $t_{f}=1$ for the HLLD fluxes
and second-order reconstruction, are shown in Figure 35.  Note that in
3D, each of the shocks, contact and rotational discontinuities have been
captured; there is excellent agreement between the profiles shown in
figure 35 and the equivalent 1D profiles shown in Figure 14.

{\em Blast wave in a strongly magnetized medium.}
As a final 3D test, we follow the growth of a strong, spherical blast wave
in a strongly magnetized medium.  The initial conditions are identical to 
those given in \S\ref{sec:2D-mhd-tests}, the only difference being that
we run the
problem on a 3D grid of size  $(L_{x},L_{y},L_{z}) =
(1,1.5,1)$, with $200 \times 300 \times 200$ grid points.   Figure 36 shows
slices of the density and magnetic pressure taken at $t=0.2$ computed
using the HLLD solver and third-order reconstruction.  The primary
difference in the solution compared to 2D is that the size of the bubble grows
more slowly in 3D, due to the increased adiabatic cooling in 3D diverging
flow.  The contours are all symmetric and smooth, with no visible asymmetries
introduced by the grid.

\section{Summary}
\label{sec:summary}

We have described Athena, a new code for astrophysical MHD.  The
code implements algorithms based on higher-order Godunov methods,
with a finite-volume discretization to evolve volume-averages of
the mass, momentum, and total energy density, and a CT algorithm
(finite-area) discretization to evolve area-averages of the
face-centered components of the magnetic field.  This combination
conserves the total mass, momentum, energy, and magnetic flux through
the grid exactly.  Such conservative algorithms are an essential
ingredient of AMR methods.

The mathematical foundation of the 2D and 3D algorithms in Athena are
described more fully in GS05 and GS08.  In this paper, we have focused on
the detailed implementation of the methods into a functioning computer
code.  Step-by-step descriptions are provided of the multidimensional
integrator for MHD in 2D and 3D (based on the CTU algorithm of Colella
1990), the 1D reconstruction algorithms (based on an extension of the
PPM algorithm of CW to multidimensional MHD), and a variety of 1D Riemann
solvers used to compute upwind fluxes.  We have emphasized the importance
of using dimensionally unsplit integrators for MHD, the advantages of
using the staggered grid formulation of CT (which requires techniques
for constructing edge-averaged, corner-centered emfs from area-averaged
face-centered electric fields returned by the Riemann solver), and the
need to test MHD codes with multidimensional problems in order to reveal
errors associated with the divergence-free constraint.

An extensive series of test problems in 1D, 2D, and 3D for both
hydrodynamics and MHD have been presented.  These tests, and others
published on the web, should be useful to others developing and testing
codes for astrophysical MHD.  The tests show Athena is second-order
accurate in space and time for smooth solutions in all MHD wave
families, even in multidimensions.  We have shown that an advantage of
directionally unsplit methods is that they preserve symmetries inherent
in the flow.  The 2D CTU+CT method described here reduces identically
to the 1D algorithm for plane-parallel grid-aligned flows.  Similarly,
the 3D CTU+CT method reduces exactly to either the 2D or 1D methods for
plane-parallel, grid-aligned flows, according to the appropriate symmetry.
We have exploited such symmetries to design a sensitive test of the
appropriate stencil for maintaining the divergence-free constraint.
A planar field loop, advected in a fully 3D velocity field, must remain
planar.  Since the evolution of the component of the field normal to
the plane of the loop is governed by a term proportional to $\nabla
\cdot {\bf B}$, the loop will only remain planar if the divergence-free
constraint is satisfied exactly on the appropriate stencil.

In addition to the CTU+CT integrator described in this paper, an
unsplit integrator based on the method described by Falle (1981) and similar
to the MUSCL-Hancock scheme described by
van Leer (2006) has
been implemented in Athena.  The details of this VL+CT method, including
tests in 3D and comparisons to the CTU+CT method described here,
are given in SG08.

The primary motivation for developing Athena has been the need to adopt
static and adaptive mesh refinement (SMR and AMR) to resolve flows
over a wide range of length scales in various astrophysical
applications of interest in our research groups (such as magnetized accretion flows, and gravitational
collapse and fragmentation in dense phases of the ISM)  In
\S\ref{sec:2D-mhd-tests} we have shown the results of tests of
AMR calculations of the Rayleigh-Taylor instability with Athena.
Both SMR and AMR add considerable complexity to the algorithms,
requiring special care to conserve mass, momentum, energy, {\em and}
magnetic flux at fine/coarse grid boundaries.  The
implementation of SMR and AMR with the CTU+CT integrator in Athena will
be given in a future communication.

Other extensions to Athena include adding gravitational source terms
for both a static gravitational potential and self-gravity (Gardiner
\& Stone, in preparation), the shearing box (Gardiner \& Stone 2006),
anisotropic heat conduction (Parrish \& Stone 2005; 2007), and transfer of
ionizing radiation (Krumholz et al. 2007).   Many more are either underway
(curvilinear coordinates, relativistic MHD, full transport radiation MHD),
or planned for the future.

Athena has moved beyond the developmental phase, and is now being used for
a variety of applications, including studies of the MRI in the
shearing box (Gardiner \& Stone 2006), colliding winds in close binaries
(Lemaster et al. 2007), decay of hydrodynamical turbulence in
the shearing box (Shen et al. 2006), the magnetic Rayleigh-Taylor 
instability (Stone \& Gardiner 2007a; 2007b), shock interactions with
magnetized clouds (Shin et al. 2007),
and the decay of supersonic turbulence
in magnetized molecular clouds (Lemaster \& Stone, in preparation).

The Athena code has been made publicly available, and can be
downloaded from the web, along with extensive documentation.  Additional
test problems beyond those presented here are also described on the
web.  We are confident that Athena will become the workhorse for
our own applications; it is hoped that the description of the
algorithms provided in this paper, along with the public version of the
code provided on the web, will be useful to others for solving many problems
in astrophysical fluid dynamics.

\acknowledgements
We have benefited from discussions with many people during the
development of Athena; in particular
we would like to acknowledge
Phil Colella, Charles Gammie, Mark Krumholz, Nicole
Lemaster, Eve Ostriker, and Ian Parrish
for their input.
Development of the Athena code was initially supported by the NSF ITR program.
JS thanks the Royal Society for financial support through the Wolfson
Research Merit scheme during 2002-2003.  Additional support was provided
by the DOE through DE-FG52-06NA26217.
Simulations were performed on
the Teragrid cluster at NCSA, the IBM Blue Gene at Princeton University, and
on computational facilities supported by NSF grant AST-0216105.



\begin{table}[h]
\caption{Performance$^{1}$ of Athena in 3D on a 2.2 GHz Opteron processor}
\begin{tabular}{cccc} \hline \hline \\
physics & Roe solver & HLLC solver & HLLD solver \\ \hline \\
isothermal hydro & 328 & 340 & - \\
adiabatic hydro & 224 & 242 & - \\
isothermal MHD & 108 & - & 124 \\
adiabatic MHD & 85.9 & - & 97.6
\end{tabular} \\
$^{1}$ measured in thousands of cell updates per cpu second.
\end{table}

\begin{table}[h]
\caption{Left- and right-states for 1D Riemann Problems}
\begin{tabular}{ccccccccccccccc} \hline \hline \\
Test 
& $\rho_{L}$ & $v_{x,L}$ & $v_{y,L}$ & $v_{z,L}$ & $P_{L}$ & $B_{y,L}$ & $B_{z,L}$
& $\rho_{R}$ & $v_{x,R}$ & $v_{y,R}$ & $v_{z,R}$ & $P_{R}$ & $B_{y,R}$ & $B_{z,R}$
\\ \hline \\
Sod & 1.0 & 0 & 0 & 0 & 1.0 & - & - & 0.125 & 0 & 0 & 0 & 0.1 & - & - \\
Shu-Osher & 3.857143 & 2.629369 & 0 & 0 & 10.3333 & - & - & $1+0.2\sin (5\pi x)$ & 0 & 0 & 0 & 1 & - & - \\
Einfeldt-1203 & 1.0 & -2.0 & 0 & 0 & 0.4 & - & - & 1.0 & 2.0 & 0 & 0 & 0.4 & - & - \\
Brio \& Wu & 1.0 & 0 & 0 & 0 & 1.0 & 1.0 & 0 & 0.125 & 0 & 0 & 0 & 0.1 & -1.0 & 0 \\
Torrilhon & 1.0 & 0 & 0 & 0 & 1.0 & 1.0 & 0 & 0.2 & 0 & 0 & 0 & 0.2 & $\cos(3)$ & $\sin(3)$ \\
RJ2a & 1.08 & 1.2 & 0.01 & 0.5 & 0.95 & 3.6 & 2 & 1 & 0 & 0 & 0 & 1 & 4 & 2 \\
RJ4d & 1 & 0 & 0 & 0 & 1 & 0 & 0 & 0.3 & 0 & 0 & 1 & 0.2 & 1 & 0 

\end{tabular}
\end{table}
\clearpage


\begin{figure}
\epsscale{0.4}
\plotone{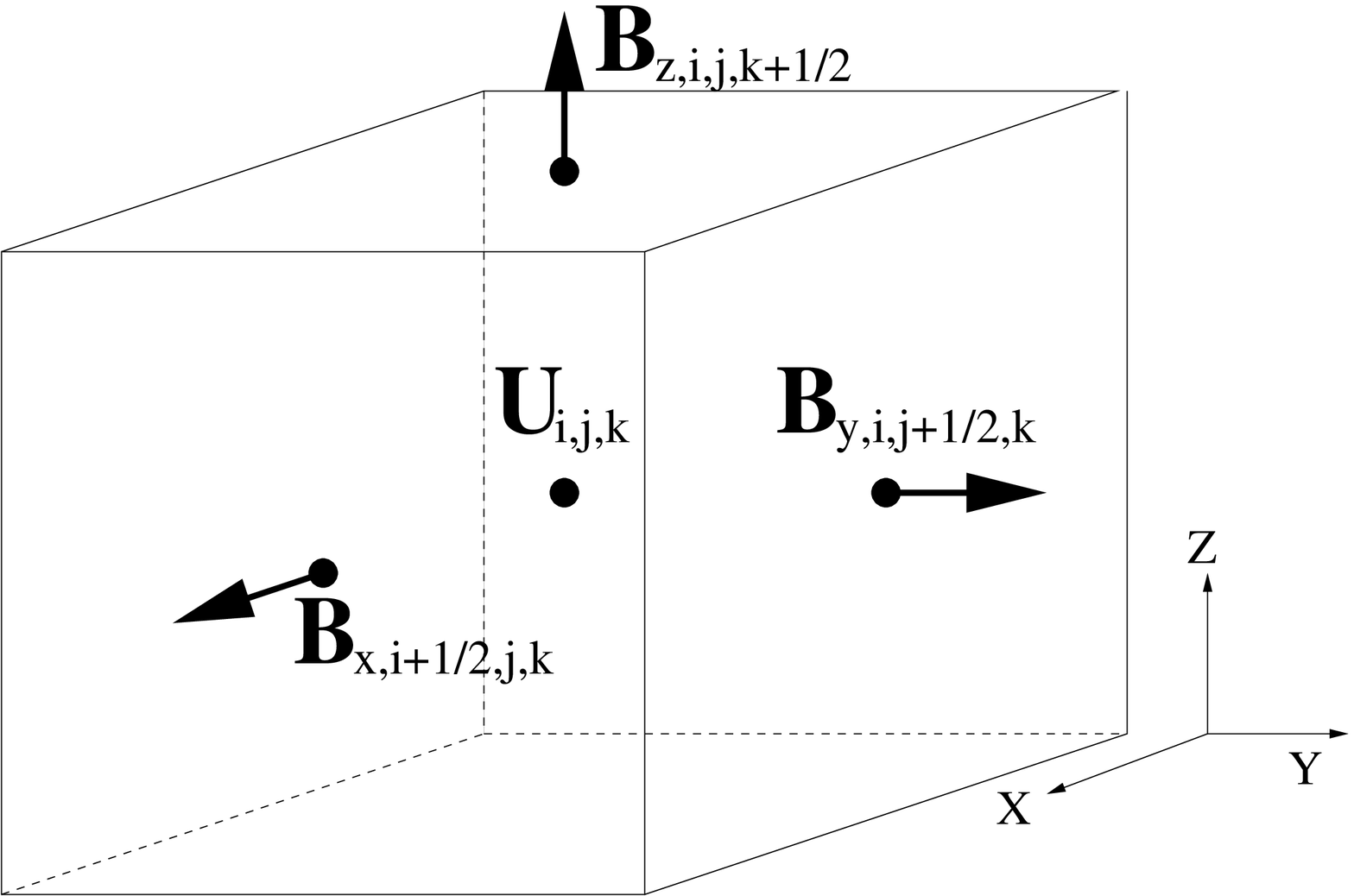}
\plotone{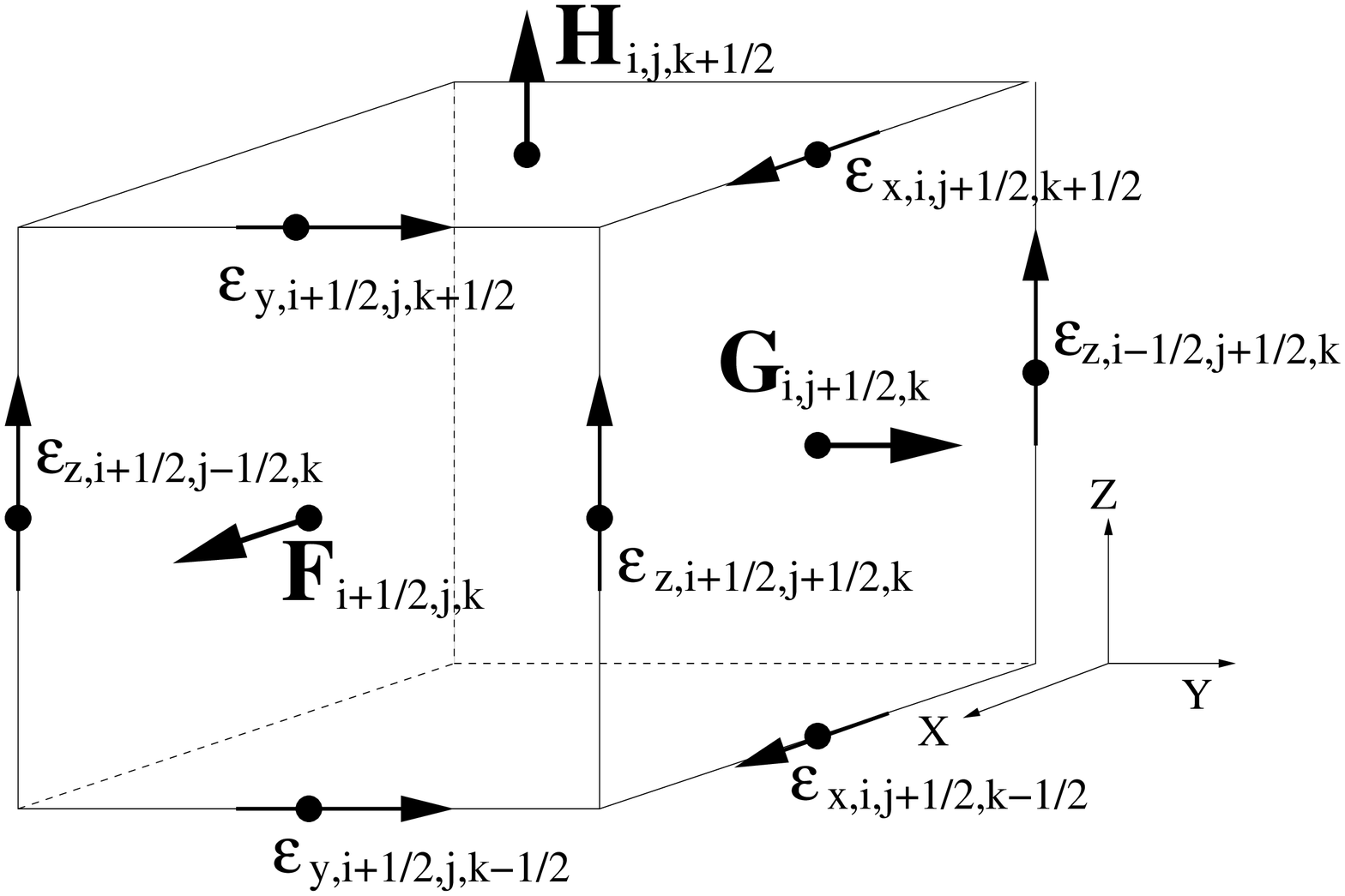}
\figcaption
{({\em left}) Centering of volume-averaged conserved variables ${\bf U}$
and area-averaged components of
magnetic field ${\bf B}$ on the grid. ({\em right}) Centering of time-
and area-averaged
components of the fluxes of ${\bf U}$, and the time- and line-averaged
emfs on the grid.}
\end{figure}
\clearpage

\begin{figure}
\epsscale{0.3}
\plotone{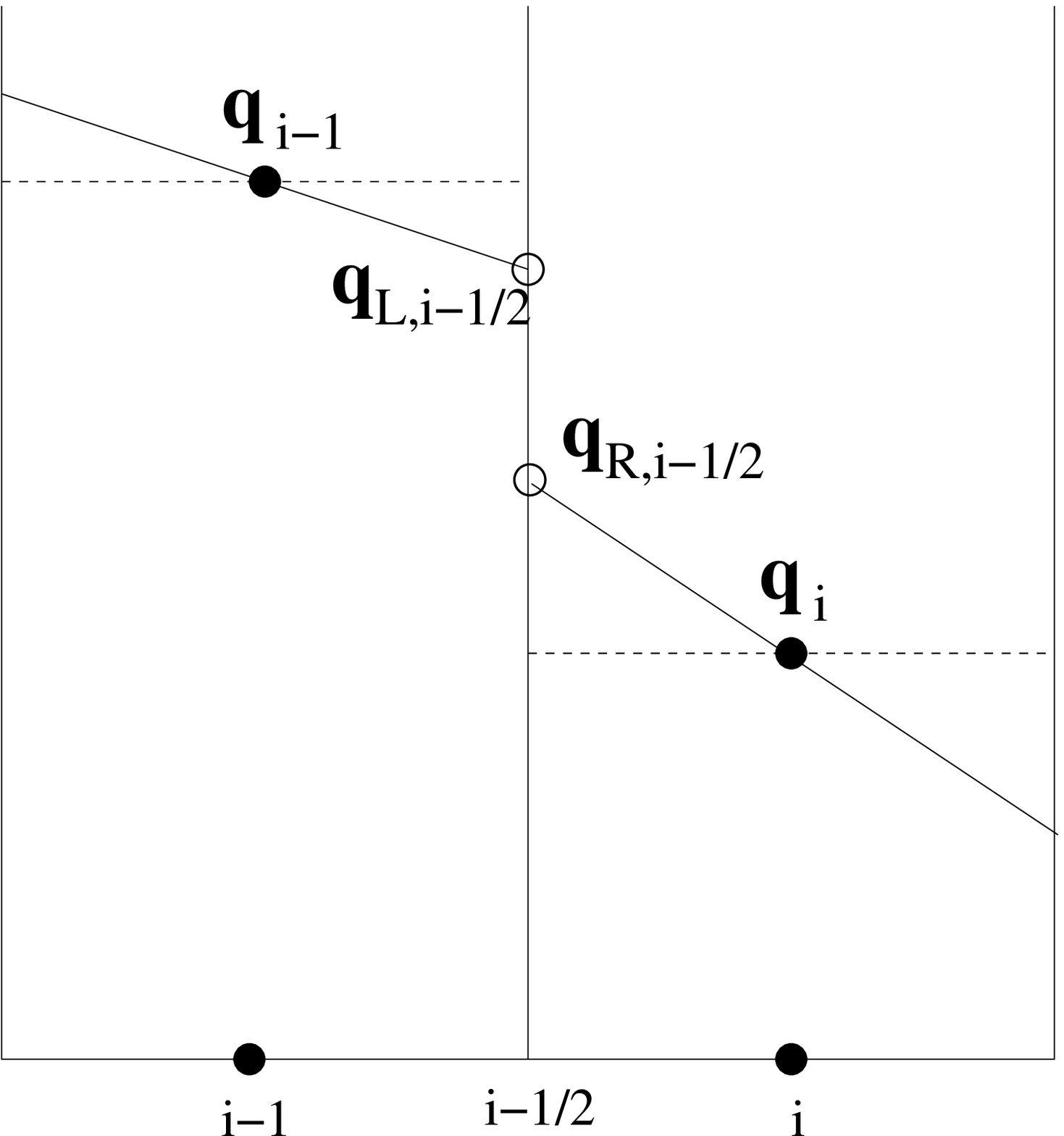}
\epsscale{0.5}
\plotone{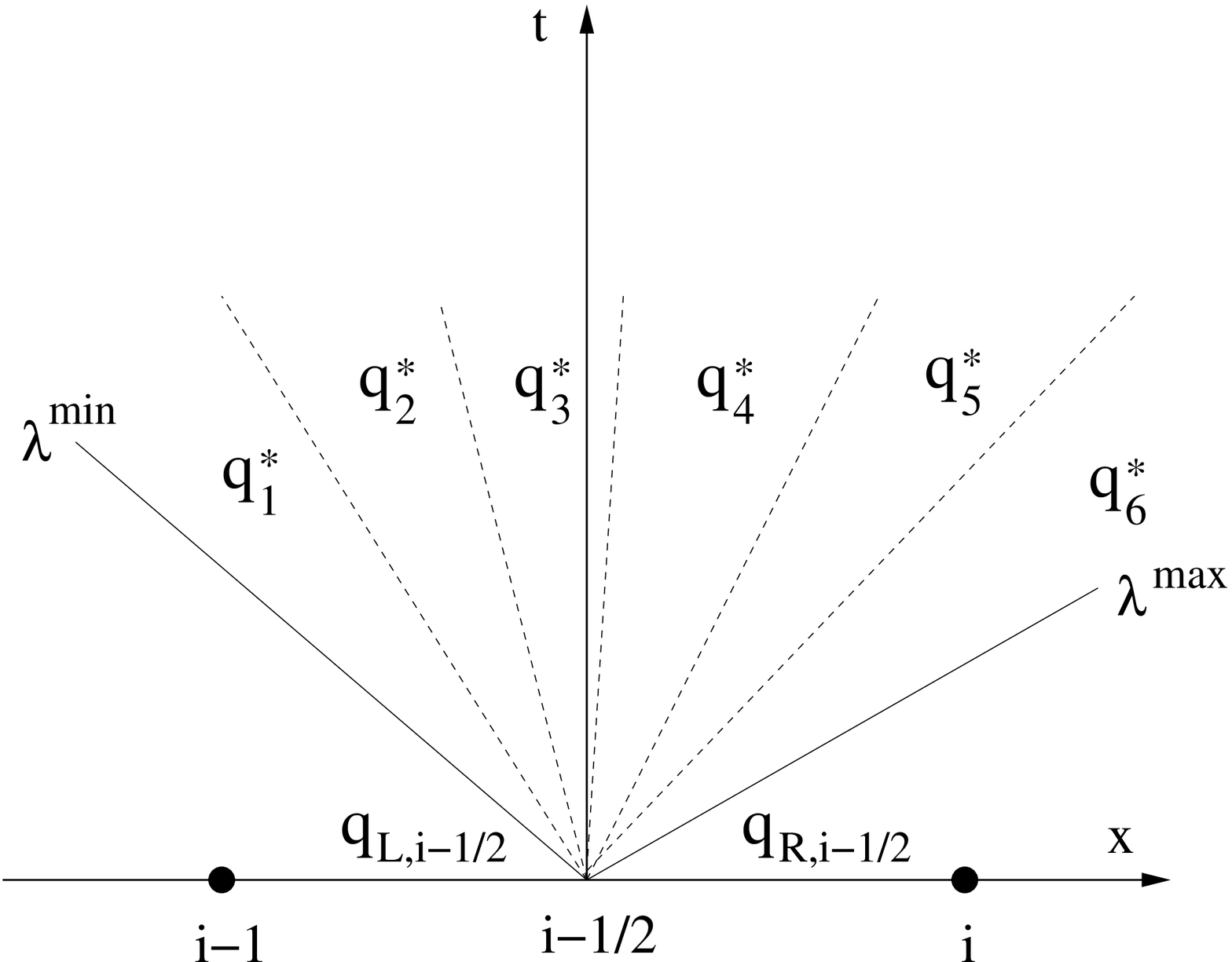}
\figcaption
{({\em left)} An example of piecewise linear reconstruction of conserved
variables within each cell to compute the left- and right-states that
define a Riemann problem at the cell interface.  The slopes chosen within each
cell are determined by limiters which depend on neighboring cell-center
values (not shown) designed to prevent the introduction of new extrema.
({\em right}) Schematic
solution of an MHD Riemann problem in spacetime, consisting of six
intermediate states bounded by the maximum and minimum wavespeeds.
The flux through the interface is the time integral of the solution
along the vertical line $x=x_{i-1/2}$, in this case given by the quantities in
state ${\bf q}^{*}_{3}$.  In MHD, some characteristics can be degenerate,
meaning that the number of intermediate states depends on the problem.}
\end{figure}

\begin{figure}
\epsscale{0.8}
\plotone{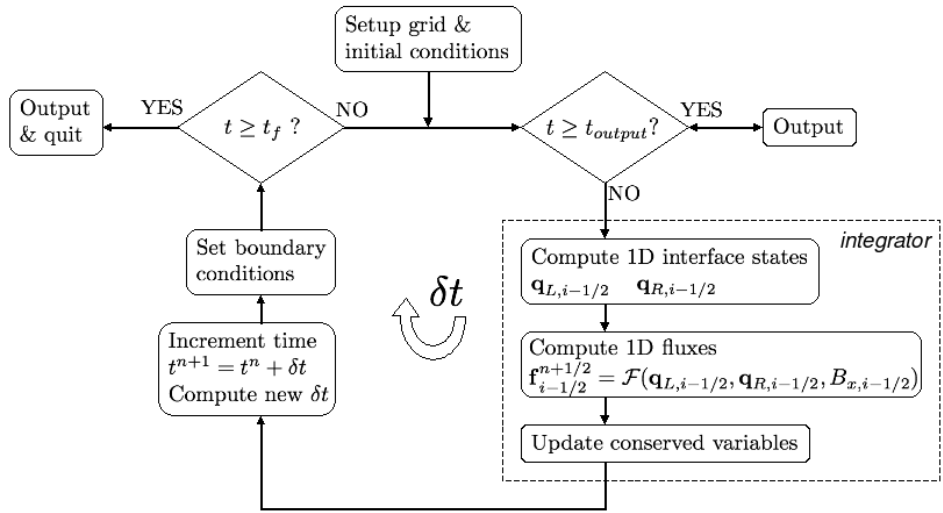}
\figcaption
{Flow chart for integration in 1D.  The dashed box groups functions that
are part of the 1D integrator (described in \S\ref{sec:oned-steps})
}
\end{figure}
\clearpage

\begin{figure}
\epsscale{0.8}
\plotone{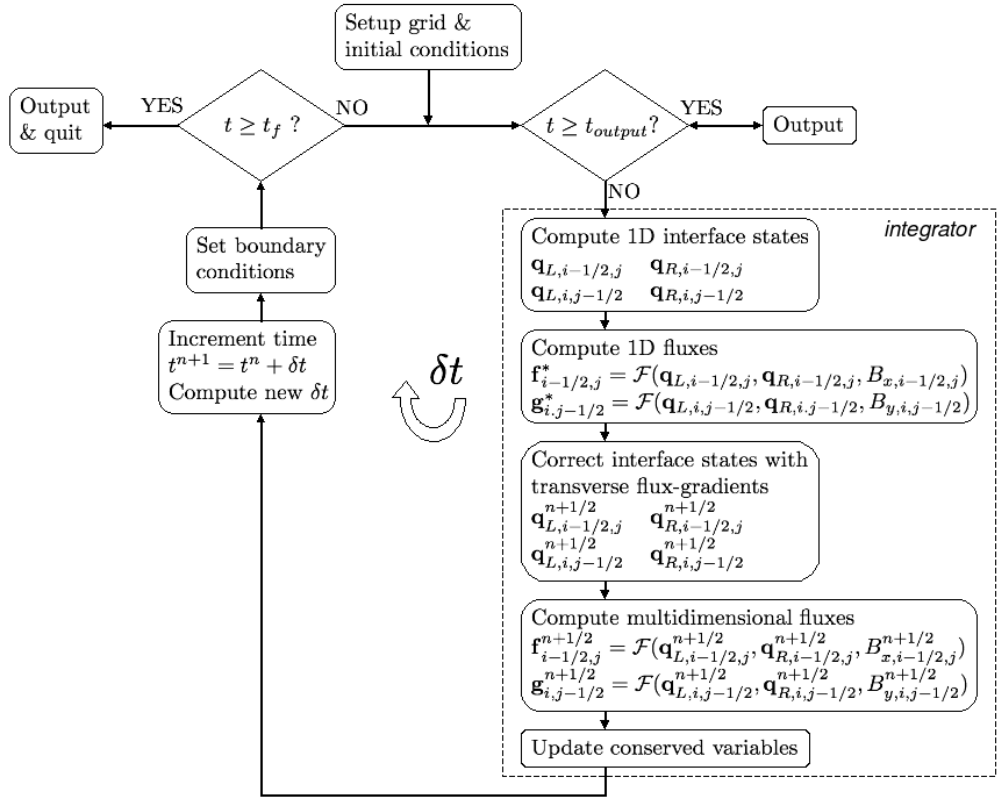}
\figcaption
{Flow chart for integration in 2D.  The dashed box groups functions that
are part of the 2D integrator (described in \ref{sec:twod-steps}).  These
steps are schematic, with many details omitted.  The flow chart for the
3D integrator is similar.
}
\end{figure}
\clearpage

\begin{figure}
\epsscale{0.6}
\plotone{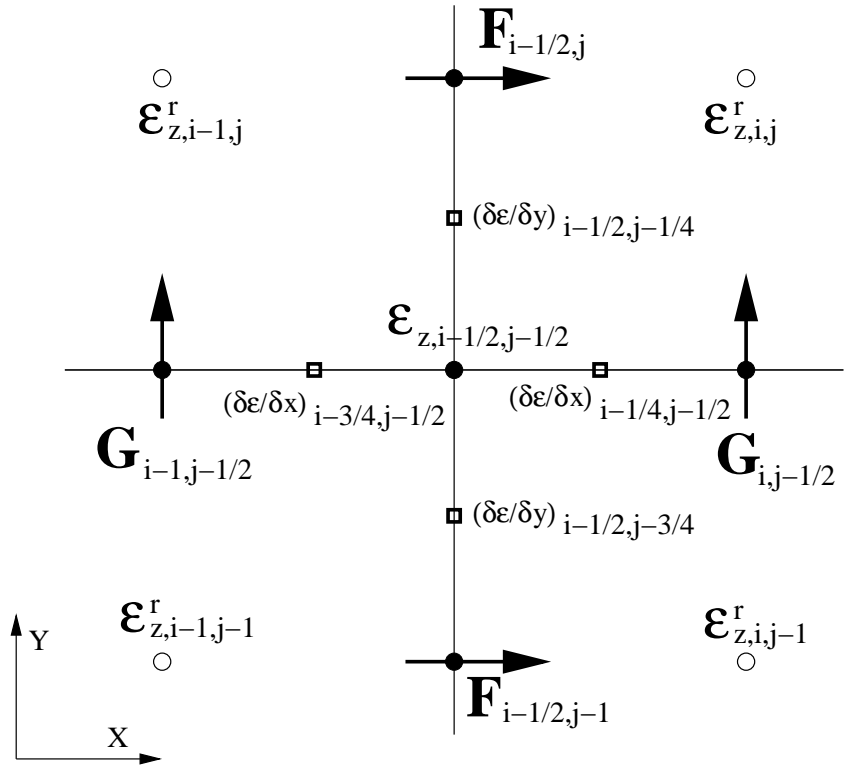}
\figcaption
{A 2D slice in the $x-y$ plane showing the centering of the fluxes of
conserved variables in the $x-$ and $y-$directions (${\bf F}$ and ${\bf
G}$ respectively), and the $z-$component of the emf centered at the cell
corner $\E_{z}$.  The CT algorithm used in Athena requires cell-centered
reference states for the emf $\E^{\rm r}_{z}$ to compute the gradients
$(\delta \E/\delta x)$ and $(\delta \E/\delta y)$ which are located
between the center of the cell faces and the cell corner.}
\end{figure}
\clearpage

\begin{figure}
\epsscale{0.8}
\plotone{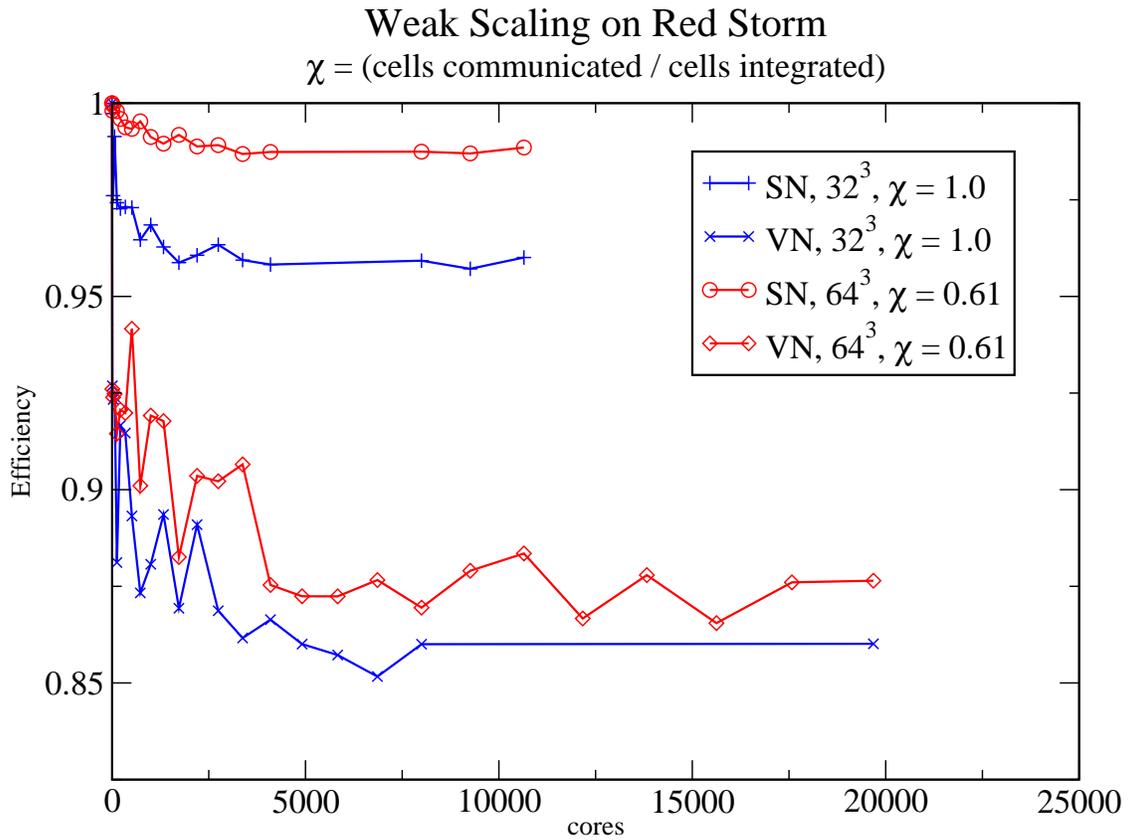}
\figcaption
{Weak scaling of the efficiency of Athena on a Cray XT-3, using
grids with either $32^{3}$ (blue lines), or $64^{3}$ (red lines) cells
per processor, and either one (SN) or two (VN) processors per node.
The quantity $\chi$ measures the ratio of the number of cells communicated
to the number updated per MPI process.
The efficiency remains flat well beyond $10^{4}$ processors, indicating
excellent weak scaling.}
\end{figure}
\clearpage

\begin{figure}
\epsscale{0.8}
\plotone{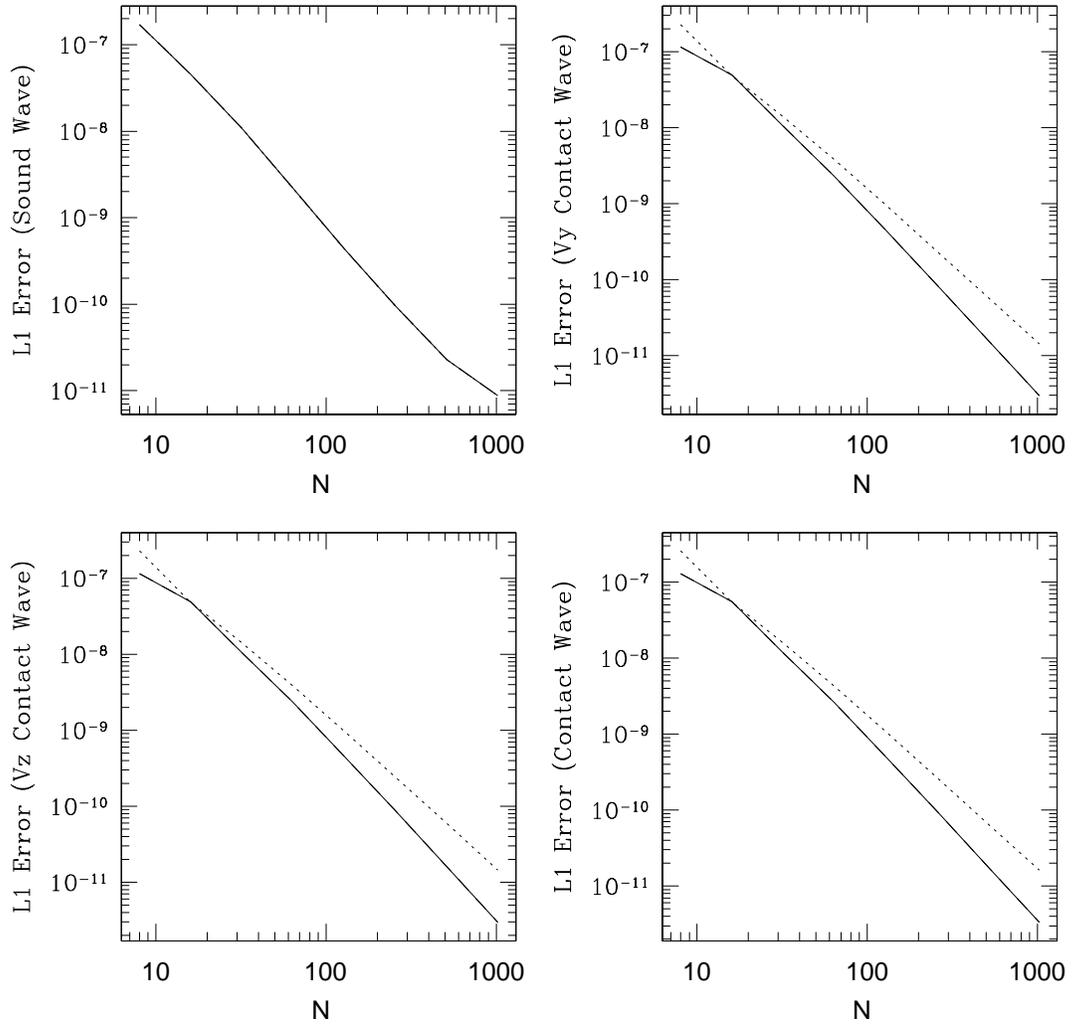}
\figcaption
{Convergence in the norm of the L$_{1}$ error vector for sound waves, shear
waves associated
with each transverse component of velocity, and the entropy (contact)
wave after
propagating a distance of one wavelength in 1D.  Solutions are computed
using third-order spatial reconstruction, and either the Roe fluxes (solid
line), or HLLE fluxes (dotted line). The errors for solutions computed with the
HLLC fluxes are identical to solutions computed with the Roe fluxes.}
\end{figure}
\clearpage

\begin{figure}
\epsscale{0.8}
\plotone{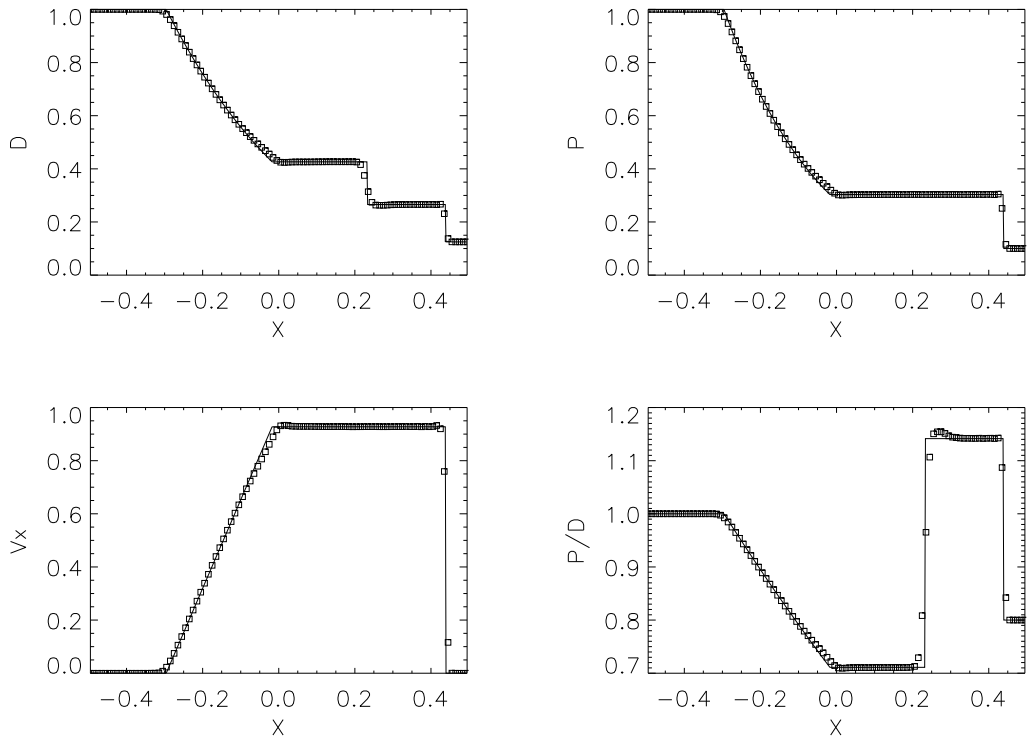}
\figcaption
{Density, pressure, velocity, and specific internal energy (scaled by
$(\gamma -1)$) for the Sod shocktube test at $t=0.25$,
computed with 100 grid points,
third-order spatial reconstruction, and the HLLC fluxes.  The solid line
is the analytic solution.}
\end{figure}

\begin{figure}
\epsscale{0.8}
\plotone{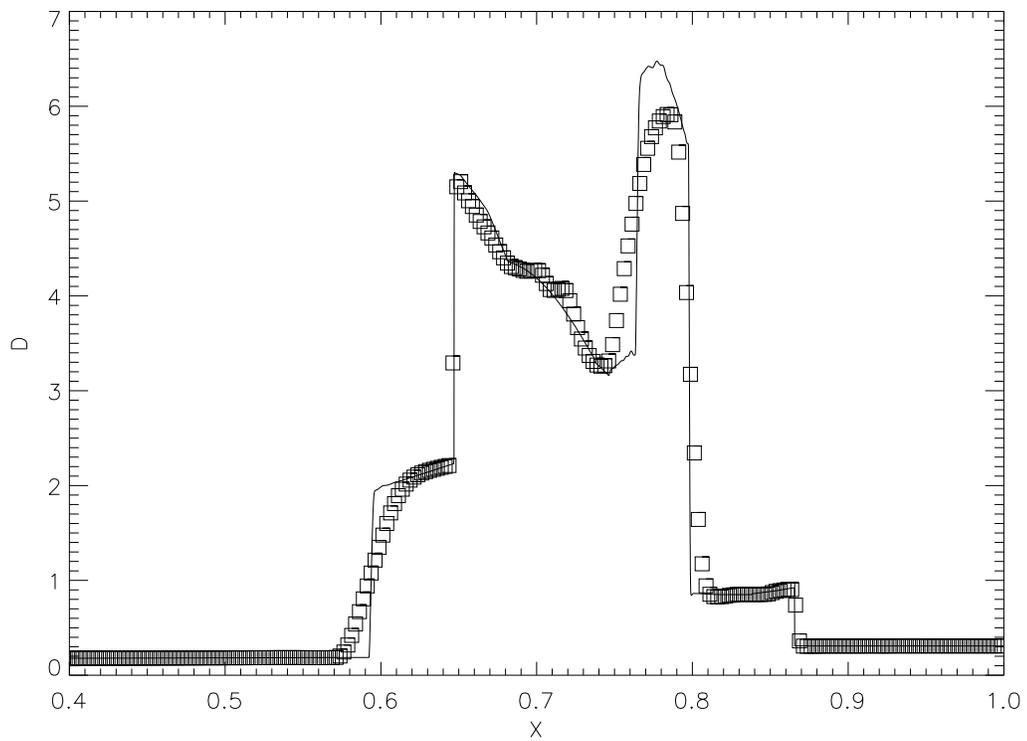}
\figcaption
{Density at $t=0.038$ in the two-interacting blast wave test,
computed with 400 grid points,
third-order spatial reconstruction, and the HLLC fluxes.  The solid line
is a reference solution computed with 9600 grid points.}
\end{figure}
\clearpage

\begin{figure}
\epsscale{0.8}
\plotone{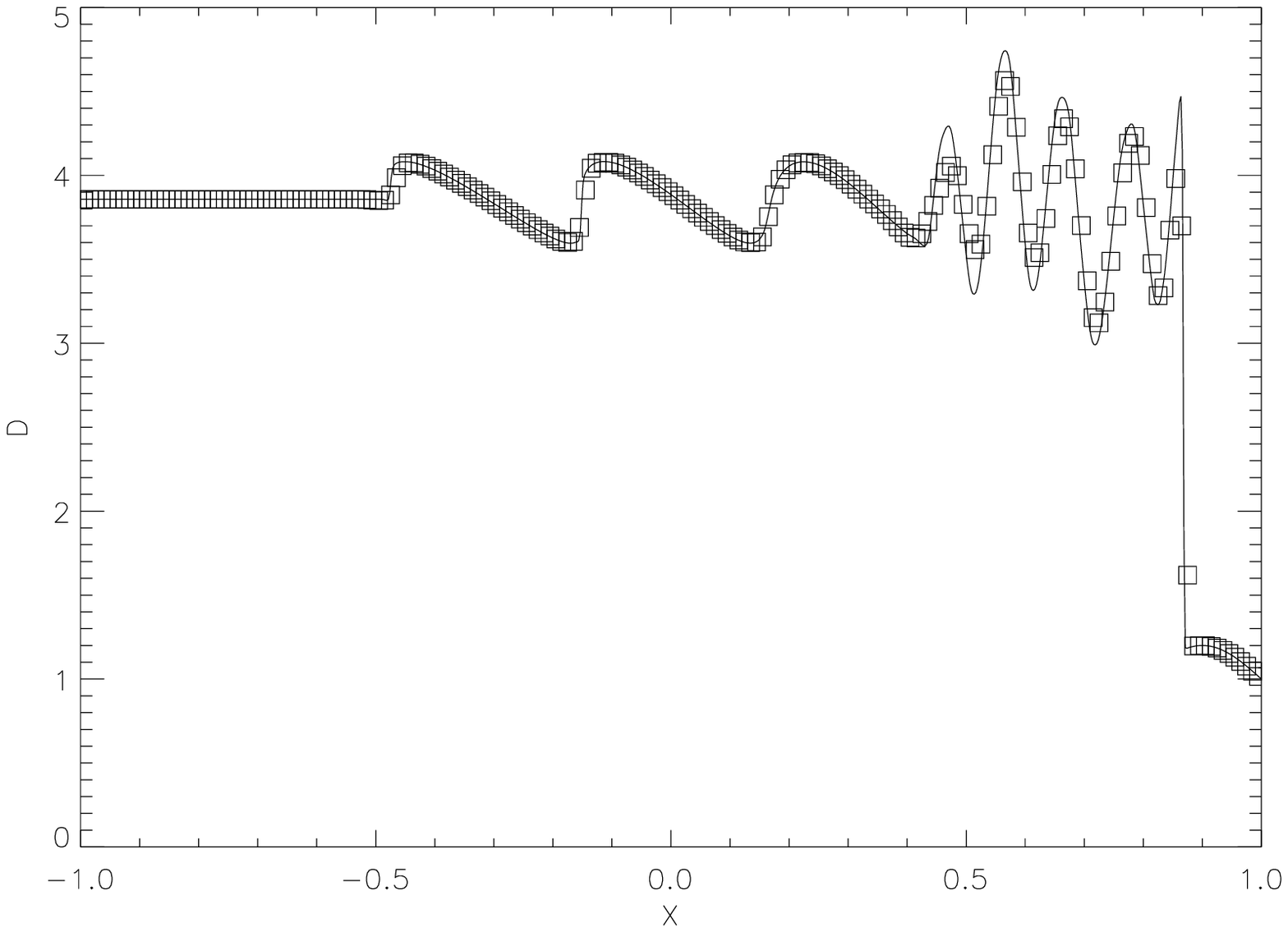}
\figcaption
{Density at $t=0.47$ in the Shu-Osher Riemann problem,
computed with 200 (squares) and 800 (solid line) grid points,
third-order spatial reconstruction, and the HLLC fluxes.}
\end{figure}

\begin{figure}
\epsscale{0.8}
\plotone{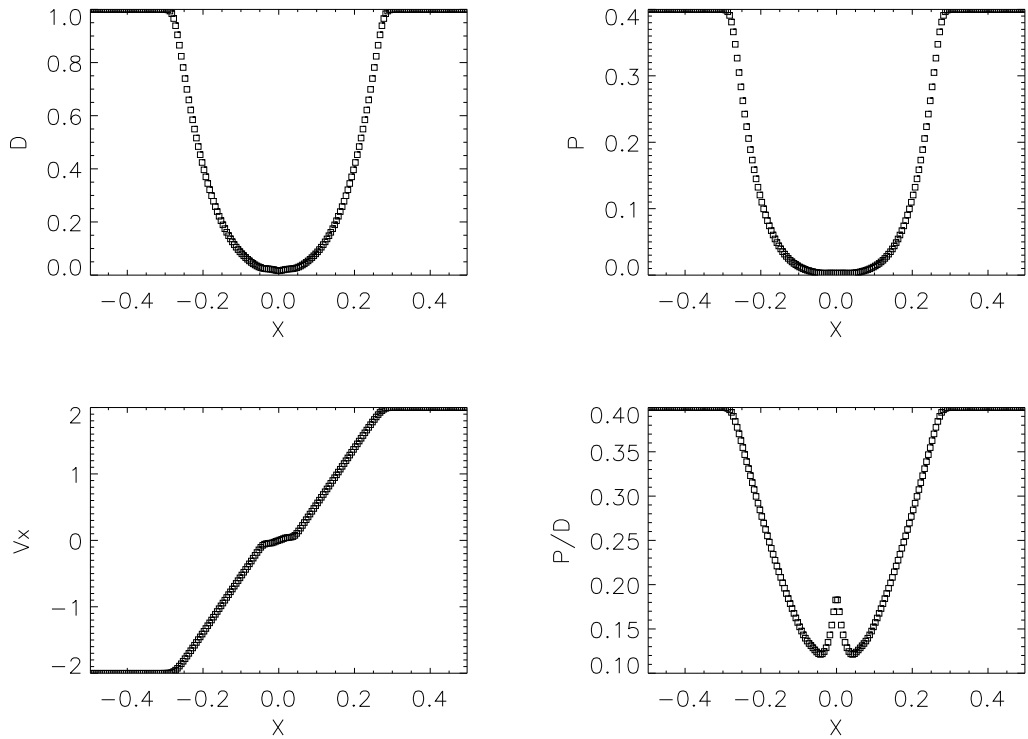}
\figcaption
{Density, pressure, velocity, and specific internal energy (scaled by
$(\gamma -1)$) for the Einfeldt strong rarefaction test at $t=0.1$,
computed with 200 grid points,
second-order spatial reconstruction, and the HLLC fluxes.}
\end{figure}
\clearpage

\begin{figure}
\epsscale{0.8}
\plotone{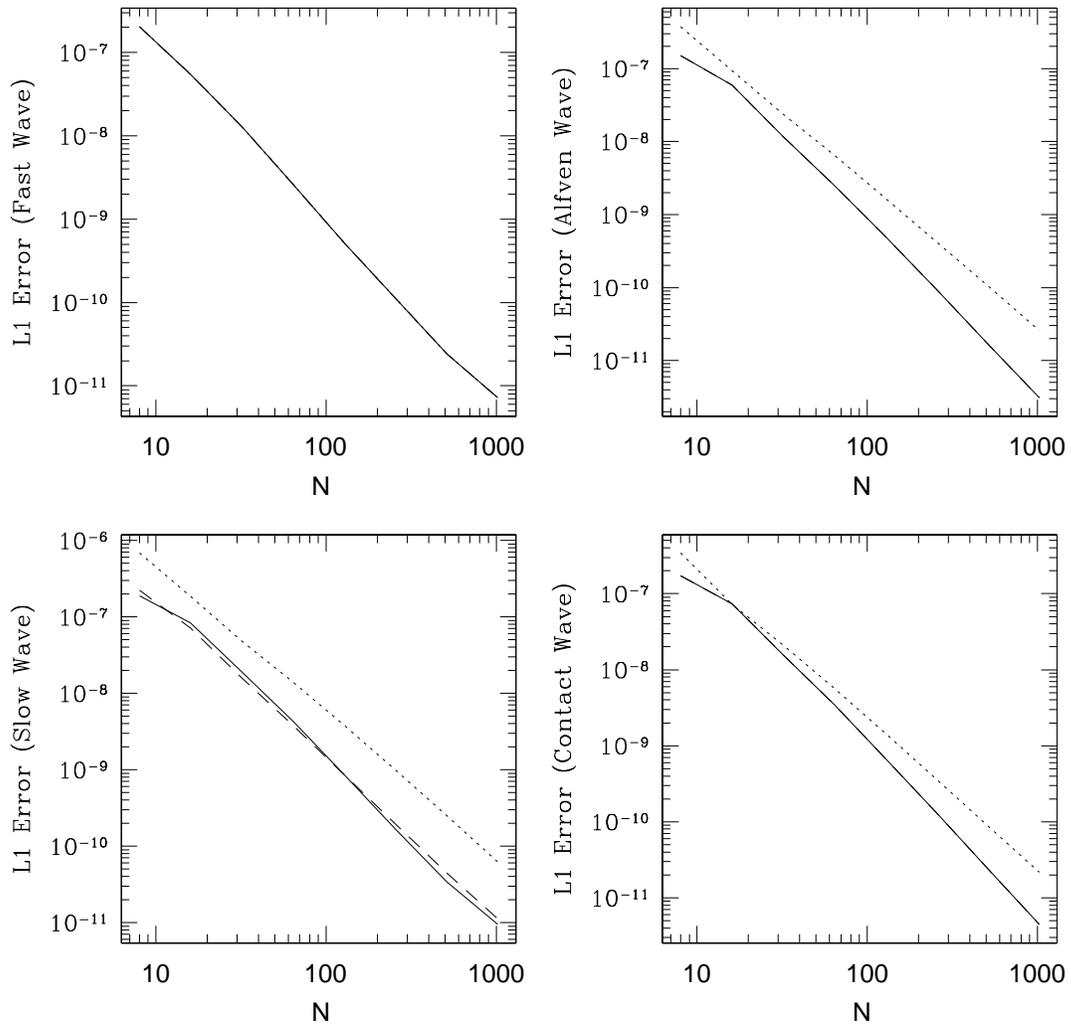}
\figcaption
{Convergence in the norm of the L$_{1}$ error vector for fast, Alfv\'{e}n, slow,
and contact waves after
propagating a distance of one wavelength in 1D.  Solutions are computed
using third-order spatial reconstruction, and either the Roe fluxes (solid
line), HLLD fluxes (dashed line), or HLLE fluxes (dotted line).}
\end{figure}
\clearpage

\begin{figure}
\epsscale{0.8}
\plotone{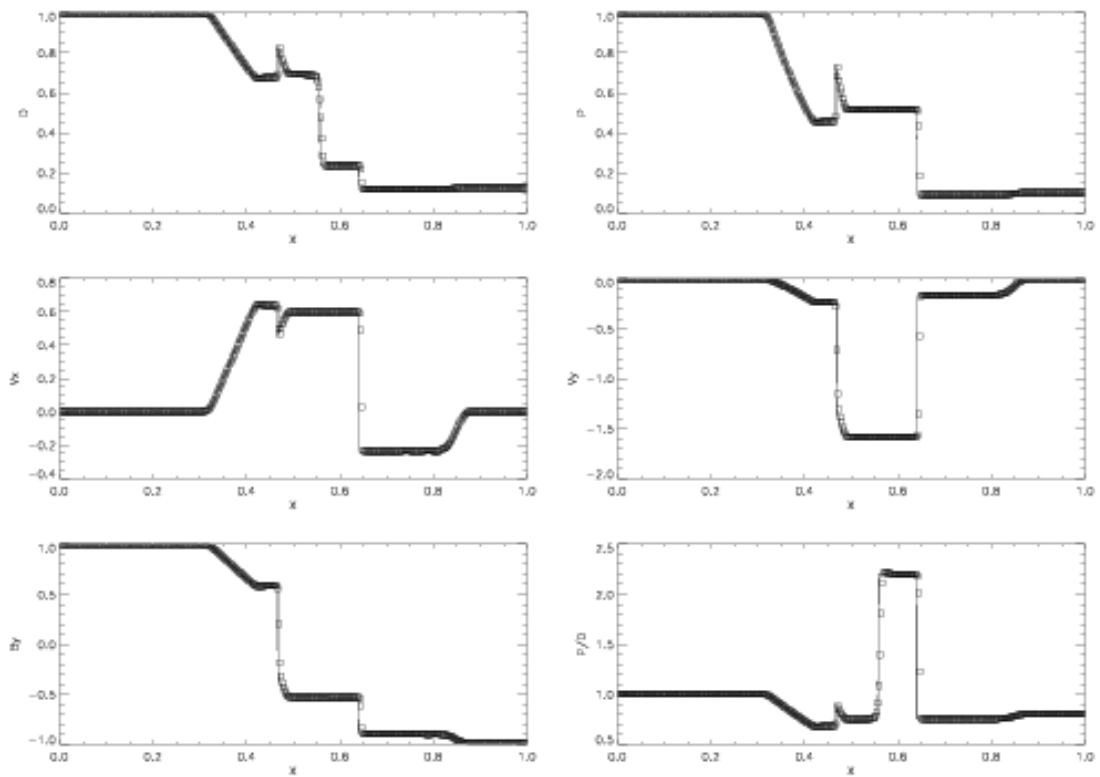}
\figcaption
{Density, pressure, velocity components, transverse component of the
magnetic field, and specific internal energy (scaled by
$(\gamma -1)$) for the Brio \& Wu shocktube problem at $t=0.08$,
computed with 400 grid points, 
second-order spatial reconstruction, and the Roe fluxes.
The solid line is a reference solution
computed with $10^{4}$ grid points.}
\end{figure}
\clearpage

\begin{figure}
\epsscale{0.8}
\plotone{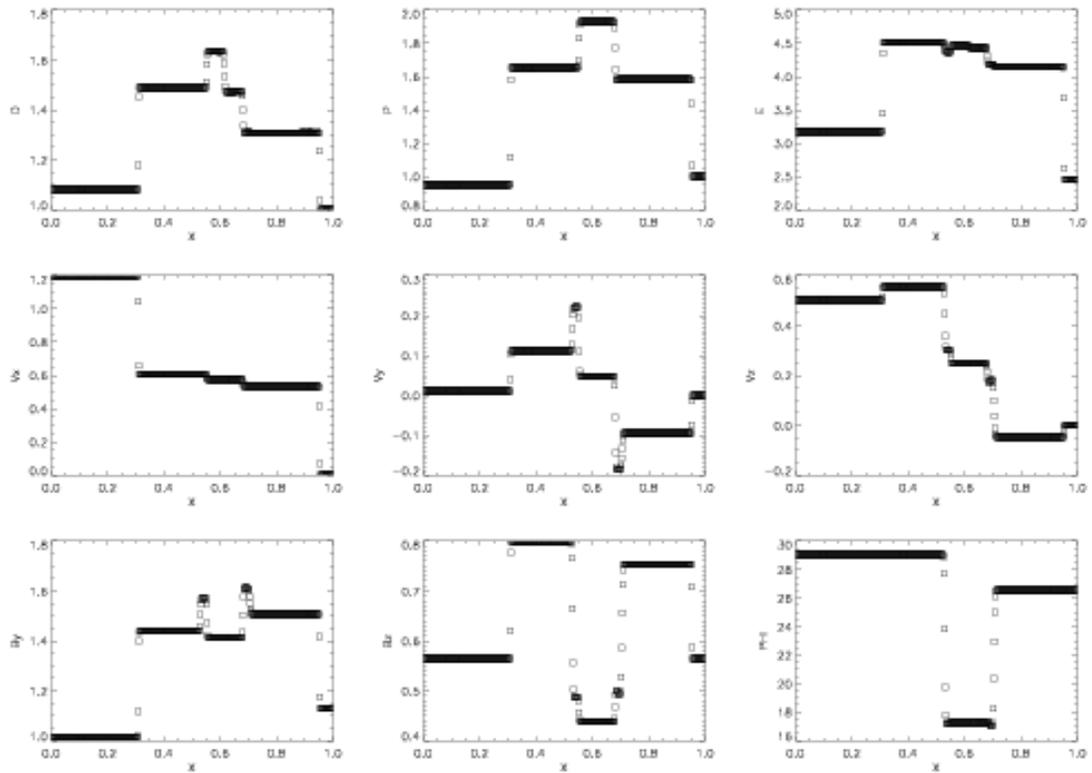}
\figcaption
{Density, pressure, total energy, all three components of velocity,
transverse components and rotation angle $\Phi = \tan^{-1}(B_{z}/B_{y})$
of the magnetic field
for the MHD Riemann problem RJ2a at $t=0.2$,
computed with 512 grid points,
third-order spatial reconstruction, and the Roe fluxes.}
\end{figure}
\clearpage

\begin{figure}
\epsscale{0.8}
\plotone{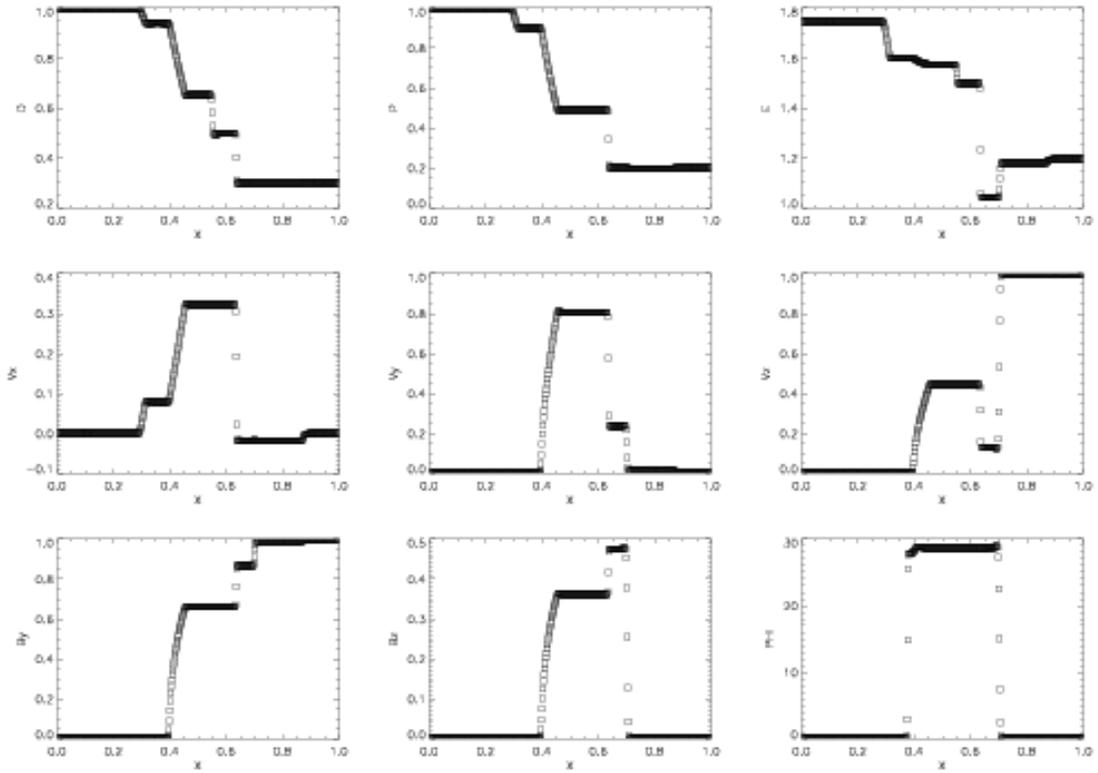}
\figcaption
{Density, pressure, total energy, all three components of velocity,
transverse components and rotation angle $\Phi = \tan^{-1}(B_{z}/B_{y})$
of the magnetic field
for the MHD Riemann problem RJ4d at $t=0.16$,
computed with 512 grid points,
third-order spatial reconstruction, and the HLLD fluxes.}
\end{figure}
\clearpage

\begin{figure}
\epsscale{0.6}
\plotone{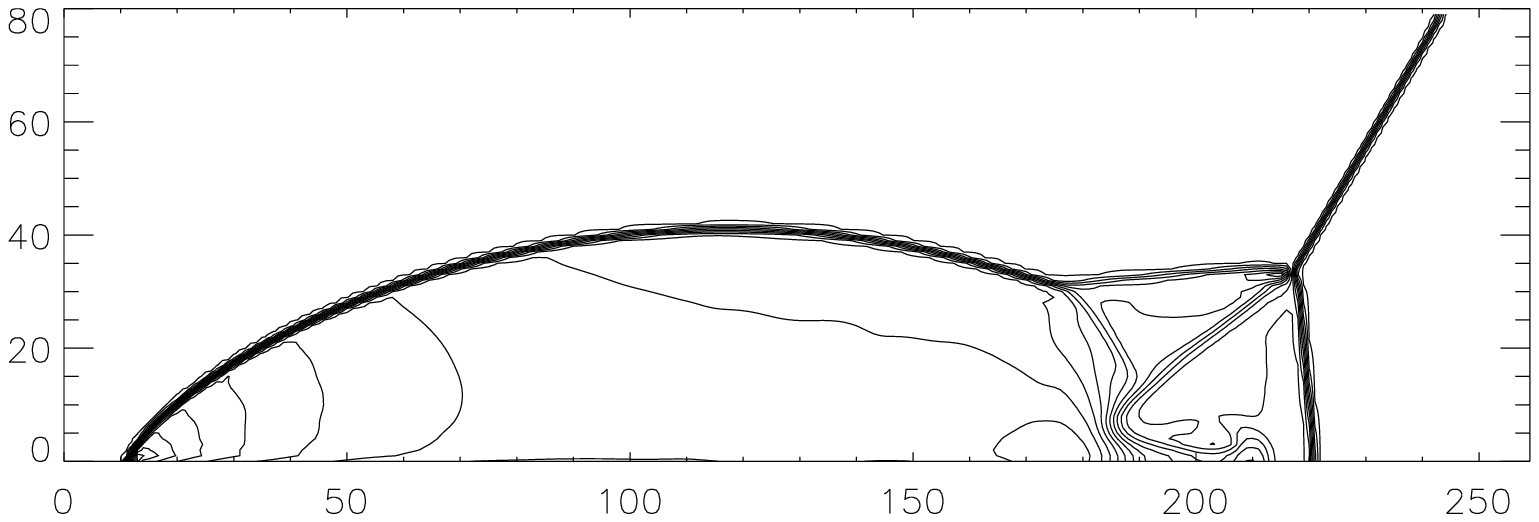} \vspace*{-1.7in}
\plotone{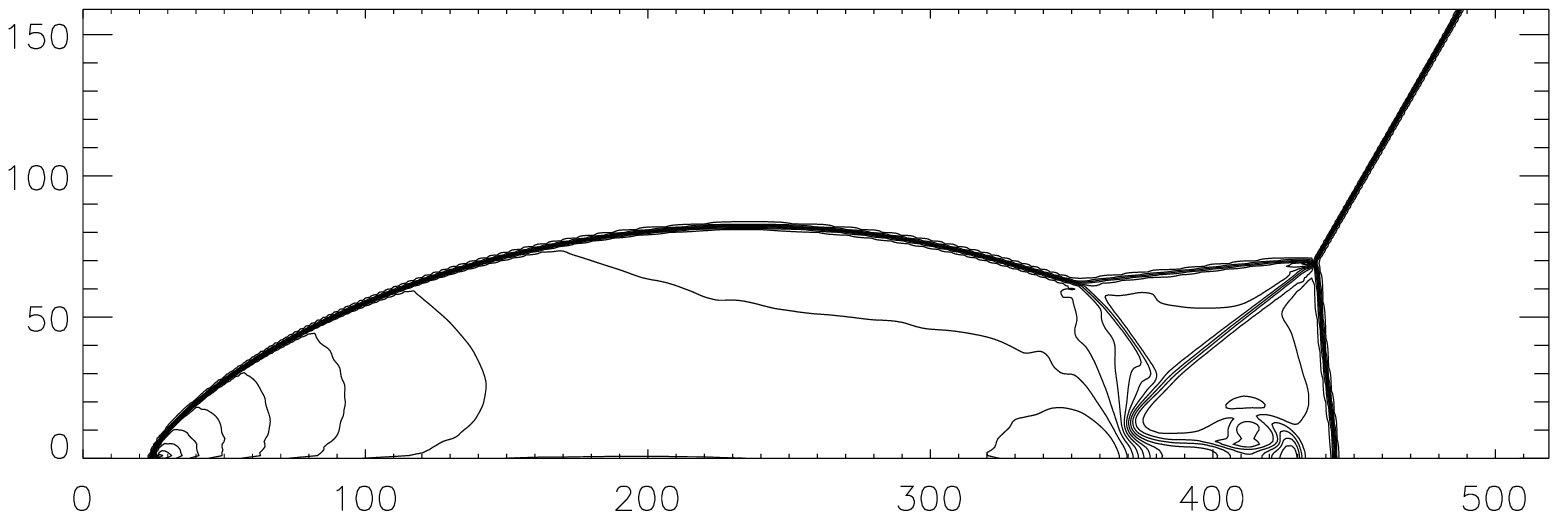} \vspace*{-1.7in}
\plotone{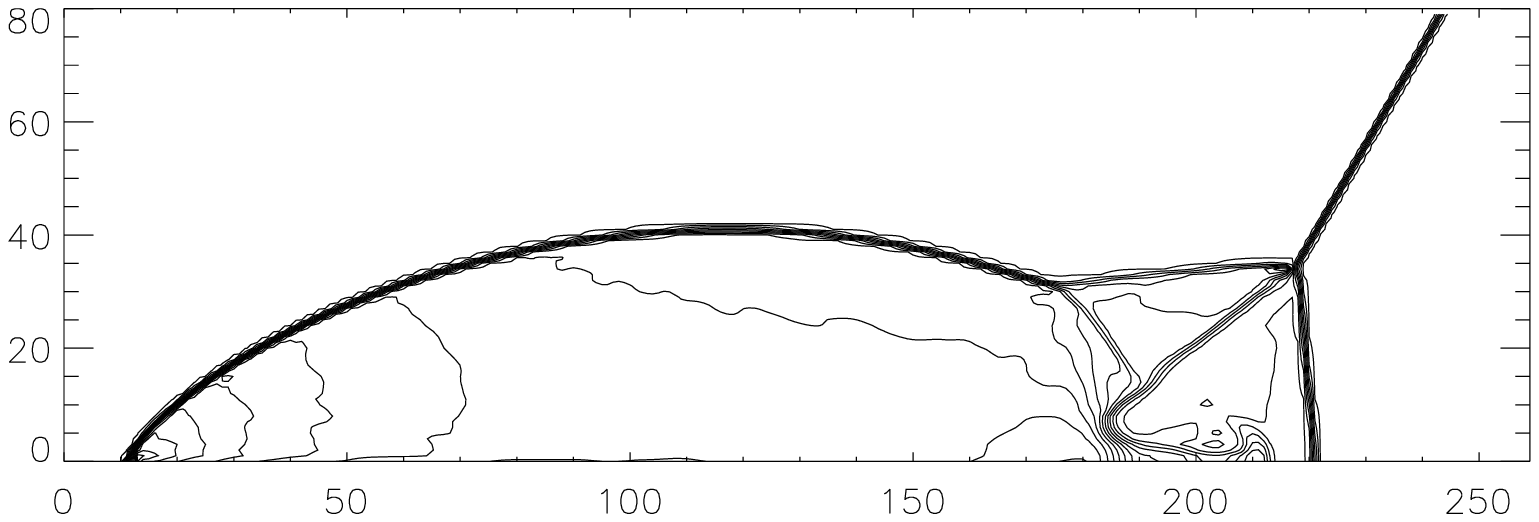} \vspace*{-1.7in}
\plotone{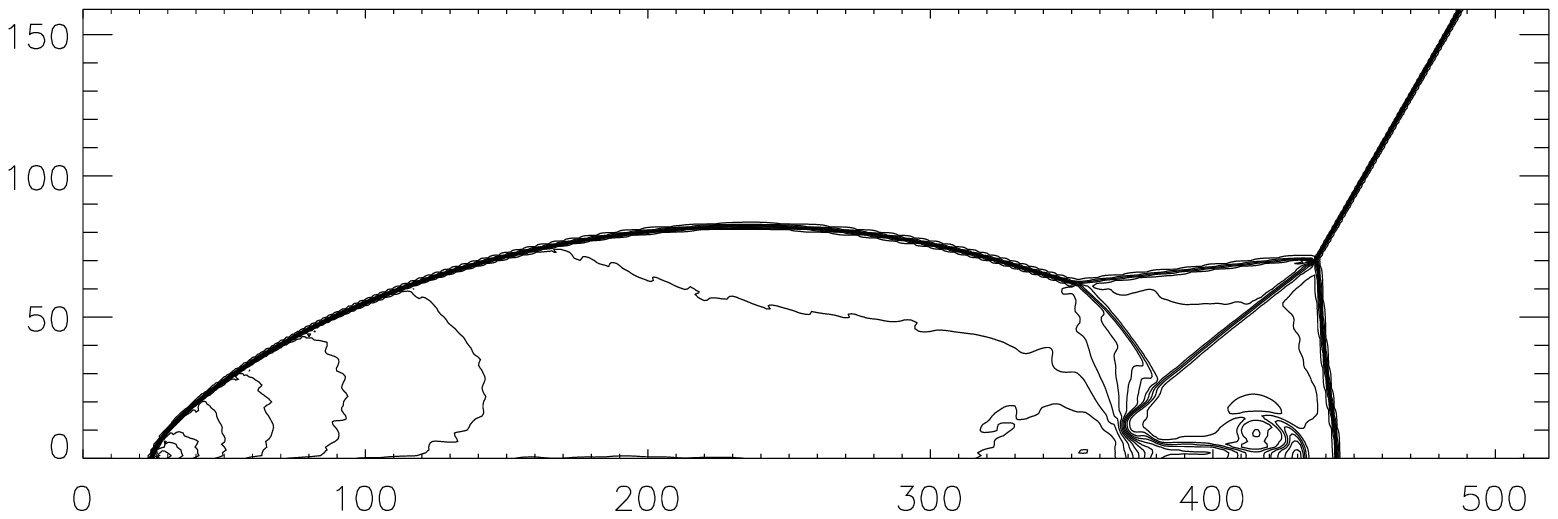}
\figcaption
{Contours of the density at $t=0.2$ for the double Mach reflection test.
From top to bottom, the solutions are computed with second order spatial
reconstruction at low and high resolution, and third order spatial
reconstruction at low and high resolution.  Here, low resolution
uses a grid of $260\times 80$ cells, and high resolution uses a grid
of $520 \times 160$ cells.  All solutions are computed with Roe fluxes
and the H-correction.}
\end{figure}
\clearpage

\begin{figure}
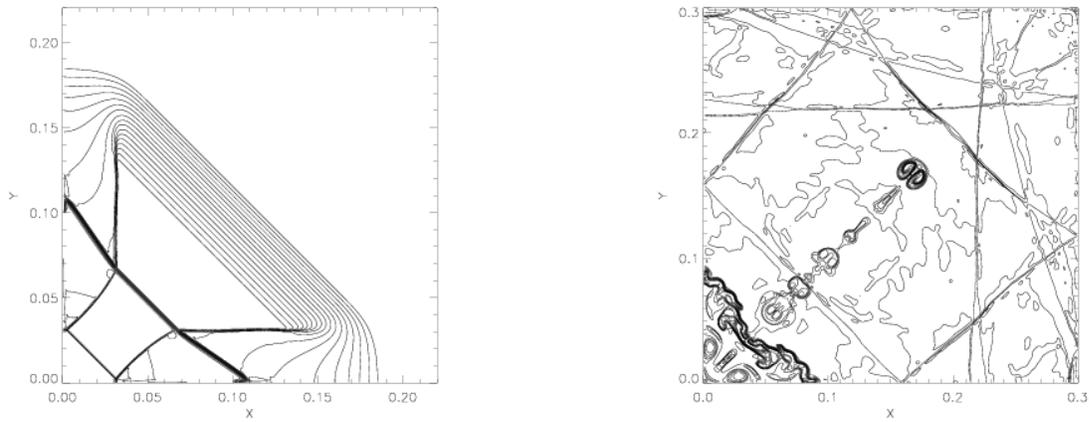

\epsscale{0.45}
\plotone{fig17a.ps3}
\plotone{fig17b.ps3}
\figcaption
{Contours of the density at $t=0.045$ ({\em left}) and $t=2.5$ ({\em right})
for the implosion test of Liska \& Wendroff.  In each case, 31 contours
are shown using a stepsize of 0.025, starting at a minimum value of
0.125 (at $t=0.045$) and 0.35 (at $t=2.5$).  The solution
is computed using third order spatial reconstruction and the HLLC fluxes,
on a grid of $400 \times 400$ cells.}
\end{figure}
\clearpage

\begin{figure}
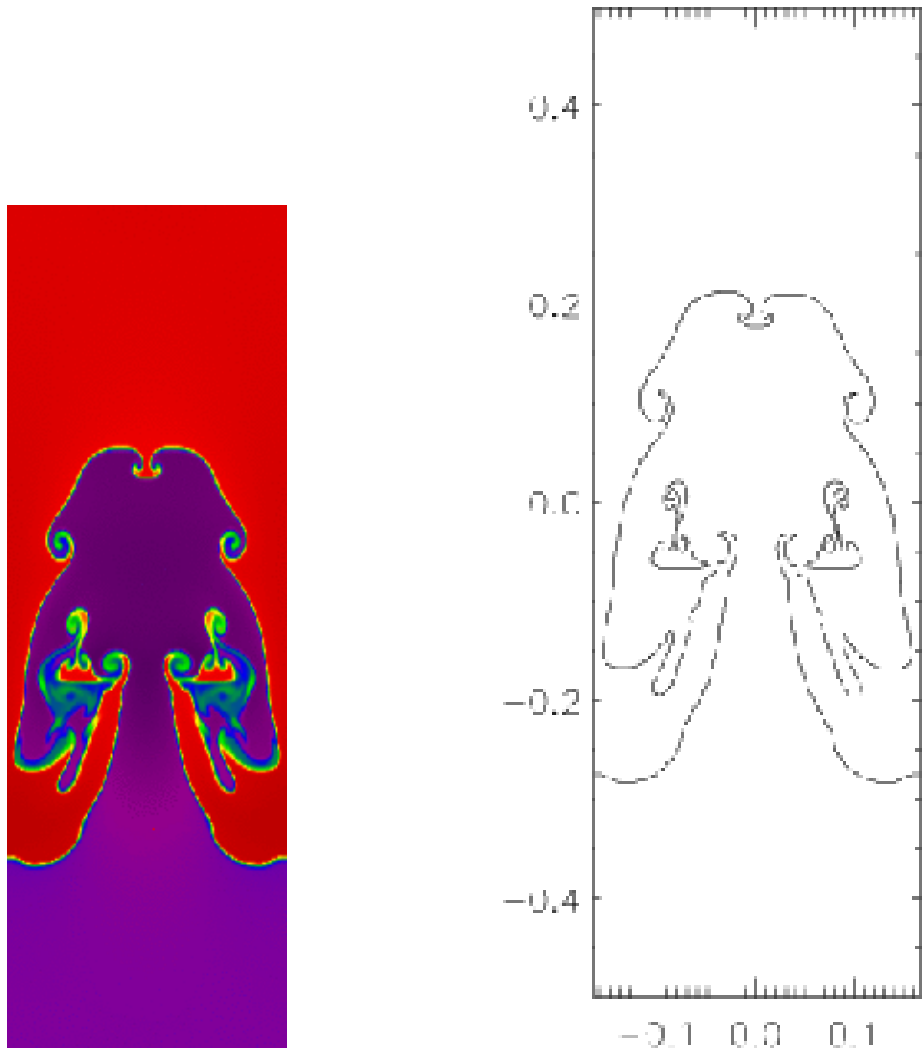

\epsscale{0.2}
\hspace*{-.5in}
\plotone{fig18a.ps3}
\epsscale{0.3}
\hspace*{1.0in}
\plotone{fig18b.ps3}
\figcaption
{Color image ({\em left}) and contours ({\em right}) of the density at
$t=8.5$ in a single mode hydrodynamic Rayleigh-Taylor instability in
2D.  Only a single contour is shown at $\rho=1.5$ in order to trace the
contact discontinuity between the heavy and light fluids.  Colors in the image
correspond to density values of 0.9 (purple) to 2.1 (red).  The solution
is computed using third order spatial reconstruction and the HLLC fluxes
on a grid of $200 \times 400$ cells.}
\end{figure}
\clearpage

\begin{figure}
\epsscale{0.4}
\plotone{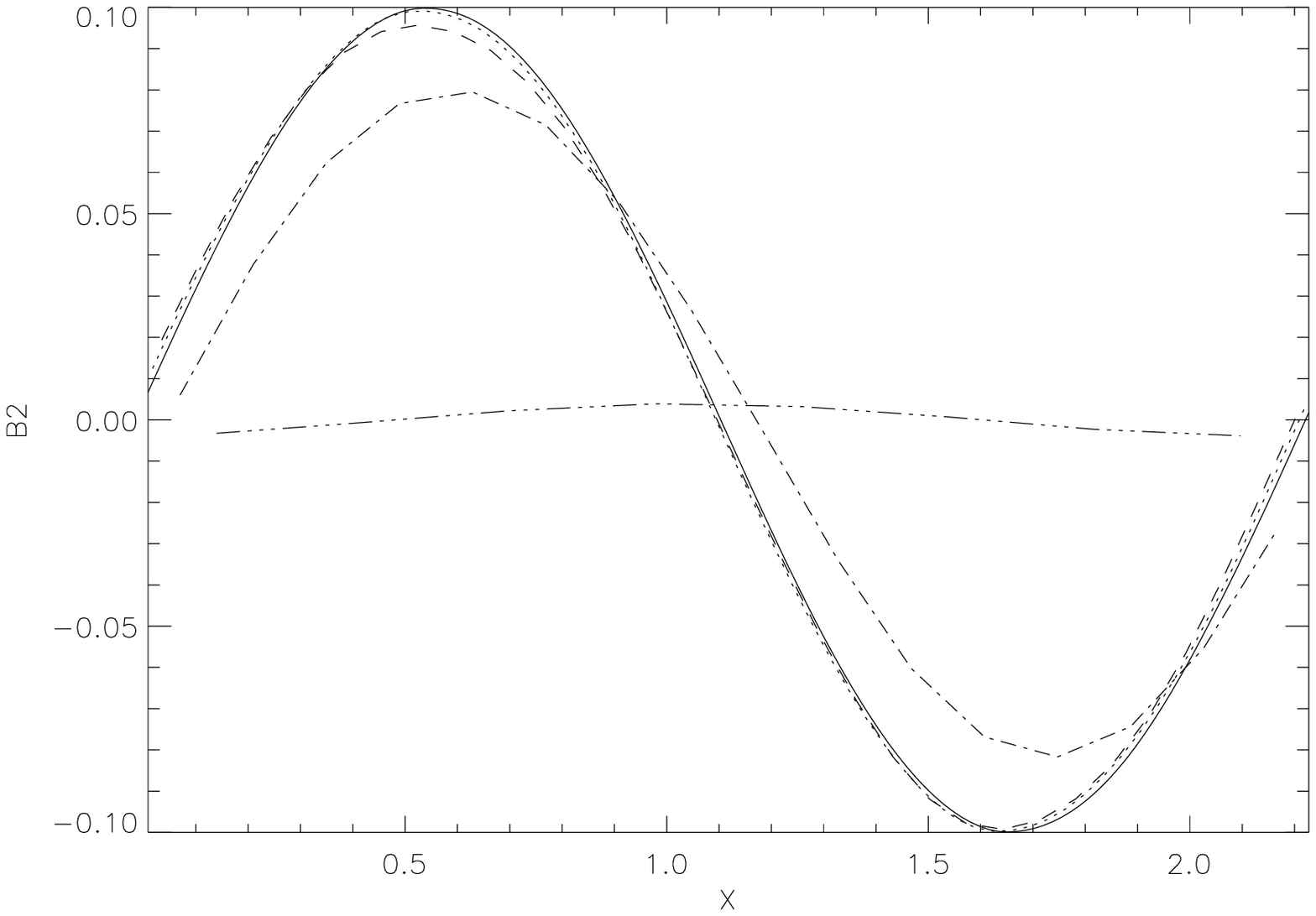}
\plotone{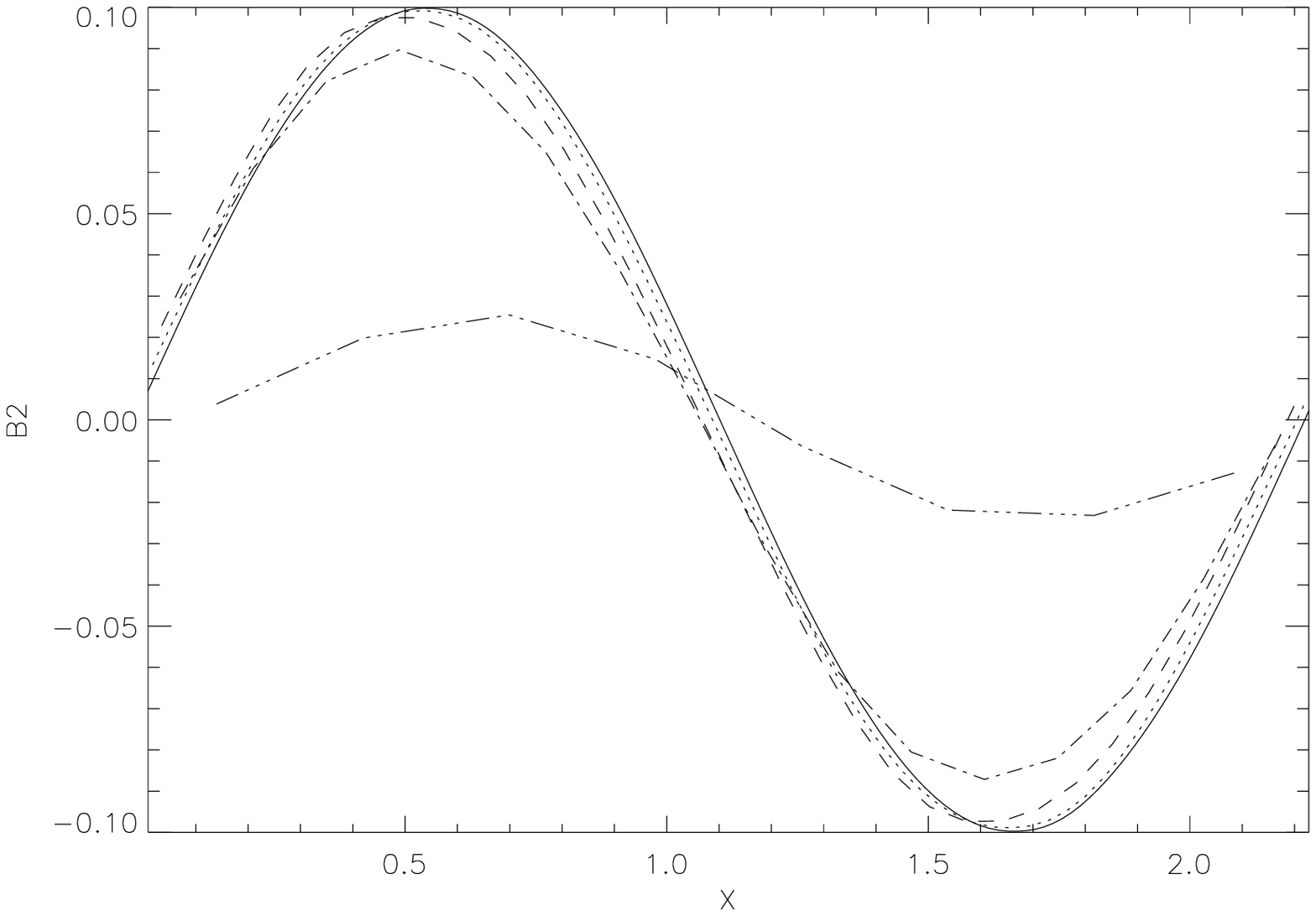}
\figcaption
{Profiles of the transverse component of the magnetic field (labelled $B2$)
for both
traveling ({\em left}) and standing ({\em right}) circularly polarized
Alfv\'{e}n waves, at a time equal to five wave periods, computed on a grid
with $2N \times N$ cells, where $N=64$ (solid line), 32 (dotted),
16 (dashed line), 8 (dot-dash line), and 4 (dot-dot-dot-dashed line).
Each solution is computed using third order spatial reconstruction
and the Roe fluxes.}
\end{figure}
\clearpage

\begin{figure}
\epsscale{0.8}
\plotone{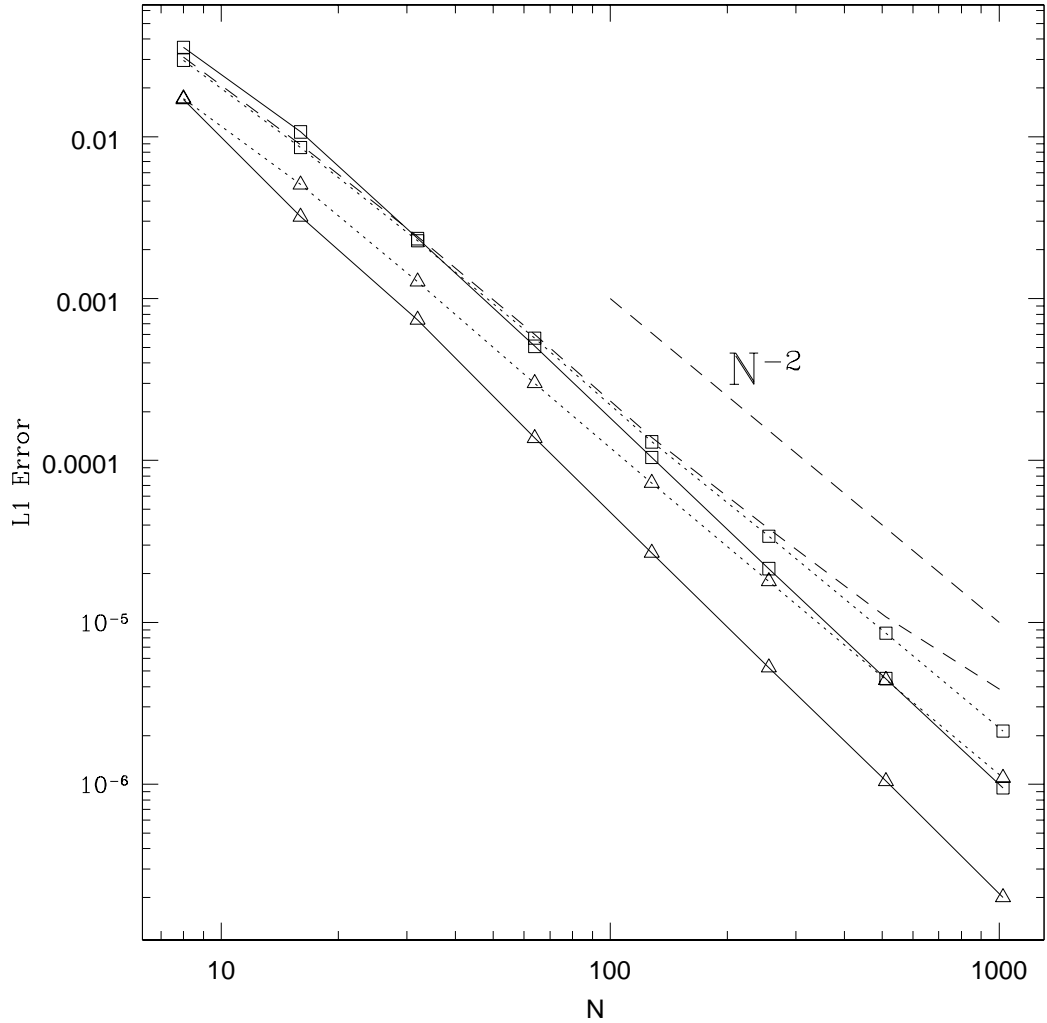}
\figcaption
{Convergence of the norm of the L$_{1}$
error vector for traveling circularly polarized
Alfv\'{e}n waves, after propagating a distance equal to one wavelength,
using an isothermal equation of state.
Points marked by squares denote second order spatial reconstruction,
triangles denote third order spatial reconstruction.  The solid lines
are solutions computed in 1D, the dotted lines are solutions computed in 2D.
The dashed line shows the norm of the L$_{1}$ error vector for a 2D
solution using
second order spatial reconstruction computed with an adiabatic equation of
state.  Also shown is a dashed line with slope -2 for comparison.}
\end{figure}
\clearpage

\begin{figure}
\epsscale{0.4}
\plotone{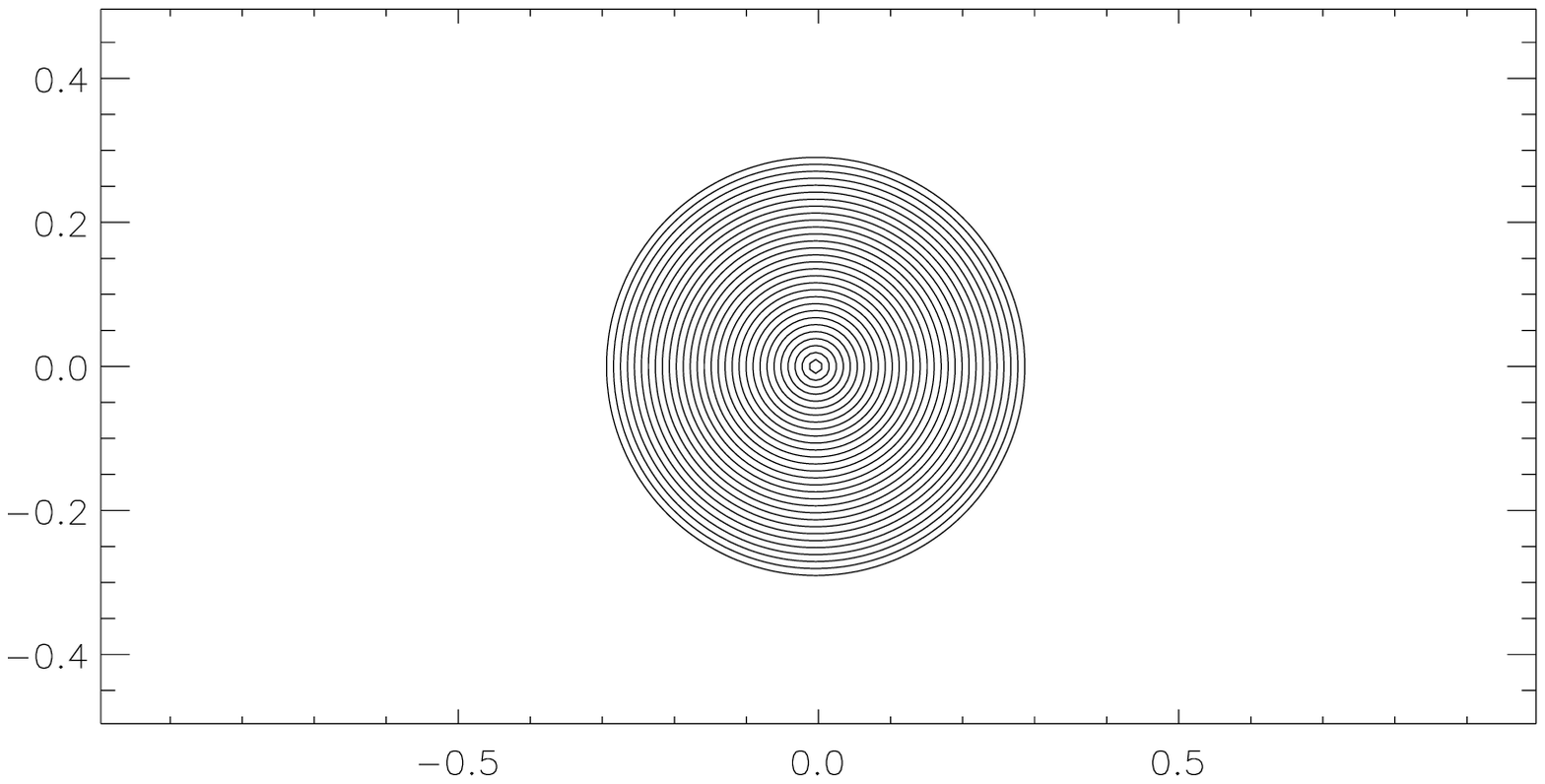}
\plotone{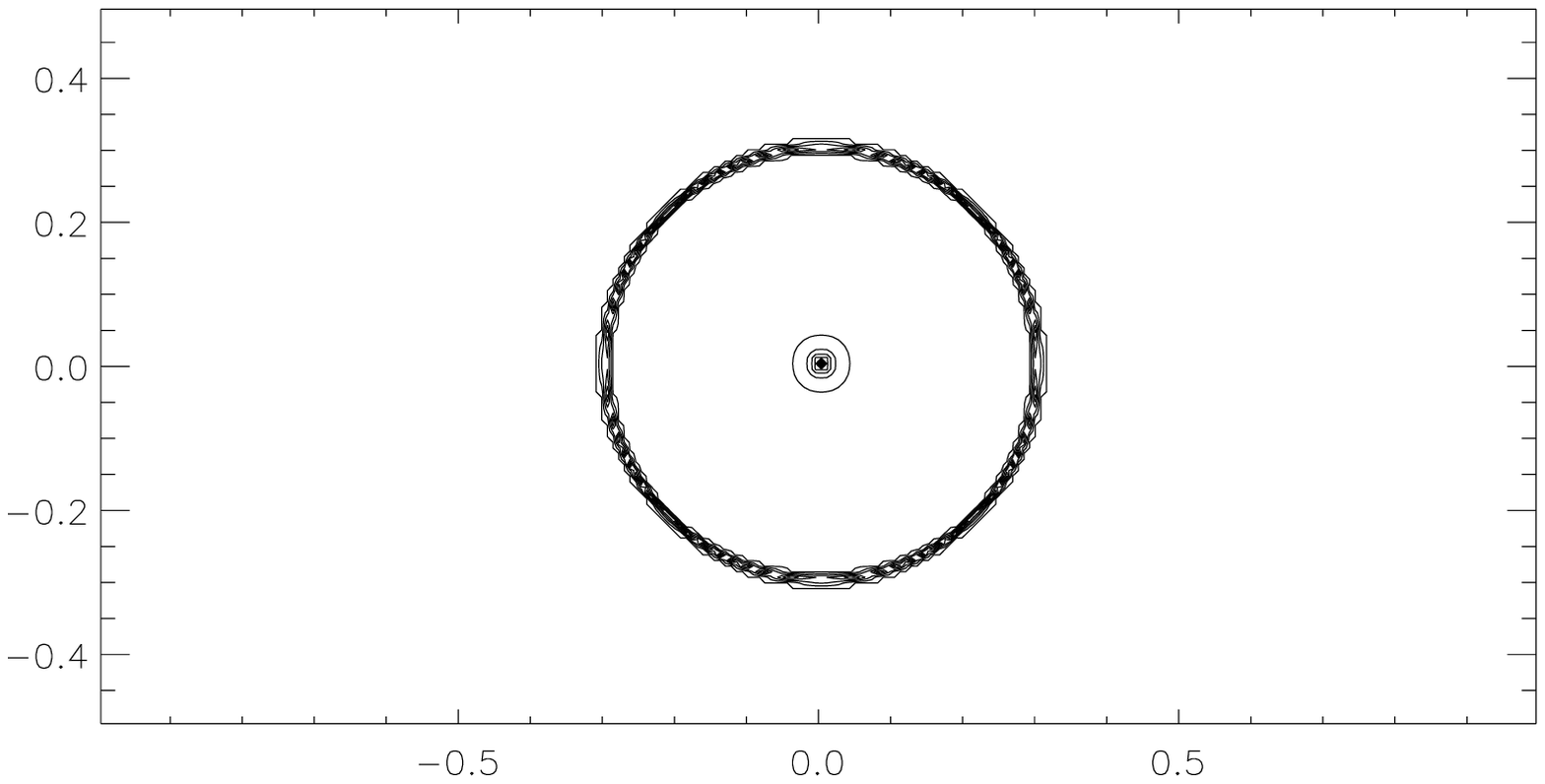}
\plotone{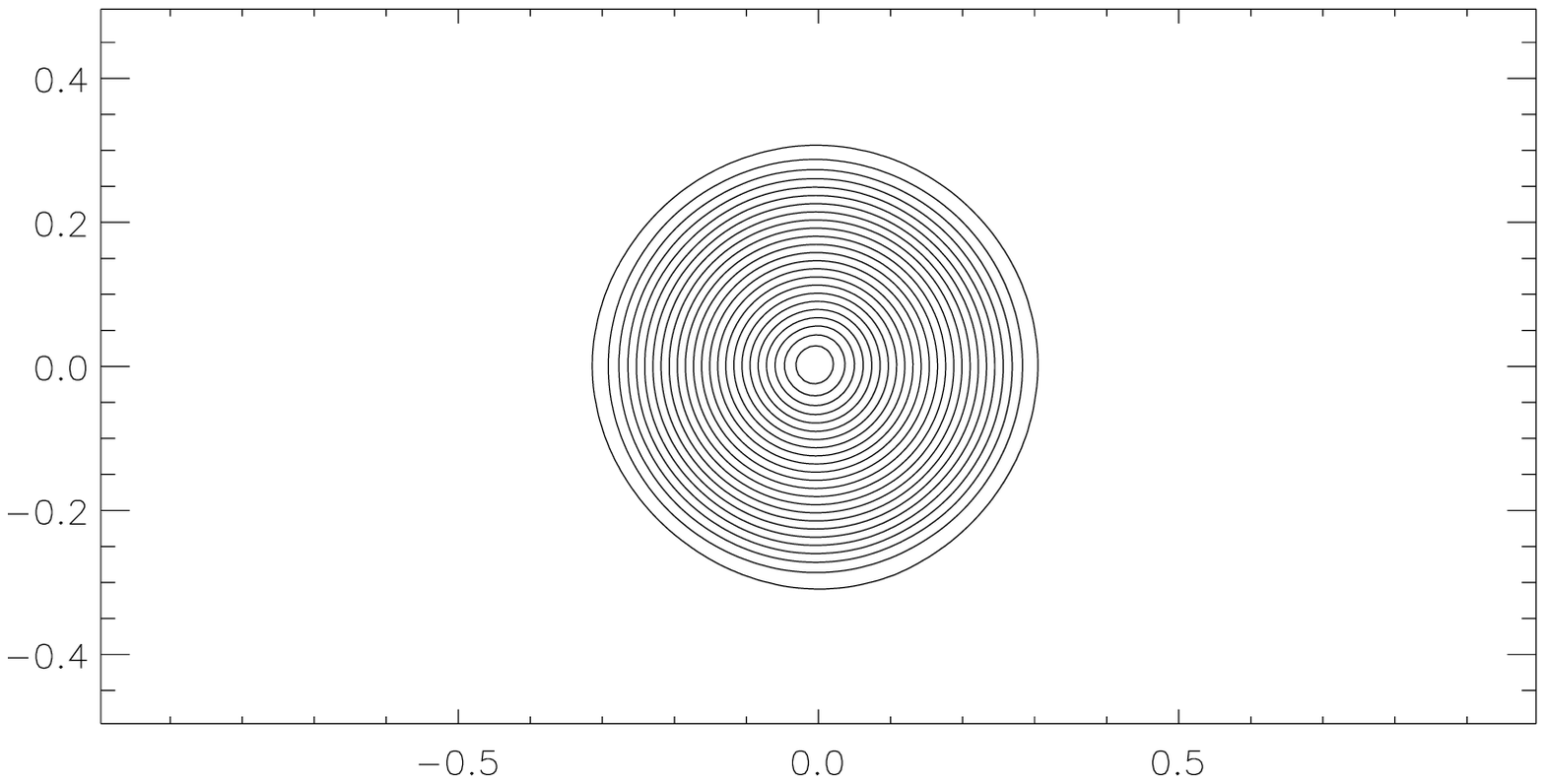}
\plotone{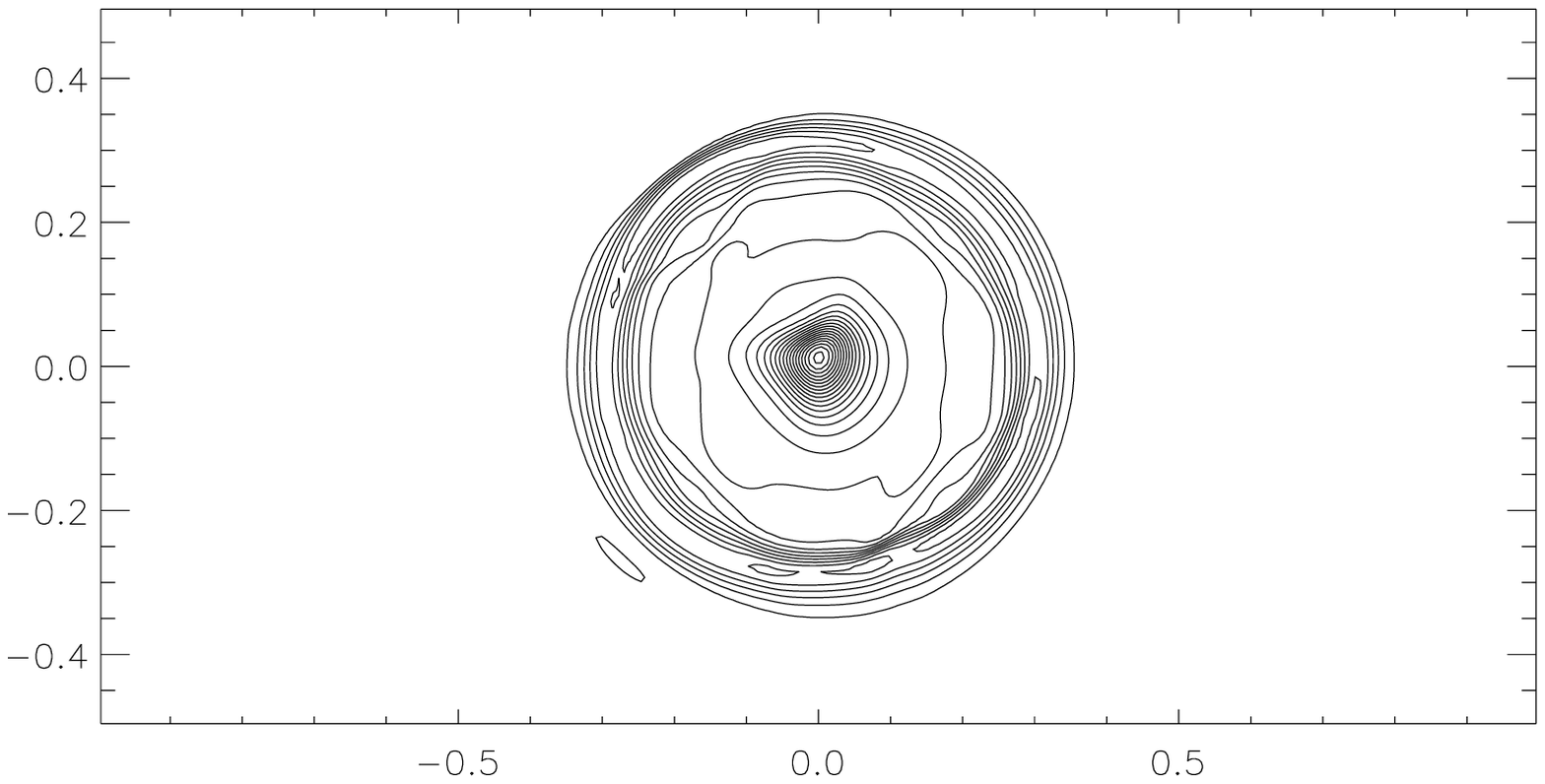}
\figcaption
{Magnetic field lines ({\em left}) and contours of the $z-$component
of the current density ({\em right}) at $t=0$ (top row) and at $t=2$
after advection of the loop twice around the grid (bottom row).
The solution is computed using second order spatial reconstruction
with the Roe fluxes on a grid of $256 \times 128$ cells.}
\end{figure}
\clearpage

\begin{figure}
\epsscale{0.8}
\plotone{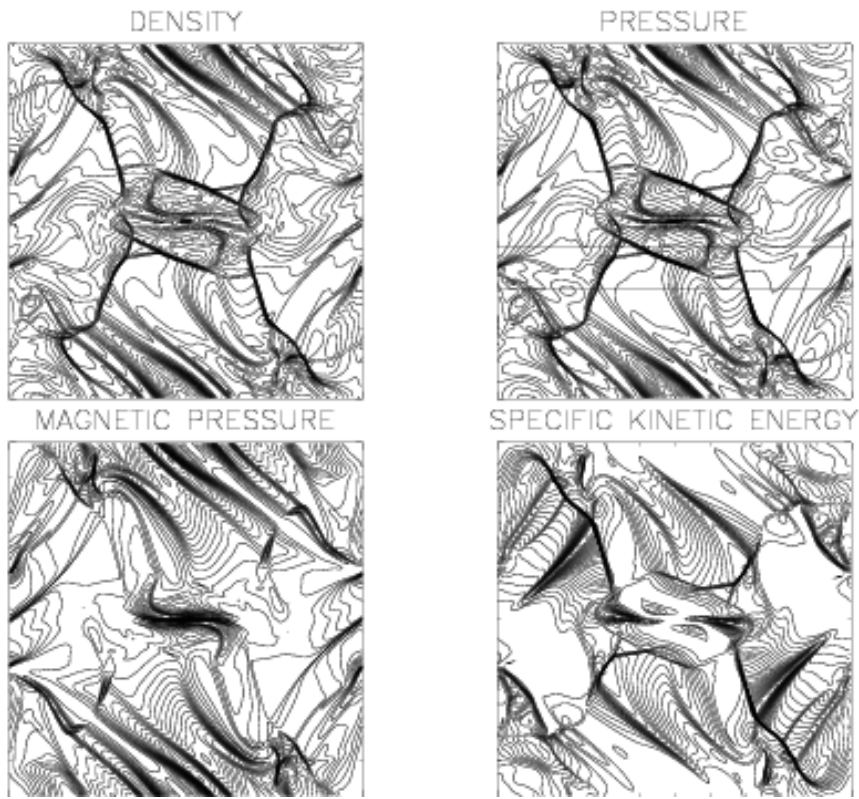}
\figcaption
{Contours of selected variables at $t_{f}=1/2$ in the adiabatic Orszag-Tang
vortex test, computed using a grid of $192 \times 192$ cells, third-order
reconstruction, and Roe fluxes.  Thirty equally spaced contours between
the minimum and maximum are used for each plot.  The horizontal lines in the
panel showing pressure correspond to the locations of the slices shown in
figure 23.}
\end{figure}
\clearpage

\begin{figure}
\epsscale{0.8}
\plotone{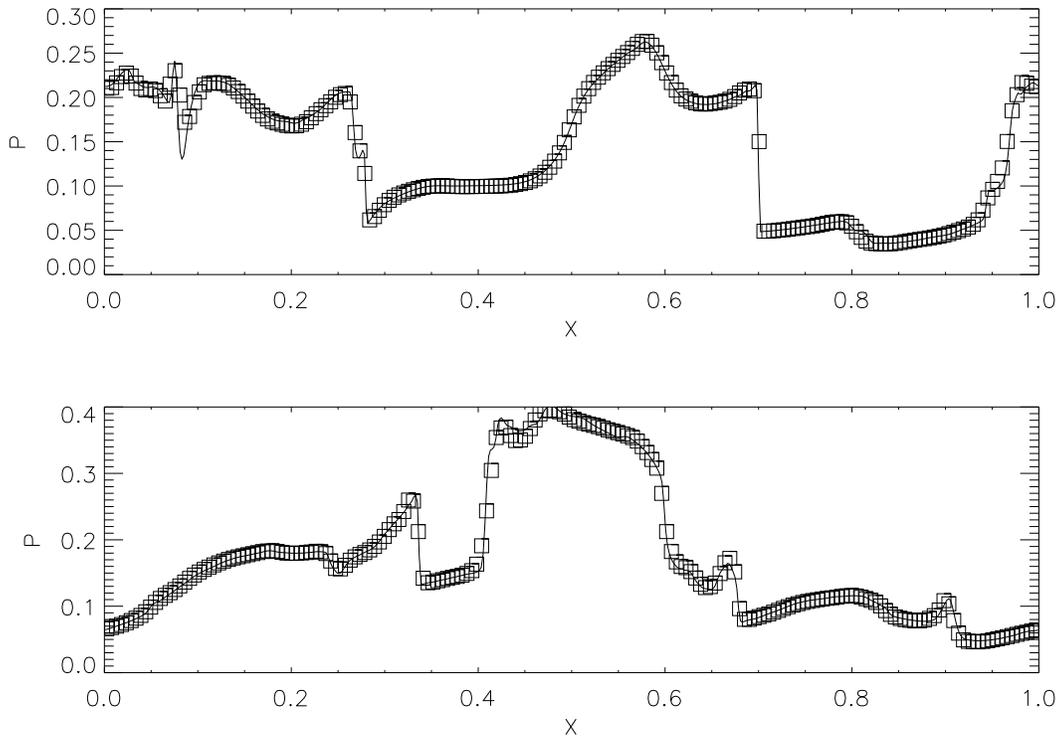}
\figcaption
{Horizontal slices of the pressure at $t_{f}=1/2$ in the adiabatic
Orszag-Tang vortex test taken at $y=0.3125$ ({\em top}) and $y=0.427$
({\em bottom}).  Squares
correspond to the solution on a $192 \times 192$ grid, while the solid
line is for a $512^{2}$ grid.}
\end{figure}
\clearpage

\begin{figure}
\epsscale{0.8}
\plotone{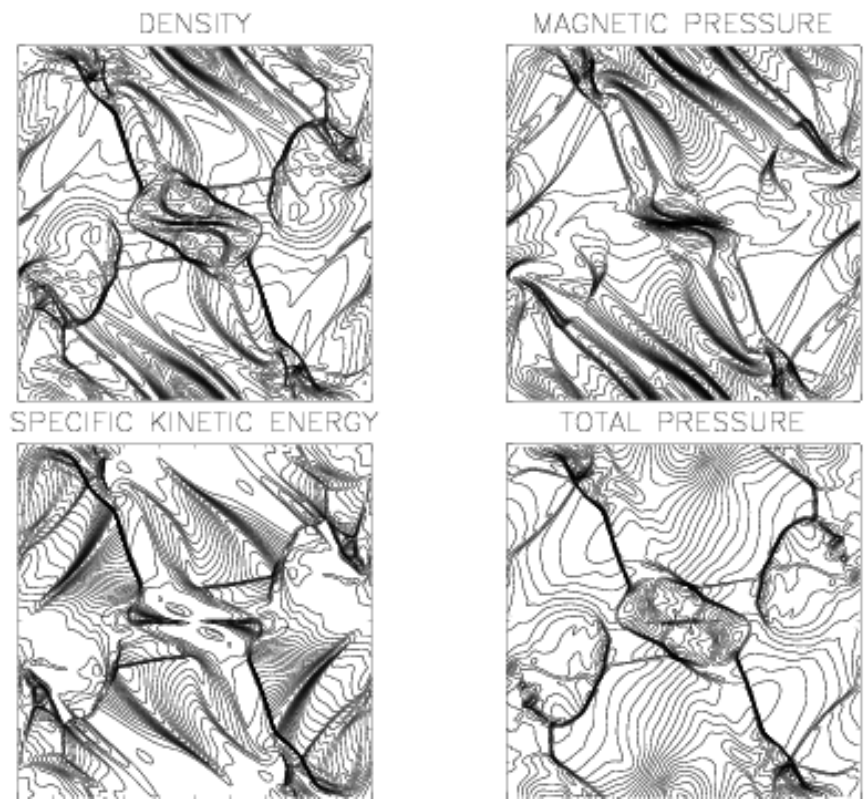}
\figcaption
{Contours of selected variables at $t_{f}=1/2$ in the isothermal Orszag-Tang
vortex test, computed using a grid of $192 \times 192$ cells, third-order
reconstruction, and Roe fluxes.  Thirty equally spaced contours between
the minimum and maximum are used for each plot.}
\end{figure}
\clearpage

\begin{figure}
\epsscale{0.8}
\plotone{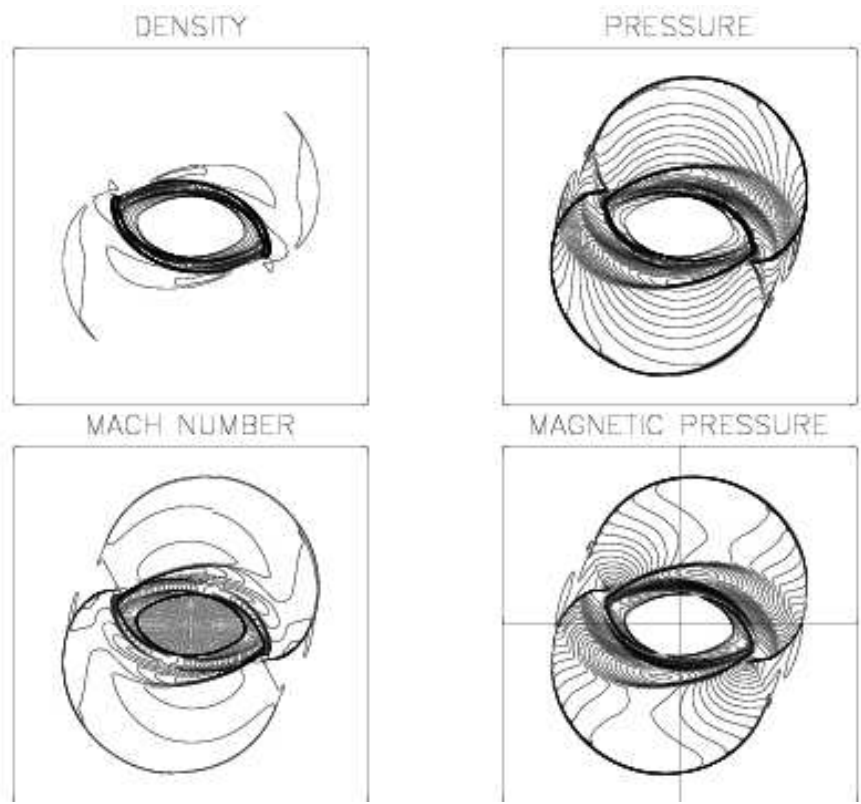}
\figcaption
{Contours of selected variables at $t_{f}=0.15$ in the adiabatic rotor
test, computed using a grid of $400 \times 400$ cells, third-order
reconstruction, and Roe fluxes.  Thirty equally spaced contours between
the minimum and maximum are used for each plot.  The horizontal and
vertical lines in the panel showing magnetic pressure correspond to the
locations of the slices shown in figure 26.}
\end{figure}
\clearpage

\begin{figure}
\epsscale{0.8}
\plotone{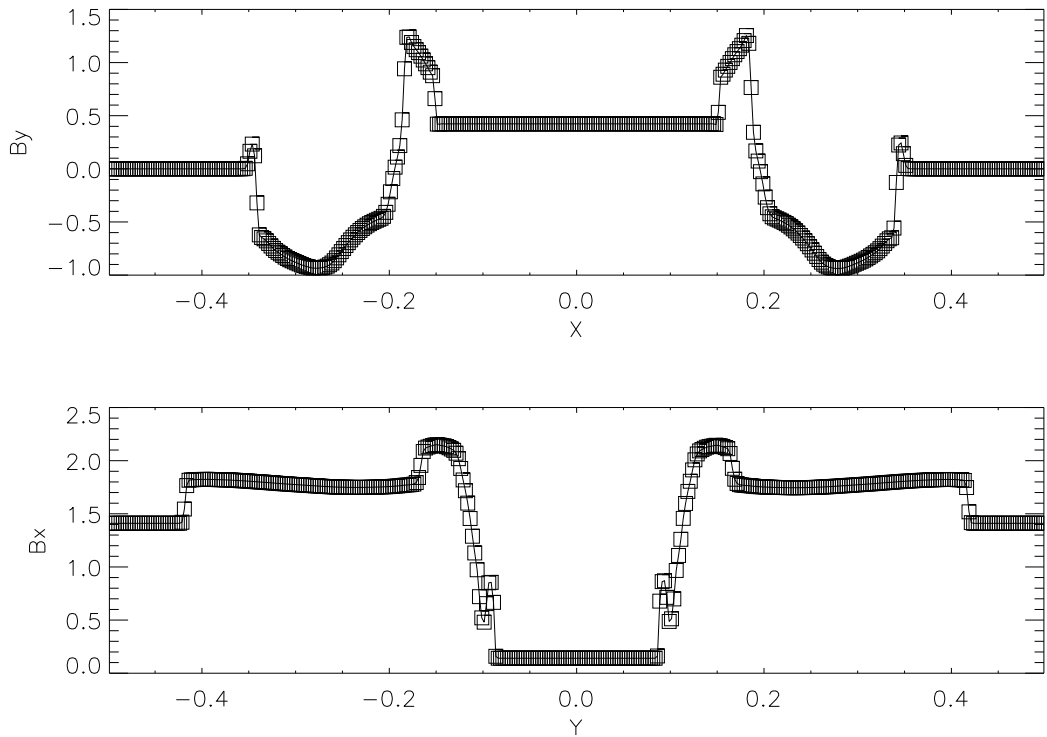}
\figcaption
{Horizontal slice of $B_{y}$ taken at $y=0$ ({\em top}), and
vertical slice of $B_{x}$ taken at $x=0$ ({\em bottom}) at $t_{f}=0.15$ in the
rotor test.  The solid line is the same data as the squares.}
\end{figure}
\clearpage

\begin{figure}
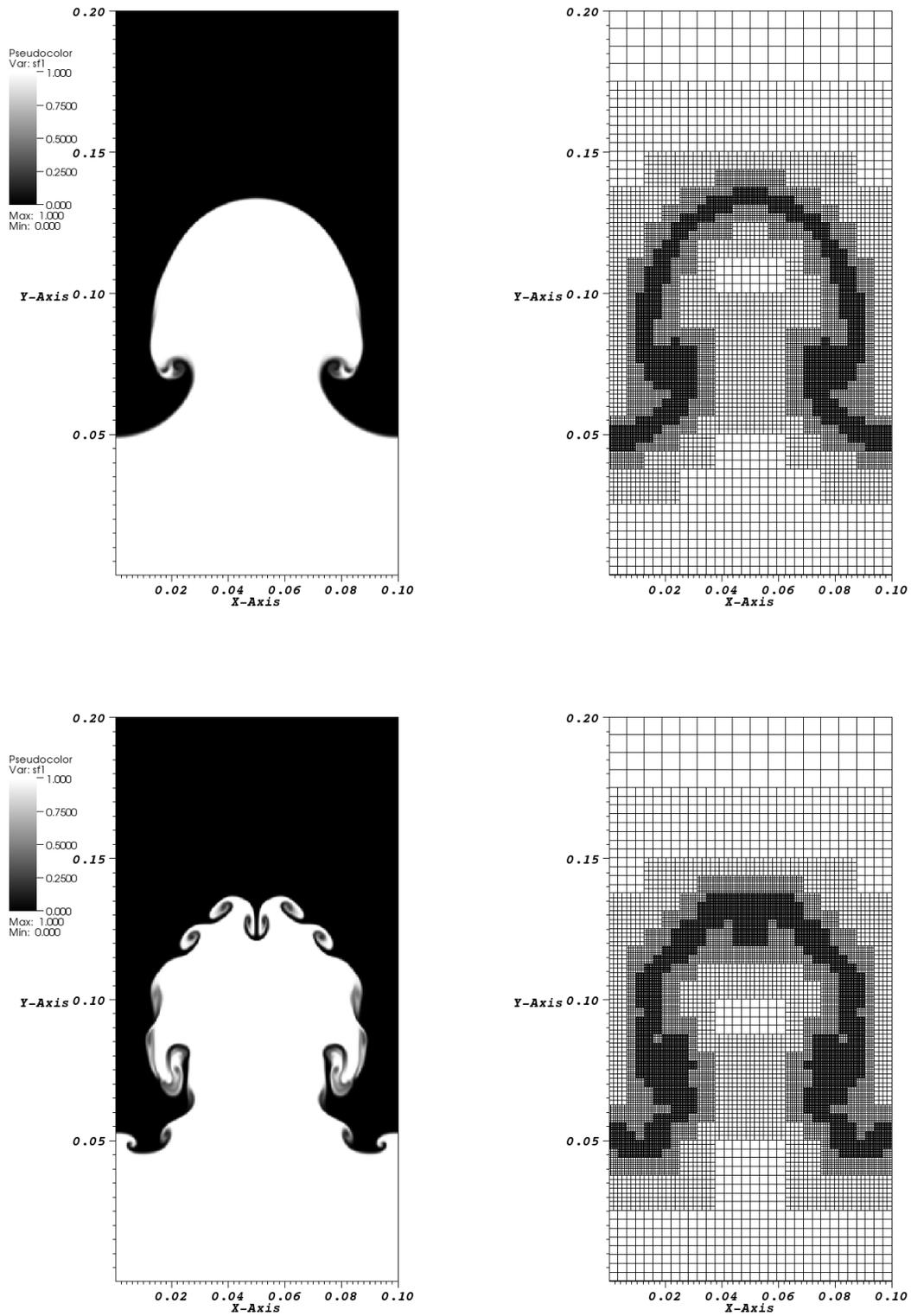

\epsscale{0.4}
\plotone{fig27a.ps3}
\plotone{fig27b.ps3} \\
\plotone{fig27c.ps3}
\plotone{fig27d.ps3}
\figcaption
{{\em (Left)} Grayscale image of the concentration of a passively-advected
contaminant at late time in the magnetic Rayleigh-Taylor instability.
{\em (Right)} Grid blocks used to resolve the interface using AMR.
The bottom row shows the same quantities, but for a calculation in which the
magnetic field strength is zero (i.e., hydrodynamics).}
\end{figure}

\begin{figure}
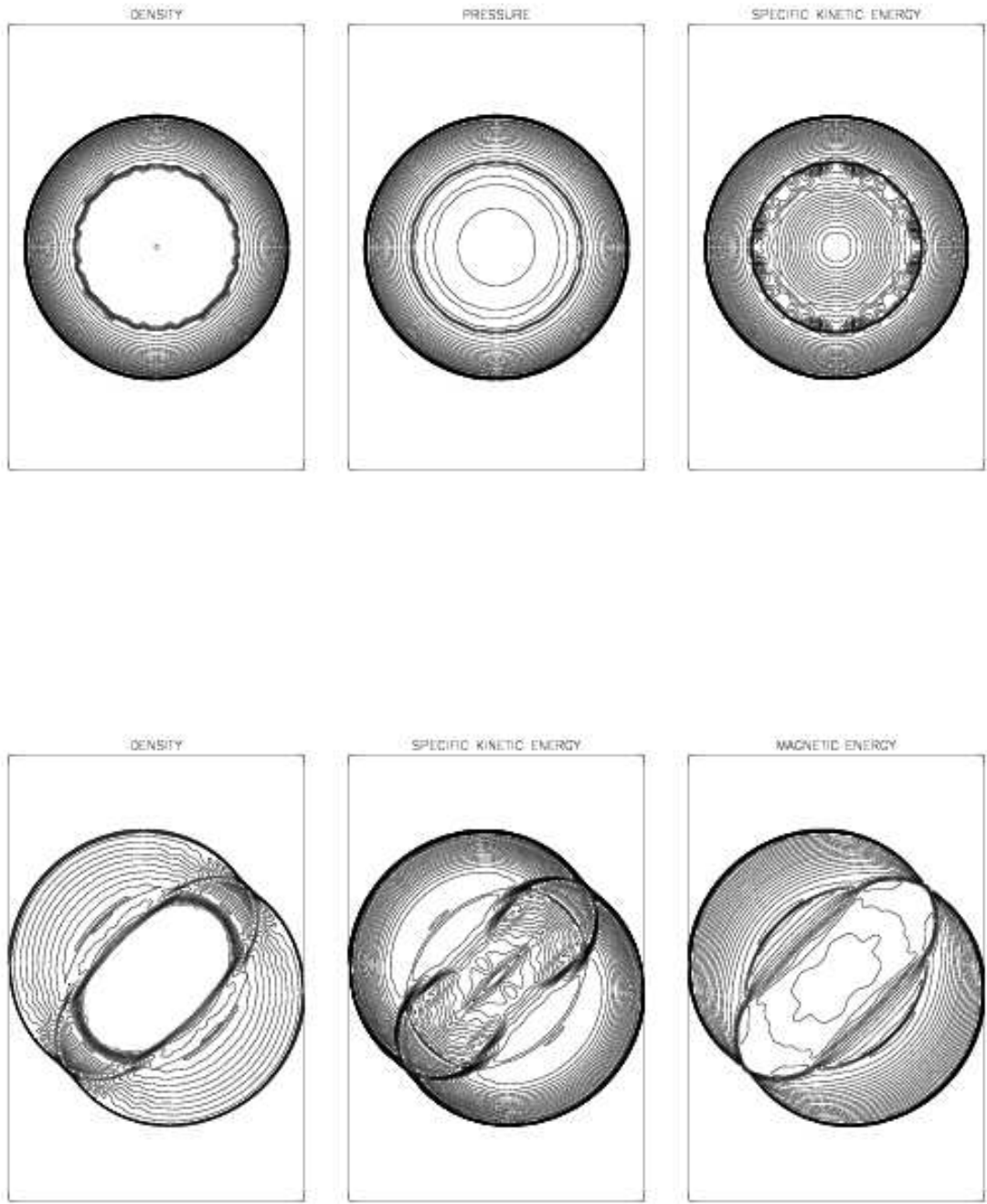

\epsscale{0.8}
\plotone{fig28a.ps3}
\plotone{fig28b.ps3}
\figcaption
{Contours of selected variables at $t_{f}=0.2$ in the adiabatic blast
wave test, computed using a grid of $200 \times 300$ cells, third-order
reconstruction, and either HLLC (hydrodynamics, top row) or HLLD (MHD
with initial $B_{0}=1$, bottom row) fluxes.
Thirty equally spaced contours between the minimum
and maximum are used for each plot.}
\end{figure}

\begin{figure}
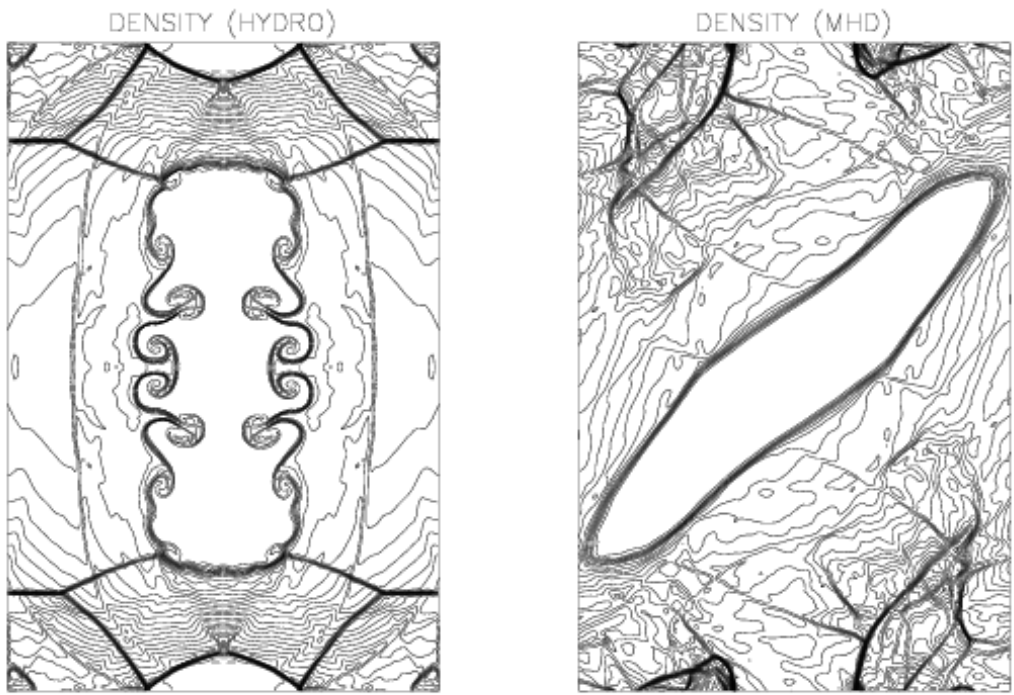

\epsscale{0.4}
\plotone{fig29a.ps3}
\plotone{fig29b.ps3}
\figcaption
{Contours of the density at $t_{f}=1$ in the hydrodynamic (left) and
MHD (right) adiabatic blast test.  Fifty equally spaced contours between
the minimum and maximum are used for each plot.}
\end{figure}
\clearpage

\begin{figure}
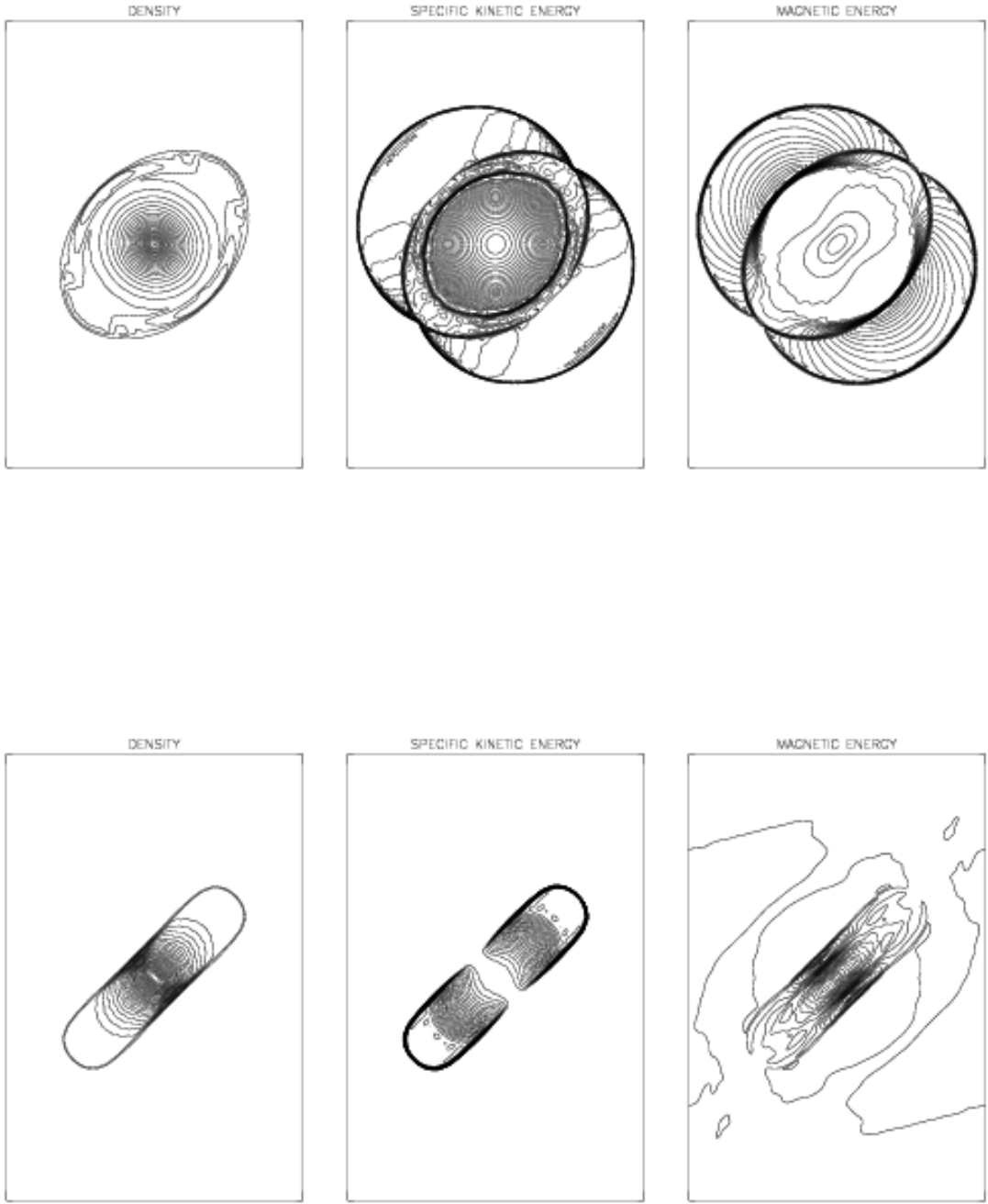

\epsscale{0.8}
\plotone{fig30a.ps3}
\plotone{fig30b.ps3}
\figcaption
{Contours of selected variables at $t_{f}=0.2$ in the isothermal blast
wave test, computed using a grid of $200 \times 300$ cells, third-order
reconstruction, and HLLD fluxes.  The top row corresponds to an initial
$B_{0}=1$, while the bottom row uses an initial $B_{0}=10$.
Thirty equally spaced contours between the minimum and maximum are used
for each plot.  Outgoing waves have already crossed and re-entered the domain 
by $t=0.2$ in the strong field case, thus the contours in the ambient medium are
due to interaction of these waves rather than oscillations introduced by the
algorithm.}
\end{figure}

\begin{figure}
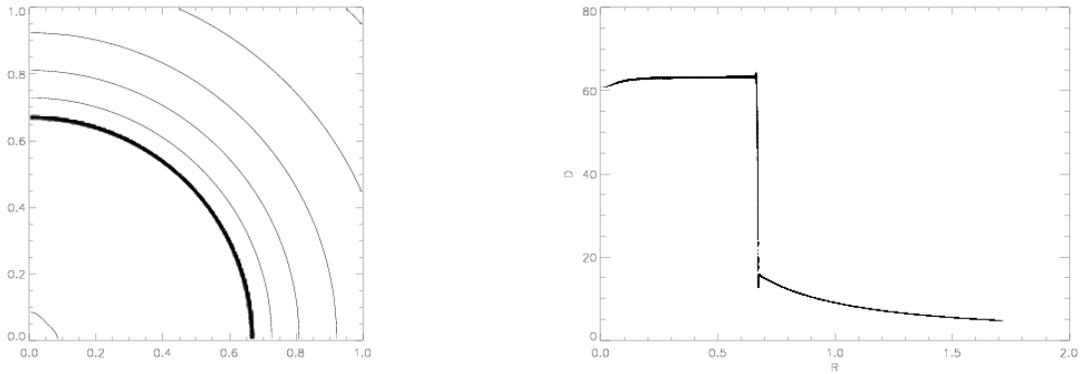

\epsscale{0.4}
\plotone{fig31a.ps3}
\plotone{fig31b.ps3}
\figcaption
{{\em (Left.)} Contours of the density in the spherical hydrodynamical Noh
strong shock test at $t=2$.  Thirty-one equally spaced contours between $\rho=4$ and
64 are shown.  {\em (Right.)} Scatter plot of the density versus spherical
radius at $t=2$.
}
\end{figure}

\begin{figure}
\epsscale{0.8}
\plotone{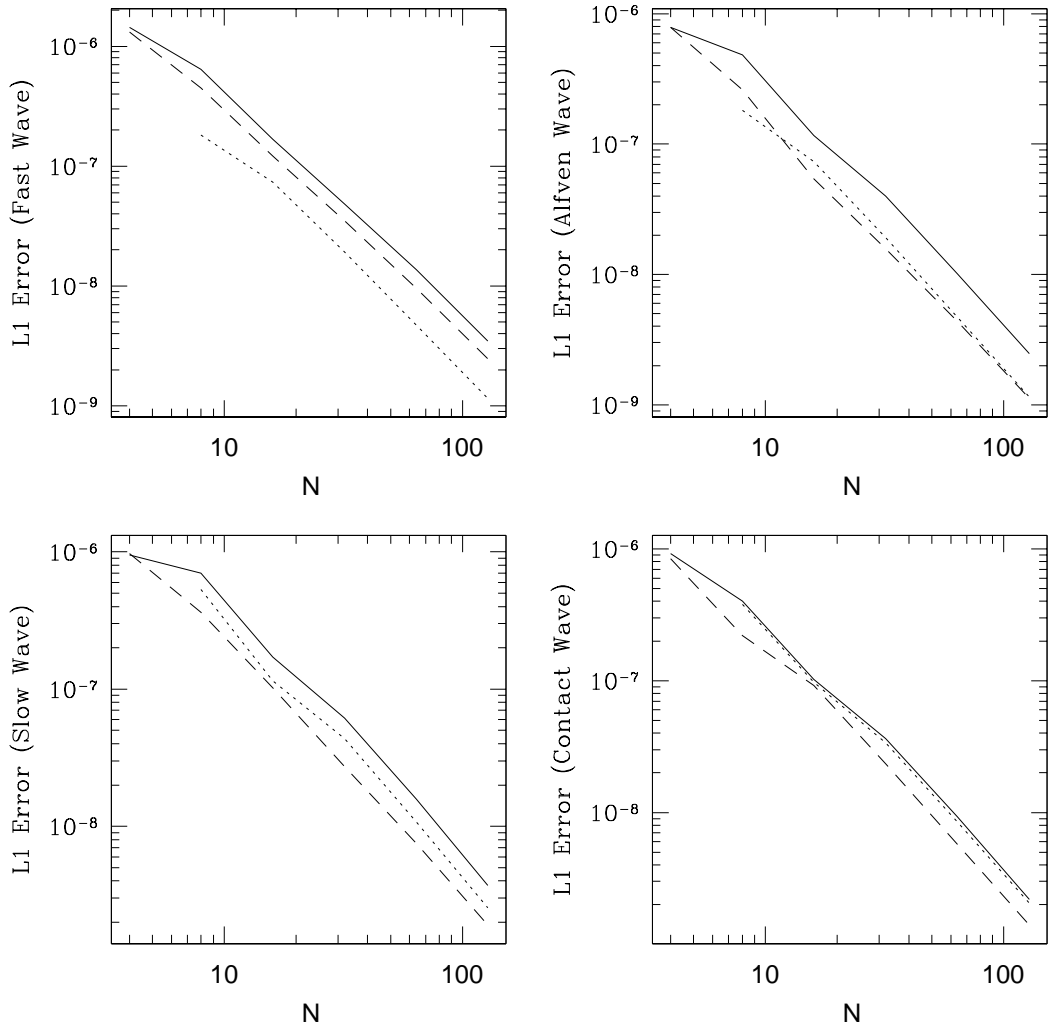}
\figcaption
{Convergence in the norm of the L$_{1}$ error vector for fast, Alfv\'{e}n, slow,
and contact waves after
propagating a distance of one wavelength at an oblique angle across a 3D grid
of size $2N\times N \times N$..
Solutions are computed
using the HLLD fluxes, and either second-order (solid line) or third-order
(dashed line) spatial reconstruction.  The dotted line shows the errors
for second-order spatial reconstruction in 1D for reference.}
\end{figure}

\begin{figure}
\epsscale{0.6}
\plotone{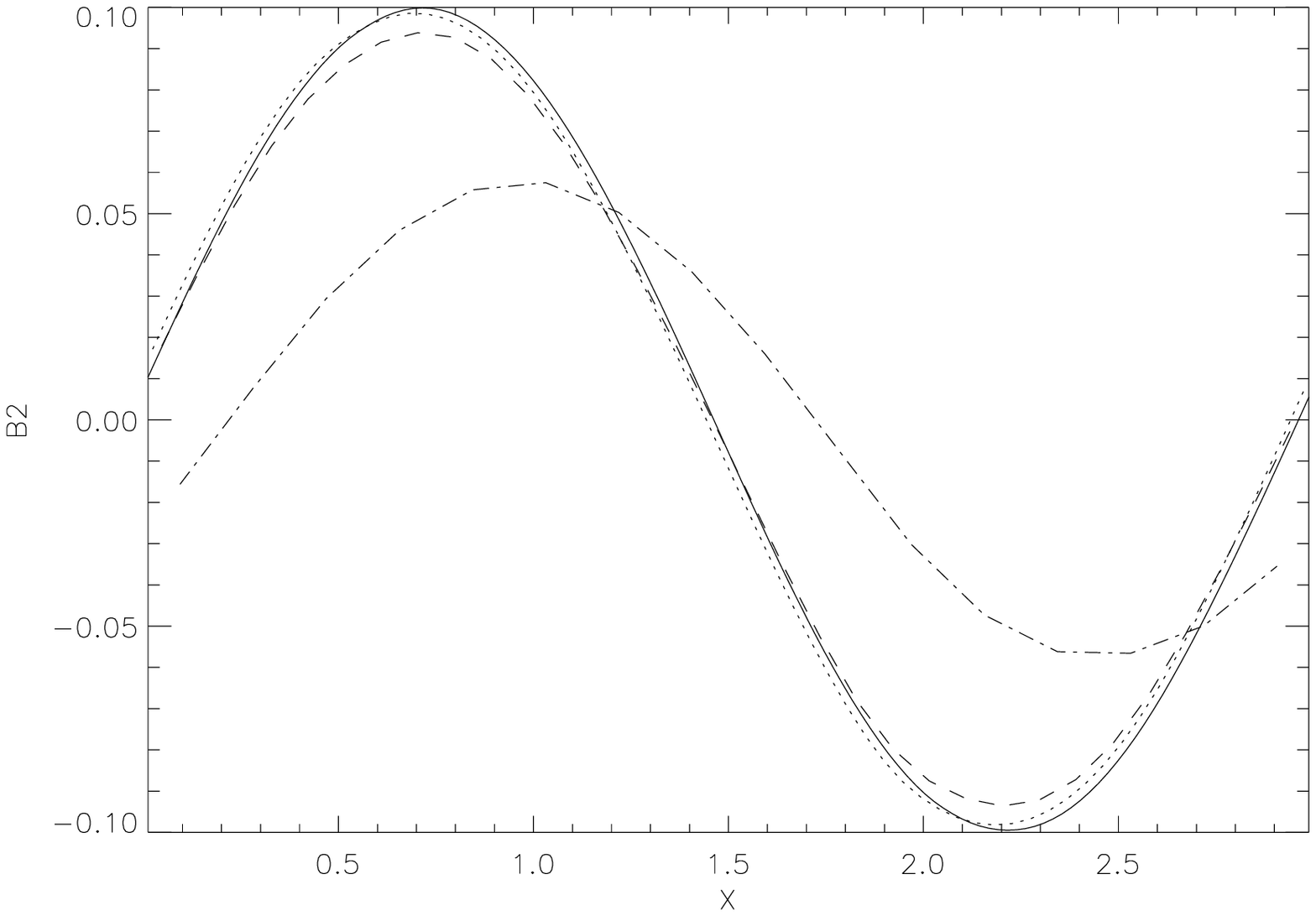}
\plotone{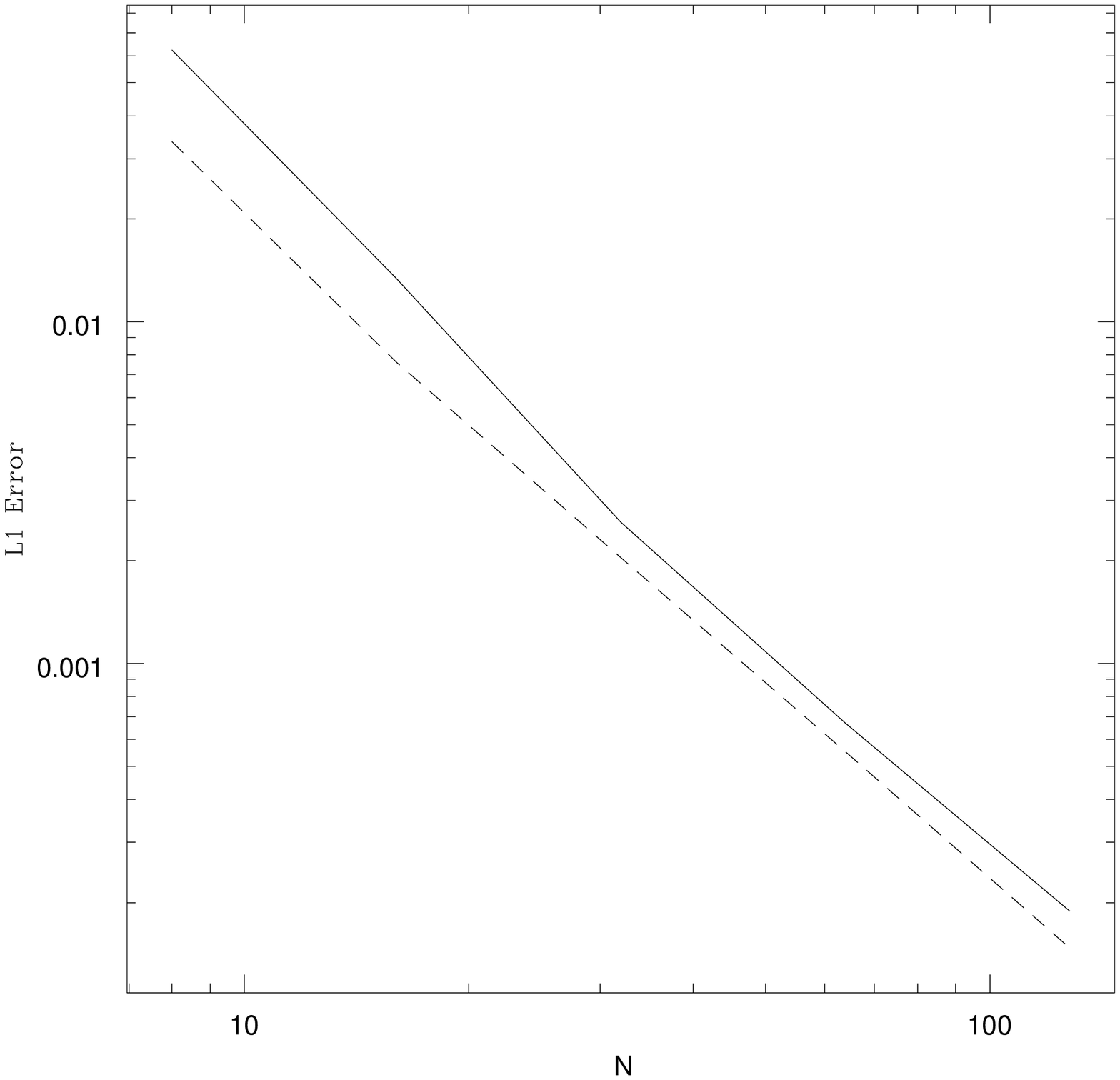}
\figcaption
{({\em Top.})  Profiles of the transverse component of the magnetic field for
traveling circularly polarized
Alfv\'{e}n waves, at a time equal to five wave periods, computed on a grid
with $2N \times N \times N$ cells, where $N=64$ (solid line), 32 (dotted),
16 (dashed line), and 8 (dot-dash line).
Each solution is computed using third order spatial reconstruction
and the HLLD fluxes.
({\em Bottom.})  Convergence of the norm of the L$_{1}$
error vector for traveling circularly polarized
Alfv\'{e}n waves, after propagating a distance equal to one wavelength,
for second-order (solid line) and third-order (dashed line) spatial 
reconstruction.
}
\end{figure}

\begin{figure}
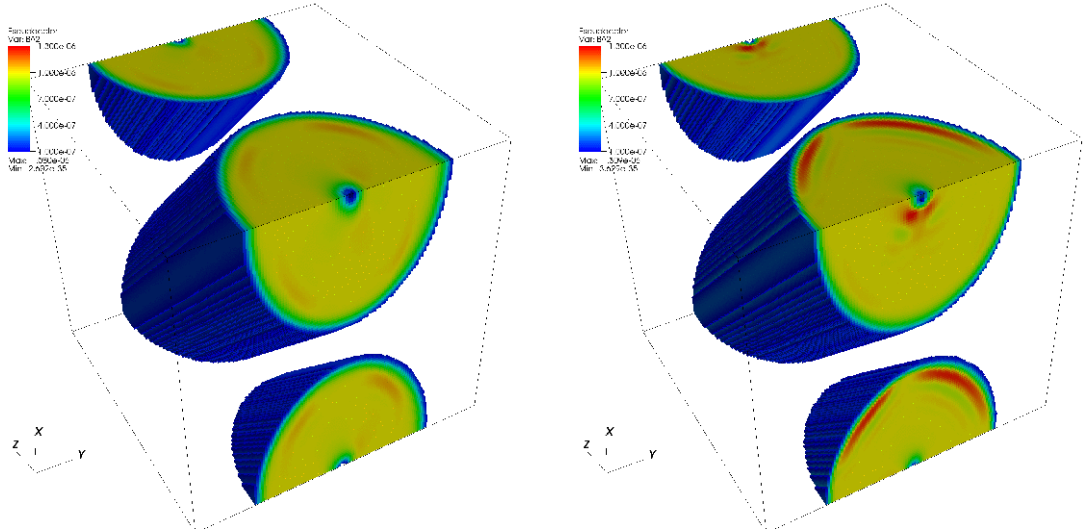

\epsscale{0.4}
\plotone{fig34a.ps3}
\plotone{fig34b.ps3}
\figcaption
{Current density in
an inclined field loop being advected along the diagonal of a 3D grid
at $t_{f}=2$ (after twice around the grid).  The left panel shows the solution
for second-order reconstruction, the right for third-order.}
\end{figure}

\begin{figure}
\epsscale{0.8}
\plotone{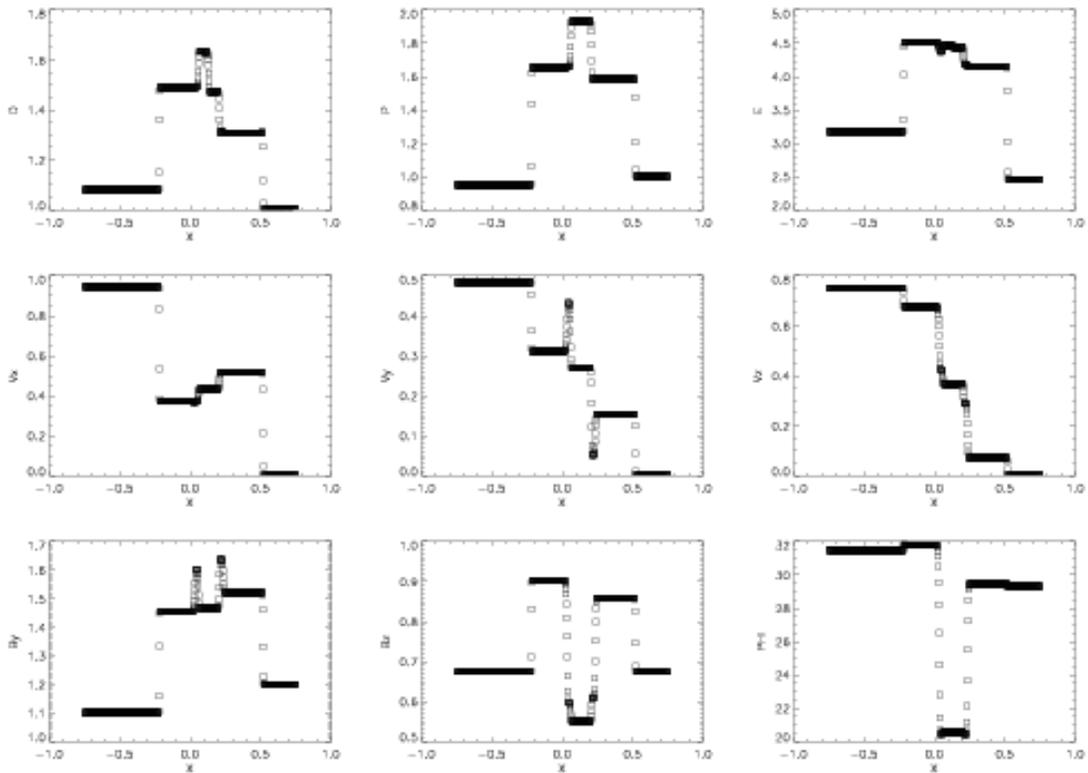}
\figcaption
{Slice through a 3D grid of selected variables for the RJ2a shocktube
initialized with the interface oblique to the grid at $t=0.2$.  This is a fully 3D version
of the 1D test shown in figure 14.}
\end{figure}

\begin{figure}
\epsscale{0.8}
\plotone{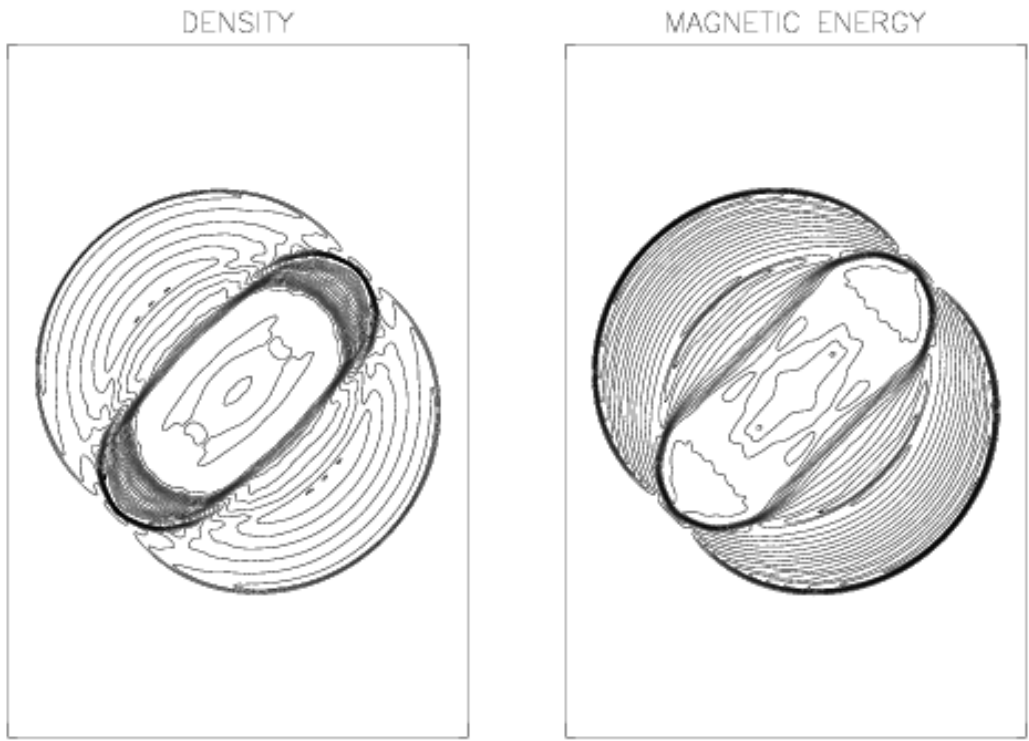}
\figcaption
{Contours of selected variables at $t_{f}=0.2$ in a 2D slice in the $x-y$
plane at $z=0$ (through the center of the grid) in the 3D adiabatic
blast wave test, computed using a grid of $200 \times 300 \times 200$
cells, third-order reconstruction, and the HLLD fluxes.  Thirty equally
spaced contours between the minimum and maximum are used for each plot.}
\end{figure}
\clearpage

\vfill
\newpage
\appendix
\section{Eigensystems in the Primitive Variables}

This appendix gives explicit forms for the eigenvalues and
eigenvectors of the matrix ${\sf A}$ resulting from linearizing the
dynamical equations as ${\bf W}_{,t} = {\sf A}({\bf W}){\bf W}_{,x}$, where
${\bf W}$ is a vector composed of the primitive variables in 1D.  These
eigensystems are needed to convert between the primitive and the characteristic
variables in the reconstruction algorithms described in
\S \ref{sec:reconstruction}.

\subsection{Adiabatic Hydrodynamics}

For adiabatic hydrodynamics, ${\bf W} = (\rho, v_{x}, v_{y}, v_{z}, P)$,
and the matrix ${\sf A}$ is
\begin{equation}
{\sf A} = \left[ \begin{array}{ccccc}
v_{x} & \rho       & 0     & 0     & 0      \\
0     & v_{x}      & 0     & 0     & 1/\rho \\
0     & 0          & v_{x} & 0     & 0      \\
0     & 0          & 0     & v_{x} & 0      \\
0     & \rho a^{2} & 0     & 0     & v_{x}  \end{array} \right] ,
\end{equation}
where $a^{2} = \gamma P/\rho$ ($a$ is the adiabatic sound speed).
The five eigenvalues of this matrix in ascending order are
\begin{equation}
{\bf \lambda} = (v_{x}-a, v_{x}, v_{x}, v_{x}, v_{x}+a).
\end{equation}
The corresponding right-eigenvectors are the columns of the matrix
\begin{equation}
{\sf R}  = \left[ \begin{array}{ccccc}
1       & 1 & 0 & 0 & 1      \\
-a/\rho & 0 & 0 & 0 & a/\rho \\
0       & 0 & 1 & 0 & 0      \\
0       & 0 & 0 & 1 & 0      \\
a^{2}   & 0 & 0 & 0 & a^{2}  \end{array} \right] ,
\end{equation}
while the left-eigenvectors are the rows of the matrix
\begin{equation}
{\sf L}  = \left[ \begin{array}{ccccc}
0 & -\rho/(2a) & 0 & 0 & 1/(2a^{2}) \\
1 &  0       & 0 & 0 & -1/a^{2} \\
0 &  0       & 1 & 0 &  0       \\
0 &  0       & 0 & 1 &  0       \\
0 & \rho/(2a)  & 0 & 0 & 1/(2a^{2}) \end{array} \right] .
\end{equation}

\subsection{Isothermal Hydrodynamics}

For isothermal hydrodynamics, ${\bf W} = (\rho, v_{x}, v_{y}, v_{z})$, and the
matrix ${\sf A}$ is
\begin{equation}
{\sf A} = \left[ \begin{array}{cccc}
v_{x}       & \rho    & 0     & 0      \\
C^{2}/\rho  & v_{x}   & 0     & 0      \\
0           & 0       & v_{x} & 0      \\
0           & 0       & 0     & v_{x}   \end{array} \right] ,
\end{equation}
where $C$ is the isothermal sound speed.
The four eigenvalues of this matrix in ascending order are
\begin{equation}
{\bf \lambda} = (v_{x}-C, v_{x}, v_{x}, v_{x}+C).
\end{equation}
The corresponding right-eigenvectors are the columns of the matrix given
in equation (A3), with the second column and fifth row dropped.  The
left-eigenvectors are the rows of the matrix
\begin{equation}
{\sf L}  = \left[ \begin{array}{cccc}
1/2 & -\rho/(2C) & 0 & 0  \\
0   &  0       & 1 & 0  \\
0   &  0       & 0 & 1  \\
1/2 & \rho/(2C)  & 0 & 0 \end{array} \right] .
\end{equation}

\subsection{Adiabatic Magnetohydrodynamics}

For adiabatic MHD, ${\bf W} = (\rho, v_{x}, v_{y}, v_{z}, P, b_{y}, b_{z})$,
where ${\bf b} = {\bf B}/\sqrt{4\pi}$, and the matrix ${\sf A}$ is
\begin{equation}
{\sf A} = \left[ \begin{array}{cccccccc}
v_x & \rho & 0 & 0 & 0 & 0 & 0 \\
0 & v_x & 0 & 0 & 1/\rho & b_y/\rho & b_z/\rho \\
0 & 0 & v_x & 0 & 0 & -b_x/\rho & 0 \\
0 & 0 & 0 & v_x & 0 & 0 & -b_x/\rho \\
0 & \rho a^{2} & 0 & 0 & v_{x} & 0 & 0 \\
0 & b_y & -b_x & 0 & 0 & v_x & 0 \\
0 & b_z & 0 & -b_x & 0 & 0 & v_x  \end{array} \right] .
\end{equation}
where $a^{2} = \gamma P/\rho$.
The seven eigenvalues of this matrix in ascending order are
\begin{equation}
{\bf \lambda} = (v_{x}-C_{f}, v_{x}-C_{Ax}, v_{x}-C_{s}, v_{x},
v_{x}+C_{s}, v_{x}+C_{Ax}, v_{x}+C_{f})
\end{equation}
where the fast- and slow-magnetosonic wave speeds are given by
\begin{equation}
C_{f,s}^2 = \frac{1}{2} \left( \left[a^{2}+C_{A}^{2}\right] \pm
\sqrt{\left[a^{2}+C_{A}^{2}\right]^{2} - 4 a^2 C_{Ax}^2}\right)
\end{equation}
(with $C_{f}[C_{s}]$ given by the $+[-]$ sign).  The Alfv\'{e}n speeds are
given by
\begin{equation}
C_{A}^{2} = (b_{x}^{2}+b_{y}^{2 }+b_{z}^{2})/\rho, \hspace*{1cm}
C_{Ax}^{2} = b_{x}^{2}/\rho.
\end{equation}
The corresponding right-eigenvectors are the columns of the matrix
\begin{equation}
{\sf R}  = \left[ \begin{array}{ccccccc}
\rho\alpha_{f} & 0 & \rho\alpha_{s} & 1 & \rho\alpha_{s} & 0 & \rho\alpha_{f} \\
-C_{ff} & 0 & -C_{ss} & 0 & C_{ss} & 0 & C_{ff} \\
Q_{s}\beta_{y} & -\beta_{z} & -Q_{f}\beta_{y} & 0 & Q_{f}\beta_{y} & \beta_{z} & -Q_{s}\beta_{y} \\
Q_{s}\beta_{z} & \beta_{y} & -Q_{f}\beta_{z} & 0 & Q_{f}\beta_{z} & -\beta_{y} & -Q_{s}\beta_{z} \\
\rho a^{2}\alpha_{f} & 0 & \rho a^{2}\alpha_{s} & 0 & \rho a^{2}\alpha_{s} & 0 & \rho a^{2}\alpha_{f} \\
A_{s}\beta_{y} & -\beta_{z}S\sqrt{\rho} & -A_{f}\beta_{y} & 0 & -A_{f}\beta_{y} & -\beta_{z}S\sqrt{\rho} & A_{s}\beta_{y} \\
A_{s}\beta_{z} & \beta_{y}S\sqrt{\rho} & -A_{f}\beta_{z} & 0 & -A_{f}\beta_{z} & \beta_{y}S\sqrt{\rho} & A_{s}\beta_{z} \end{array} \right] ,
\end{equation}
where $S={\rm sign}(b_{x})$, and
\begin{equation}
C_{ff} = C_{f}\alpha_{f}, \hspace*{1cm}
C_{ss} = C_{s}\alpha_{s},
\end{equation}
\begin{equation}
Q_{f} = C_{f}\alpha_{f}S, \hspace*{1cm}
Q_{s} = C_{s}\alpha_{s}S,
\end{equation}
\begin{equation}
A_{f} = a\alpha_{f}\sqrt{\rho}, \hspace*{1cm}
A_{s} = a\alpha_{s}\sqrt{\rho},
\end{equation}
\begin{equation}
\alpha_{f}^{2} = \frac{a^{2} - C_{s}^{2}}{C_{f}^{2} - C_{s}^{2}}, \hspace*{1cm}
\alpha_{s}^{2} = \frac{C_{f}^{2} - a^{2}}{C_{f}^{2} - C_{s}^{2}},
\end{equation}
\begin{equation}
\beta_{y} = \frac{b_{y}}{\sqrt{b_{y}^{2} + b_{z}^{2}}}, \hspace*{1cm}
\beta_{z} = \frac{b_{z}}{\sqrt{b_{y}^{2} + b_{z}^{2}}}.
\end{equation}
In the degenerate case where $C_{A}=C_{Ax}=a$, so that $C_{f}=C_{s}$, then
equation (A16) becomes $\alpha_{f}=1$ and $\alpha_{s}=0$.
The left-eigenvectors are the rows of the matrix
\begin{equation}
{\sf L}  = \left[ \begin{array}{ccccccc}
0 & -N_{f}C_{ff} & N_{f}Q_{s}\beta_{y} & N_{f}Q_{s}\beta_{z} & N_{f}\alpha_{f}/\rho & N_{f}A_{s}\beta_{y}/\rho & N_{f}A_{s}\beta_{z}/\rho \\
0 & 0 & -\beta_{z}/2 & \beta_{y}/2 & 0 & -\beta_{z}S/(2\sqrt{\rho}) & \beta_{y}S/(2\sqrt{\rho}) \\
0 & -N_{s}C_{ss} & -N_{s}Q_{f}\beta_{y} & -N_{s}Q_{f}\beta_{z} & N_{s}\alpha_{s}/\rho & -N_{s}A_{f}\beta_{y}/\rho & -N_{s}A_{f}\beta_{z}/\rho \\
1 & 0 & 0 & 0 & -1/a^{2} & 0 & 0 \\
0 & N_{s}C_{ss} & N_{s}Q_{f}\beta_{y} & N_{s}Q_{f}\beta_{z} & N_{s}\alpha_{s}/\rho & -N_{s}A_{f}\beta_{y}/\rho & -N_{s}A_{f}\beta_{z}/\rho \\
0 & 0 & \beta_{z}/2 & -\beta_{y}/2 & 0 & -\beta_{z}S/(2\sqrt{\rho}) & \beta_{y}S/(2\sqrt{\rho}) \\
0 & N_{f}C_{ff} & -N_{f}Q_{s}\beta_{y} & -N_{f}Q_{s}\beta_{z} & N_{f}\alpha_{f}/\rho & N_{f}A_{s}\beta_{y}/\rho & N_{f}A_{s}\beta_{z}/\rho \end{array} \right] ,
\end{equation}
where
\begin{equation}
N_{f} = N_{s} = \frac{1}{2a^{2}}
\end{equation}
are normalization factors for the eigenvectors corresponding to the fast-
and slow-magnetosonic waves respectively.

\subsection{Isothermal Magnetohydrodynamics}

For isothermal MHD, ${\bf W} = (\rho, v_{x}, v_{y}, v_{z}, b_{y}, b_{z})$,
where ${\bf b} = {\bf B}/\sqrt{4\pi}$, and the matrix ${\sf A}$ is
\begin{equation}
{\sf A} = \left[ \begin{array}{ccccccc}
v_x & \rho & 0 & 0 & 0 & 0 \\
C^{2}/\rho & v_x & 0 & 0 & b_y/\rho & b_z/\rho \\
0 & 0 & v_x & 0 & -b_x/\rho & 0 \\
0 & 0 & 0 & v_x & 0 & -b_x/\rho \\
0 & b_y & -b_x & 0 & v_x & 0 \\
0 & b_z & 0 & -b_x & 0 & v_x  \end{array} \right] .
\end{equation}
The six eigenvalues of this matrix in ascending order are
\begin{equation}
{\bf \lambda} = (v_{x}-C_{f}, v_{x}-C_{Ax}, v_{x}-C_{s}, 
v_{x}+C_{s}, v_{x}+C_{Ax}, v_{x}+C_{f}), 
\end{equation}
where the fast and slow-magnetosonic wave speeds are given by equation
(A10) (with $a$ replaced by the isothermal sound speed $C$ here and throughout),
and the Alfv\'{e}n speeds are given by equation (A11).  The
corresponding right-eigenvectors are the columns of the matrix given in
equation (A10), with the fifth row and fourth column dropped.  The
left-eigenvectors are the rows of the matrix
\begin{equation}
{\sf L}  = \left[ \begin{array}{cccccc}
N_{f}\alpha_{f}C^{2}/\rho & -N_{f}C_{ff} & N_{f}Q_{s}\beta_{y} & N_{f}Q_{s}\beta_{z} & N_{f}A_{s}\beta_{y}/\rho & N_{f}A_{s}\beta_{z}/\rho \\
0 & 0 & -\beta_{z}/2 & \beta_{y}/2 & -\beta_{z}S/(2\sqrt{\rho}) & \beta_{y}S/(2\sqrt{\rho}) \\
N_{s}\alpha_{s}C^{2}/\rho & -N_{s}C_{ss} & -N_{s}Q_{f}\beta_{y} & -N_{s}Q_{f}\beta_{z} & -N_{s}A_{f}\beta_{y}/\rho & -N_{s}A_{f}\beta_{z}/\rho \\
N_{s}\alpha_{s}C^{2}/\rho & N_{s}C_{ss} & N_{s}Q_{f}\beta_{y} & N_{s}Q_{f}\beta_{z} & -N_{s}A_{f}\beta_{y}/\rho & -N_{s}A_{f}\beta_{z}/\rho \\
0 & 0 & \beta_{z}/2 & -\beta_{y}/2 & -\beta_{z}S/(2\sqrt{\rho}) & \beta_{y}S/(2\sqrt{\rho}) \\
N_{f}\alpha_{f}C^{2}/\rho & N_{f}C_{ff} & -N_{f}Q_{s}\beta_{y} & -N_{f}Q_{s}\beta_{z} & N_{f}A_{s}\beta_{y}/\rho & N_{f}A_{s}\beta_{z}/\rho \end{array} \right] ,\end{equation}
where
\begin{equation}
N_{f} = N_{s} = \frac{1}{2C^{2}}
\end{equation}
are normalization factors for the eigenvectors corresponding to the fast-
and slow-magnetosonic waves respectively.


\setcounter{equation}{0}
\section{Eigensystems in the Conserved Variables}

This appendix gives explicit forms for the eigenvalues and
eigenvectors of the matrix ${\sf A}$ resulting from linearizing the
dynamical equations as ${\bf U}_{,t} = {\sf A}{\bf U}_{,x}$, where
${\bf U}$ is a vector composed of the conserved variables.  These
eigensystems are needed to construct the fluxes of the conserved
quantities using Roe's method (see \S \ref{sec:Roe}).

\subsection{Adiabatic Hydrodynamics}

For adiabatic hydrodynamics, ${\bf U} = (\rho, \rho v_{x}, \rho v_{y},
\rho v_{z}, E)$, and the matrix ${\sf A}$ is
\begin{equation}
{\sf A}  = \left[ \begin{array}{ccccc}
0 & 1 & 0 & 0 & 0 \\
-v_{x}^{2} + \gamma^{\prime}v^{2}/2 & -(\gamma -3)v_{x} & -\gamma^{\prime}v_{y} & -\gamma^{\prime}v_{z} & \gamma^{\prime} \\
-v_{x}v_{y} & v_{y} & v_{x} & 0 & 0 \\
-v_{x}v_{z} & v_{z} & 0 & v_{x} & 0 \\
-v_{x}H + \gamma^{\prime}v_{x}v^{2}/2 & -\gamma^{\prime}v_{x}^{2} + H & -\gamma^{\prime}v_{x}v_{y} & -\gamma^{\prime}v_{x}v_{z} & \gamma v_{x} \end{array} \right]
\end{equation}
where the enthalpy $H = (E+P)/\rho$, $v^{2} = {\bf v}\cdot{\bf v}$, and
$\gamma^{\prime} = (\gamma -1)$.  The five eigenvalues of this matrix
in ascending order are
\begin{equation}
{\bf \lambda} = (v_{x}-a, v_{x}, v_{x}, v_{x}, v_{x}+a),
\end{equation}
where $a^{2} = (\gamma-1)(H-v^{2}/2) = \gamma P/\rho$ ($a$ is the
adiabatic sound speed).  The corresponding right-eigenvectors are the
columns of the matrix
\begin{equation}
{\sf R}  = \left[ \begin{array}{ccccc}
1         & 0     & 0     & 1       & 1         \\ 
v_{x} - a & 0     & 0     & v_{x}   & v_{x} + a \\
v_{y}     & 1     & 0     & v_{y}   & v_{y}     \\
v_{z}     & 0     & 1     & v_{z}   & v_{z}     \\
H-v_{x}a  & v_{y} & v_{z} & v^{2}/2 & H+v_{x}a  \end{array} \right] ,
\end{equation}
The left-eigenvectors are the rows of the matrix
\begin{equation}
{\sf L} =  \left[ \begin{array}{ccccc}
N_{a}(\gamma^{\prime}v^{2}/2 + v_{x}a) & -N_{a}(\gamma^{\prime}v_{x} + a) & -N_{a}\gamma^{\prime}v_{y} & -N_{a}\gamma^{\prime}v_{z} & N_{a}\gamma^{\prime} \\
-v_{y} & 0 & 1 & 0 & 0 \\
-v_{z} & 0 & 0 & 1 & 0 \\
1-N_{a}\gamma^{\prime}v^{2} & \gamma^{\prime}v_{x}/a^{2} & \gamma^{\prime}v_{y}/a^{2} & \gamma^{\prime}v_{z}/a^{2} & -\gamma^{\prime}/a^{2} \\
N_{a}(\gamma^{\prime}v^{2}/2 - v_{x}a) & -N_{a}(\gamma^{\prime}v_{x} - a) & -N_{a}\gamma^{\prime}v_{y} & -N_{a}\gamma^{\prime}v_{z} & N_{a}\gamma^{\prime}
\end{array} \right] ,
\end{equation}
where $N_{a} = 1/(2a^{2})$.  These are identical to those given by Roe (1981),
except the second and third eigenvectors (corresponding to the transport of
shear motion) have been rescaled to avoid singularities.

\subsection{Isothermal Hydrodynamics}

For isothermal hydrodynamics, ${\bf U} = (\rho, \rho v_{x}, \rho v_{y},
\rho v_{z})$, and the matrix ${\sf A}$ is
\begin{equation}
{\sf A}  = \left[ \begin{array}{cccc}
0 & 1 & 0 & 0 \\
-v_{x}^{2} + C^{2} & 2v_{x} & 0 & 0 \\
-v_{x}v_{y} & v_{y} & v_{x} & 0 \\
-v_{x}v_{z} & v_{z} & 0 & v_{x} \end{array} \right]
\end{equation}
where $C$ is the isothermal sound speed.
The four eigenvalues of this matrix in ascending order are
\begin{equation}
{\bf \lambda} = (v_{x}-C, v_{x}, v_{x}, v_{x}+C).
\end{equation}
The corresponding right-eigenvectors are the columns of the matrix given
in equation (B3) with the fifth row and fourth column dropped, and $a$ replaced
by $C$ throughout.  The left-eigenvectors are the rows of the matrix
\begin{equation}
{\sf L}  = \left[ \begin{array}{cccc}
(1 + v_{x}/C)/2 & -1/(2C) & 0 & 0 \\
-v_{y}          &  0    & 1 & 0 \\
-v_{z}          &  0    & 0 & 1 \\
(1 - v_{x}/C)/2 & 1/(2C)  & 0 & 0 \end{array} \right] .
\end{equation}

\subsection{Adiabatic Magnetohydrodynamics}

For adiabatic MHD, ${\bf U} = (\rho, \rho v_{x}, \rho v_{y},
\rho v_{z}, E, b_{y}, b_{z})$, where ${\bf b} = {\bf B}/\sqrt{4\pi}$,
and the matrix ${\sf A}$ is
\begin{equation}
{\sf A} = \left[ \begin{array}{ccccccc}
0 & 1 & 0 & 0 & 0 & 0 & 0 \\
-v_{x}^{2} + \gamma^{\prime}v^{2}/2 - X^{\prime} & -(\gamma-3)v_{x} & -\gamma^{\prime}v_{y} & -\gamma^{\prime}v_{z} & \gamma^{\prime} & -b_{y}Y^{\prime} & -b_{z}Y^{\prime} \\
- v_{x} v_{y} & v_{y} & v_{x} & 0 & 0 & -b_{x} & 0 \\
- v_{x} v_{z} & v_{z} & 0 & v_{x} & 0 & 0 & -b_{x} \\
A_{51} & A_{52} & A_{53} & A_{54} & \gamma v_{x} & A_{56} & A_{57} \\
(b_{x}v_{y} - b_{y}v_{x})/\rho & b_{y}/\rho & -b_{x}/\rho & 0 & 0 & v_{x} & 0 \\
(b_{x}v_{z} - b_{z}v_{x})/\rho & b_{z}/\rho & 0 & -b_{x}/\rho & 0 & 0 & v_{x}
\end{array} \right]
\end{equation}
where $v^{2} = {\bf v}\cdot{\bf v}$, and
\begin{equation}
A_{51} = -v_{x}H + \gamma^{\prime}v_{x}v^{2}/2 + b_{x}(b_{x}v_{x} + b_{y}v_{y}
+ b_{z}v_{z})/\rho - v_{x}X^{\prime}
\end{equation}
\begin{equation}
A_{52} = -\gamma^{\prime}v_{x}^{2} + H - b_{x}^{2}/\rho
\end{equation}
\begin{equation}
A_{53} = -\gamma^{\prime}v_{x}v_{y} - b_{x}b_{y}/\rho
\end{equation}
\begin{equation}
A_{54} = -\gamma^{\prime}v_{x}v_{z} - b_{x}b_{z}/\rho
\end{equation}
\begin{equation}
A_{56} = -(b_{x}v_{y} + b_{y}v_{x}Y^{\prime})
\end{equation}
\begin{equation}
A_{57} = -(b_{x}v_{z} + b_{z}v_{x}Y^{\prime}) 
\end{equation}
\begin{equation}
X=\left[(b_{y,L}-b_{y,R})^2+(b_{z,L}-b_{z,R})^2\right]/(2(\sqrt{\rho_L} +
\sqrt{\rho_R}))
\end{equation}
\begin{equation}
Y = \frac{\rho_{L} + \rho_{R}}{2\rho}.
\end{equation}
In these equations
$\gamma^{\prime} = (\gamma - 1)$, $X^{\prime} = (\gamma-2)X$,
$Y^{\prime} = (\gamma-2)Y$, and $H=(E+P+b^{2}/2)/\rho$.
The factors $X$ and $Y$ are introduced by the averaging scheme defined by
equation (56); thus the matrix ${\sf A}$ and its eigenvectors depend
explicitly on our choice of the Roe averaging scheme.
The seven eigenvalues of this matrix in ascending order are
\begin{equation}
{\bf \lambda} = (v_{x}-C_{f}, v_{x}-C_{Ax}, v_{x}-C_{s}, v_{x}
v_{x}+C_{s}, v_{x}+C_{Ax}, v_{x}+C_{f})
\end{equation}
where the fast and slow-magnetosonic wave speeds are given by
\begin{equation}
C_{f,s}^2 = \frac{1}{2} \left( \left[\tilde{a}^{2}+\tilde{C}_{A}^{2}\right] \pm
\sqrt{\left[\tilde{a}^{2}+\tilde{C}_{A}^{2}\right]^{2} - 
4\tilde{a}^2 C_{Ax}^2}\right)
\end{equation}
(with $C_{f}[C_{s}]$ given by the $+[-]$ sign), and
\begin{equation}
\tilde{a}^{2} = \gamma^{\prime}\left( H-v^{2}/2-b^{2}/\rho \right) - X^{\prime}
\end{equation}
\begin{equation}
\tilde{C}_{A}^{2} = C_{Ax}^{2} + b_{\perp}^{\ast 2}/\rho \hspace*{1cm}
C_{Ax}^{2} = b_{x}^{2}/\rho \hspace*{1cm}
b_{\perp}^{\ast 2} = (\gamma^{\prime} - Y^{\prime})(b_{y}^{2} + b_{z}^{2}).
\end{equation}
The corresponding right-eigenvectors are the columns of the matrix
\begin{equation}
{\sf R} = \left[ \begin{array}{ccccccc}
\alpha_{f} & 0 & \alpha_{s} & 1 &\alpha_{s} & 0 & \alpha_{f} \\
V_{xf}-C_{ff} & 0 & V_{xs}-C_{ss} & v_{x} & V_{xs}+C_{ss} & 0 & V_{xf}+C_{ff} \\
V_{yf}+Q_{s}\beta_{y}^{\ast} & -\beta_{z} & V_{ys}-Q_{f}\beta_{y}^{\ast} & v_{y} & V_{ys}+Q_{f}\beta_{y}^{\ast} & \beta_{z} & V_{yf}-Q_{s}\beta_{y}^{\ast} \\
V_{zf}+Q_{s}\beta_{z}^{\ast} & \beta_{y} & V_{zs}-Q_{f}\beta_{z}^{\ast} & v_{z} & V_{zs}+Q_{f}\beta_{z}^{\ast} & -\beta_{y} & V_{zf}-Q_{s}\beta_{z}^{\ast} \\
R_{51} & R_{52} & R_{53} & R_{54} & R_{55} & R_{56} & R_{57} \\ 
A_{s}\beta_{y}^{\ast}/\rho & -\beta_{z}S/\sqrt{\rho} & -A_{f}\beta_{y}^{\ast}/\rho & 0 & -A_{f}\beta_{y}^{\ast}/\rho & -\beta_{z}S/\sqrt{\rho} & A_{s}\beta_{y}^{\ast}/\rho \\
A_{s}\beta_{z}^{\ast}/\rho & \beta_{y}S/\sqrt{\rho} & -A_{f}\beta_{z}^{\ast}/\rho & 0 & -A_{f}\beta_{z}^{\ast}/\rho & \beta_{y}S/\sqrt{\rho} & A_{s}\beta_{z}^{\ast}/\rho \end{array} \right]
\end{equation}
where the $C_{ff,ss}, Q_{f,s}, A_{f,s}, \alpha_{f,s}$ and $\beta_{y,z}$ are
given by equation (A13) to (A17) (with $a$ replaced by $\tilde{a}$), 
$V_{if,s} = v_{i}\alpha_{f,s}$ $(i=x,y,z)$, and
\begin{equation}
R_{51} = \alpha_{f} \left( H^{\prime}- v_{x}C_{f} \right)
+Q_{s}(v_{y}\beta_{y}^{\ast} + v_{z}\beta_{z}^{\ast})
+A_{s}b_{\perp}^{\ast}\beta_{\perp}^{\ast 2}/\rho,
\end{equation}
\begin{equation}
R_{52} = -(v_{y}\beta_{z} - v_{z}\beta_{y}) = -R_{56},
\end{equation}
\begin{equation}
R_{53} = \alpha_{s} \left( H^{\prime} - v_{x}C_{s} \right)
-Q_{f}(v_{y}\beta_{y}^{\ast} + v_{z}\beta_{z}^{\ast})
-A_{f}b_{\perp}^{\ast}\beta_{\perp}^{\ast 2}/\rho,
\end{equation}
\begin{equation}
R_{54} = v^{2}/2 + X^{\prime}/\gamma^{\prime}
\end{equation}
\begin{equation}
R_{55} = \alpha_{s} \left( H^{\prime}+ v_{x}C_{s} \right)
+Q_{f}(v_{y}\beta_{y}^{\ast} + v_{z}\beta_{z}^{\ast})
-A_{f}b_{\perp}^{\ast}\beta_{\perp}^{\ast 2}/\rho,
\end{equation}
\begin{equation}
R_{57} = \alpha_{f} \left( H^{\prime} + v_{x}C_{f} \right)
-Q_{s}(v_{y}\beta_{y}^{\ast} + v_{z}\beta_{z}^{\ast})
+A_{s}b_{\perp}^{\ast}\beta_{\perp}^{\ast 2}/\rho.
\end{equation}
where $H^{\prime} = H-b^{2}/\rho$.  In these equations
\begin{equation}
\beta_{y}^{\ast} = b_{y}/ \vert b_{\perp}^{\ast} \vert , \hspace*{1cm}
\beta_{z}^{\ast} = b_{z}/ \vert b_{\perp}^{\ast} \vert , \hspace*{1cm}
\beta_{\perp}^{\ast 2} = \beta_{y}^{\ast 2} + \beta_{z}^{\ast 2}.
\end{equation}
The left-eigenvectors are the rows of the matrix
\begin{equation}
{\sf L} = \left[ \begin{array}{ccccccc}
L_{11} & -\bar{V}_{xf}-\hat{C}_{ff} & -\bar{V}_{yf} + \hat{Q}_{s}Q_{y}^{\ast} & -\bar{V}_{zf} + \hat{Q}_{s}Q_{z}^{\ast} & \bar{\alpha}_{f} & \hat{A}_{s}{Q}_{y}^{\ast} - \bar{\alpha}_{f}b_{y} & \hat{A}_{s}{Q}_{z}^{\ast} - \bar{\alpha}_{f}b_{z} \\
L_{21} & 0 & -\beta_{z}/2 & \beta_{y}/2 & 0 & -\beta_{z}S\sqrt{\rho}/2 & \beta_{y}S\sqrt{\rho}/2 \\
L_{31} & -\bar{V}_{xs}-\hat{C}_{ss} & -\bar{V}_{ys} - \hat{Q}_{f}Q_{y}^{\ast} & -\bar{V}_{zs} - \hat{Q}_{f}Q_{z}^{\ast} & \bar{\alpha}_{s} & -\hat{A}_{f}{Q}_{y}^{\ast} - \bar{\alpha}_{s}b_{y} & -\hat{A}_{f}{Q}_{z}^{\ast} - \bar{\alpha}_{s}b_{z} \\
L_{41} & 2\bar{v}_{x} & 2\bar{v}_{y} & 2\bar{v}_{z} & -\gamma^{\prime}/a^{2} & 2\bar{b}_{y} & 2\bar{b}_{z} \\
L_{51} & -\bar{V}_{xs}+\hat{C}_{ss} & -\bar{V}_{ys} + \hat{Q}_{f}Q_{y}^{\ast} & -\bar{V}_{zs} + \hat{Q}_{f}Q_{z}^{\ast} & \bar{\alpha}_{s} & -\hat{A}_{f}{Q}_{y}^{\ast} - \bar{\alpha}_{s}b_{y} & -\hat{A}_{f}{Q}_{z}^{\ast} - \bar{\alpha}_{s}b_{z} \\
L_{61} & 0 & \beta_{z}/2 & -\beta_{y}/2 & 0 & -\beta_{z}S\sqrt{\rho}/2 & \beta_{y}S\sqrt{\rho}/2 \\
L_{71} & -\bar{V}_{xf}+\hat{C}_{ff} & -\bar{V}_{yf} - \hat{Q}_{s}Q_{y}^{\ast} & -\bar{V}_{zf} - \hat{Q}_{s}Q_{z}^{\ast} & \bar{\alpha}_{f} & \hat{A}_{s}{Q}_{y}^{\ast} - \bar{\alpha}_{f}b_{y} & \hat{A}_{s}{Q}_{z}^{\ast} - \bar{\alpha}_{f}b_{z}
\end{array} \right]
\end{equation}
where a symbol over the quantity $q$ denotes normalization via
$\bar{q} = \gamma^{\prime}q/(2a^{2})$ or $\hat{q} = q/(2a^{2})$.  In addition,
\begin{equation}
Q_{y}^{\ast} = \beta_{y}^{\ast}/\beta_{\perp}^{\ast 2}, \hspace*{1cm}
Q_{z}^{\ast} = \beta_{z}^{\ast}/\beta_{\perp}^{\ast 2},
\end{equation}
and
\begin{equation}
L_{11} = \bar{\alpha}_{f}(v^{2} - H^{\prime}) + \hat{C}_{ff}(C_{f}+v_{x})
- \hat{Q}_{s}(v_{y}Q_{y}^{\ast} + v_{z}Q_{z}^{\ast})
- \hat{A}_{s} \vert b_{\perp} \vert /\rho,
\end{equation}
\begin{equation}
L_{21} = (v_{y}\beta_{z} - v_{z}\beta_{y})/2 = -L_{61}
\end{equation}
\begin{equation}
L_{31} = \bar{\alpha}_{s}(v^{2} - H^{\prime}) + \hat{C}_{ss}(C_{s}+v_{x})
+ \hat{Q}_{f}(v_{y}Q_{y}^{\ast} + v_{z}Q_{z}^{\ast})
+ \hat{A}_{f} \vert b_{\perp} \vert /\rho,
\end{equation}
\begin{equation}
L_{41} = 1 - \bar{v}^{2} + 2\hat{X}^{\prime}
\end{equation}
\begin{equation}
L_{51} = \bar{\alpha}_{s}(v^{2} - H^{\prime}) + \hat{C}_{ss}(C_{s}-v_{x})
- \hat{Q}_{f}(v_{y}Q_{y}^{\ast} + v_{z}Q_{z}^{\ast})
+ \hat{A}_{f} \vert b_{\perp} \vert /\rho,
\end{equation}
\begin{equation}
L_{71} = \bar{\alpha}_{f}(v^{2} - H^{\prime}) + \hat{C}_{ff}(C_{f}-v_{x})
+ \hat{Q}_{s}(v_{y}Q_{y}^{\ast} + v_{z}Q_{z}^{\ast})
- \hat{A}_{s} \vert b_{\perp} \vert /\rho,
\end{equation}

\subsection{Isothermal Magnetohydrodynamics}

For isothermal MHD, ${\bf U} = (\rho, \rho v_{x}, \rho v_{y},
\rho v_{z}, b_{y}, b_{z})$, where ${\bf b} = {\bf B}/\sqrt{4\pi}$,
and the matrix ${\sf A}$ is
\begin{equation}
{\sf A} = \left[ \begin{array}{cccccc}
0 & 1 & 0 & 0 & 0 & 0 \\
- v_{x}^2 + C^2 + X & 2 v_{x} & 0 & 0 & b_{y}Y & b_{z}Y  \\
- v_{x} v_{y} & v_{y} & v_{x} & 0 & -b_{x} & 0 \\
- v_{x} v_{z} & v_{z} & 0 & v_{x} & 0 & -b_{x} \\
(b_{x}v_{y} - b_{y}v_{x})/\rho & b_{y}/\rho & -b_{x}/\rho & 0 & v_{x} & 0 \\
(b_{x}v_{z} - b_{z}v_{x})/\rho & b_{z}/\rho & 0 & -b_{x}/\rho & 0 & v_{x}  
\end{array} \right]
\end{equation}
where $C$ is the isothermal sound speed, and $X$ and $Y$ are given by equations
(B15) and (B16).  The six eigenvalues of this matrix in ascending order are
\begin{equation}
{\bf \lambda} = (v_{x}-C_{f}, v_{x}-C_{Ax}, v_{x}-C_{s},
v_{x}+C_{s}, v_{x}+C_{Ax}, v_{x}+C_{f})
\end{equation}
where the fast- and slow-magnetosonic wave speeds are given by
\begin{equation}
C_{f,s}^2 = \frac{1}{2} \left( \left[\tilde{C}^{2}+\tilde{C}_{A}^{2}\right] \pm
\sqrt{\left[\tilde{C}^{2}+\tilde{C}_{A}^{2}\right]^{2} - 
4\tilde{C}^2 C_{Ax}^2}\right)
\end{equation}
(with $C_{f}[C_{s}]$ given by the $+[-]$ sign), where $\tilde{C}^{2} = C^{2}+X$,
and the Alfv\'{e}n speeds are 
\begin{equation}
\tilde{C}_{A}^{2} = C_{Ax}^{2} + b_{\perp}^{\ast 2}/\rho, \hspace*{1cm}
C_{Ax}^{2} = b_{x}^{2}/\rho, \hspace*{1cm}
b_{\perp}^{\ast 2} = Y(b_{y}^{2} + b_{z}^{2}).
\end{equation}
The corresponding right-eigenvectors are the columns of the matrix given by
equation (B21) with the fifth row and fourth column dropped, and $a$ replaced
by $\tilde{C}$ in the definitions given in equations (A15)-(A16).
The left-eigenvectors are the rows of the matrix
\begin{equation}
{\sf L} = \left[ \begin{array}{cccccc}
L_{11} & -\hat{C}_{ff} & \hat{Q}_{s}Q_{y}^{\ast} & \hat{Q}_{s}Q_{z}^{\ast} & \hat{A}_{s}Q_{y}^{\ast} & \hat{A}_{s}Q_{z}^{\ast} \\
(v_{y}\beta_{z} - v_{z}\beta_{y})/2 & 0 & -\beta_{z}/2 & \beta_{y}/2 & -\beta_{z}S\sqrt{\rho}/2 & \beta_{y}S\sqrt{\rho}/2 \\
L_{31} & -\bar{C}_{ss} & -\bar{Q}_{f}Q_{y}^{\ast} & -\bar{Q}_{f}Q_{z}^{\ast} & -\bar{A}_{f}Q_{y}^{\ast} & -\bar{A}_{f}Q_{z}^{\ast} \\
L_{41} & \bar{C}_{ss} & \bar{Q}_{f}Q_{y}^{\ast} & \bar{Q}_{f}Q_{z}^{\ast} & -\bar{A}_{f}Q_{y}^{\ast} & -\bar{A}_{f}Q_{z}^{\ast} \\
-(v_{y}\beta_{z} - v_{z}\beta_{y})/2 & 0 & \beta_{z}/2 & -\beta_{y}/2 & -\beta_{z}S\sqrt{\rho}/2 & \beta_{y}S\sqrt{\rho}/2 \\
L_{61} & \hat{C}_{ff} & -\hat{Q}_{s}Q_{y}^{\ast} & -\hat{Q}_{s}Q_{z}^{\ast} & \hat{A}_{s}Q_{y}^{\ast} & \hat{A}_{s}Q_{z}^{\ast} \\
\end{array} \right]
\end{equation}
where $C_{ff,ss}, Q_{f,s}$, and $A_{f,s}$ are given by equations (A13)-(A15)
(with $a$ replaced by $C$), $\beta_{y,z}$ are given by equation (A17),
$Q_{y,z}^{\ast}$ are given by equation (B30), and
\begin{equation}
L_{11} = \hat{C}_{ff}(C_{f} + v_{x}) 
- \hat{Q}_{s}(v_{y}Q_{y}^{\ast} + v_{z}Q_{z}^{\ast})
- \hat{A}_{s} \vert b^{\ast}_{\perp} \vert /\rho,
\end{equation}
\begin{equation}
L_{31} = \bar{C}_{ss}(C_{s} + v_{x}) 
+ \bar{Q}_{f}(v_{y}Q_{y}^{\ast} + v_{z}Q_{z}^{\ast})
+ \bar{A}_{f} \vert b^{\ast}_{\perp} \vert /\rho,
\end{equation}
\begin{equation}
L_{41} = \bar{C}_{ss}(C_{s} - v_{x}) 
- \bar{Q}_{f}(v_{y}Q_{y}^{\ast} + v_{z}Q_{z}^{\ast})
+ \bar{A}_{f} \vert b^{\ast}_{\perp} \vert /\rho,
\end{equation}
\begin{equation}
L_{61} = \hat{C}_{ff}(C_{f} - v_{x}) 
+ \hat{Q}_{s}(v_{y}Q_{y}^{\ast} + v_{z}Q_{z}^{\ast})
- \hat{A}_{s} \vert b^{\ast}_{\perp} \vert /\rho.
\end{equation}
In these equations, a symbol over the quantity $q$ denotes normalization via
$\bar{q} = q/(C^{2}[1+\alpha_{f}^{2}])$ and
$\hat{q} = q/(C^{2}[1+\alpha_{s}^{2}])$.


\setcounter{equation}{0}
\section{The H-correction: Fixing the Carbuncle Problem}

For strong, planar shocks in multidimensions propagating along a
grid direction, higher-order Godunov methods can be subject to a
numerical instability (Quirk 1994) that grows into large amplitude
perturbations of the shock front at the grid scale.  This ``carbuncle"
instability can easily be demonstrated with a simple 2D test: a uniform
high Mach number flow in the $+x$-direction is initialized everywhere in
the domain, with inflow boundary conditions on the right boundary, and
reflecting everywhere else.  If zone-to-zone perturbations in the density
with small amplitude ($\delta \rho/\rho = 10^{-4}$) are added, the reflected
shock will develop the carbuncle instability as it propagates to the
left across the grid.  Radiative cooling in the postshock gas can amplify
the effect (Sutherland et al. 2003).

The source of the instability is the use of 1D Riemann solvers to compute
fluxes in a multidimensional flow.  When a planar shock is
grid aligned, there is too little dissipation added to the fluxes in directions
perpendicular to the shock front.  Thus, small amplitude perturbations in the
transverse direction grow, rather than being damped.  The solution is
to identify grid-aligned shocks and add extra dissipation to the transverse
fluxes (e.g. Sutherland et al. 2003).  In Athena, we use one form of the 
{\em ``H-correction"} technique described in Sanders et al. (1998) to
identify shocks and to add the appropriate dissipation.

The H-correction is most easily described when used in combination with
the Roe fluxes.  Consider the calculation of the flux at the interface
located at $(i-1/2,j)$ in 2D.  When the H-correction is used
the absolute value of the eigenvalues $\vert \lambda^{\alpha} \vert$
in the Roe flux formula (equation \ref{eq:roe-flux}) are replaced with 
$\vert {\bar \lambda}^{\alpha} \vert$, where for each component $\alpha$
\begin{equation}
\vert {\bar \lambda}^{\alpha} \vert = {\rm max}(\vert \lambda^{\alpha} \vert, {\bar \eta}_{i-1/2,j}).
\end{equation}
Note the {\tt max} is taken over each $\vert \lambda^{\alpha} \vert$
independently in a pairwise fashion with ${\bar \eta}_{i-1/2,j}$, rather
than over all $\alpha$ eigenvalues at once.  Here, ${\bar \eta}_{i-1/2,j}$
comes from a 2D average using a five-point stencil in the shape of the
letter `H', that is
\begin{equation}
{\bar \eta}_{i-1/2,j} = {\rm max}(\eta_{i-1,j+1/2},\eta_{i-1,j-1/2},\eta_{i-1/2,j},\eta_{i,j+1/2},\eta_{i,j-1/2})
\end{equation}
where $\eta_{i-1/2,j} = \frac{1}{2} \vert (u_{i,j}+C_{f,i,j})
- (u_{i-1,j}-C_{f,i-1,j}) \vert$, $u_{i,j}$ is the component
of the velocity normal to the interface, and $C_{f,i,j}$ is the fast
magnetosonic speed (for MHD) in the direction normal to the interface.
This correction is only added to the final multidimensional fluxes (computed
in step 6 in 2D, and step 7 in 3D).
It only becomes important in shocks, and for grid-aligned
shocks it results in the dissipation in the transverse directions being
comparable to that added in the direction of shock propagation.  In 3D
the H-correction generalizes to a 9-point average (one `H' in each
orthogonal plane).  We find the HLL-type fluxes are less susceptible to the
carbuncle instability, but are still affected by it in some circumstances.
The H-correction can be added to HLL-type solvers by making the
appropriate modification to the
wavespeeds $b^{+}$ and $b^{-}$ defined in equations \ref{eq:HLL+speed} and
\ref{eq:HLL-speed}.

Use of the H-correction is only required for flows with strong,
grid-aligned shocks (for most applications with Athena it is not needed).
The results of the Noh strong shock test described
in \S\ref{sec:3D-hydro-tests} show the H-correction is extremely effective
at eliminating the carbuncle instability.  In fact a variety of forms for
the correction were proposed by Sanders et al. (see their equation
9).  Tests using planar shocks in 2D subject to the carbuncle instability
showed little difference between the formulations suggested by Sanders
et al., thus we have chosen to adopt only the version described above.

\end{document}